\documentclass[11pt]{article}
\usepackage{amsfonts,amssymb,amsmath,amscd,theorem,oldgerm,epsfig,units} 
\usepackage[mathscr]{eucal}
\usepackage{mathrsfs}

\makeatletter

\def\diagram{\m@th\leftwidth=\z@ \rightwidth=\z@ \topheight=\z@
\botheight=\z@ \setbox\@picbox\hbox\bgroup}

\def\enddiagram{\egroup\wd\@picbox\rightwidth\unitlength
\ht\@picbox\topheight\unitlength \dp\@picbox\botheight\unitlength
\hskip\leftwidth\unitlength\box\@picbox}

\def\bfig{\begin{diagram}}
\def\efig{\end{diagram}}
\newcount\wideness \newcount\leftwidth \newcount\rightwidth
\newcount\highness \newcount\topheight \newcount\botheight

\def\ratchet#1#2{\ifnum#1<#2 \global #1=#2 \fi}

\def\putbox(#1,#2)#3{%
\horsize{\wideness}{#3} \divide\wideness by 2
{\advance\wideness by #1 \ratchet{\rightwidth}{\wideness}}
{\advance\wideness by -#1 \ratchet{\leftwidth}{\wideness}}
\vertsize{\highness}{#3} \divide\highness by 2
{\advance\highness by #2 \ratchet{\topheight}{\highness}}
{\advance\highness by -#2 \ratchet{\botheight}{\highness}}
\put(#1,#2){\makebox(0,0){$#3$}}}

\def\putlbox(#1,#2)#3{%
\horsize{\wideness}{#3}
{\advance\wideness by #1 \ratchet{\rightwidth}{\wideness}}
{\ratchet{\leftwidth}{-#1}}
\vertsize{\highness}{#3} \divide\highness by 2
{\advance\highness by #2 \ratchet{\topheight}{\highness}}
{\advance\highness by -#2 \ratchet{\botheight}{\highness}}
\put(#1,#2){\makebox(0,0)[l]{$#3$}}}

\def\putrbox(#1,#2)#3{%
\horsize{\wideness}{#3}
{\ratchet{\rightwidth}{#1}}
{\advance\wideness by -#1 \ratchet{\leftwidth}{\wideness}}
\vertsize{\highness}{#3} \divide\highness by 2
{\advance\highness by #2 \ratchet{\topheight}{\highness}}
{\advance\highness by -#2 \ratchet{\botheight}{\highness}}
\put(#1,#2){\makebox(0,0)[r]{$#3$}}}

\def\adjust[#1]{} 

\newcount \coefa
\newcount \coefb
\newcount \coefc
\newcount\tempcounta
\newcount\tempcountb
\newcount\tempcountc
\newcount\tempcountd
\newcount\xext
\newcount\yext
\newcount\xoff
\newcount\yoff
\newcount\gap%
\newcount\arrowtypea
\newcount\arrowtypeb
\newcount\arrowtypec
\newcount\arrowtyped
\newcount\arrowtypee
\newcount\height
\newcount\width
\newcount\xpos
\newcount\ypos
\newcount\run
\newcount\rise
\newcount\arrowlength
\newcount\halflength
\newcount\arrowtype
\newdimen\tempdimen
\newdimen\xlen
\newdimen\ylen
\newsavebox{\tempboxa}%
\newsavebox{\tempboxb}%
\newsavebox{\tempboxc}%

\newdimen\w@dth

\def\setw@dth#1#2{\setbox\z@\hbox{\m@th$#1$}\w@dth=\wd\z@
\setbox\@ne\hbox{\m@th$#2$}\ifnum\w@dth<\wd\@ne \w@dth=\wd\@ne \fi
\advance\w@dth by 1.2em}

\def\t@^#1_#2{\allowbreak\def\n@one{#1}\def\n@two{#2}\mathrel
{\setw@dth{#1}{#2}
\mathop{\hbox to \w@dth{\rightarrowfill}}\limits
\ifx\n@one\empty\else ^{\box\z@}\fi
\ifx\n@two\empty\else _{\box\@ne}\fi}}
\def\t@@^#1{\@ifnextchar_{\t@^{#1}}{\t@^{#1}_{}}}
\def\to{\@ifnextchar^{\t@@}{\t@@^{}}}

\def\t@left^#1_#2{\def\n@one{#1}\def\n@two{#2}\mathrel{\setw@dth{#1}{#2}
\mathop{\hbox to \w@dth{\leftarrowfill}}\limits
\ifx\n@one\empty\else ^{\box\z@}\fi
\ifx\n@two\empty\else _{\box\@ne}\fi}}
\def\t@@left^#1{\@ifnextchar_{\t@left^{#1}}{\t@left^{#1}_{}}}
\def\toleft{\@ifnextchar^{\t@@left}{\t@@left^{}}}

\def\two@^#1_#2{\allowbreak
\def\n@one{#1}\def\n@two{#2}\mathrel{\setw@dth{#1}{#2}
\mathop{\vcenter{\lineskip\z@\baselineskip\z@
                 \hbox to \w@dth{\rightarrowfill}%
                 \hbox to \w@dth{\rightarrowfill}}%
       }\limits
\ifx\n@one\empty\else ^{\box\z@}\fi
\ifx\n@two\empty\else _{\box\@ne}\fi}}
\def\tw@@^#1{\@ifnextchar _{\two@^{#1}}{\two@^{#1}_{}}}
\def\two{\@ifnextchar ^{\tw@@}{\tw@@^{}}}

\def\tofr@^#1_#2{\def\n@one{#1}\def\n@two{#2}\mathrel{\setw@dth{#1}{#2}
\mathop{\vcenter{\hbox to \w@dth{\rightarrowfill}\kern-1.7ex
                 \hbox to \w@dth{\leftarrowfill}}%
       }\limits
\ifx\n@one\empty\else ^{\box\z@}\fi
\ifx\n@two\empty\else _{\box\@ne}\fi}}
\def\t@fr@^#1{\@ifnextchar_ {\tofr@^{#1}}{\tofr@^{#1}_{}}}
\def\tofro{\@ifnextchar^ {\t@fr@}{\t@fr@^{}}}

\def\mon{\mathop{\m@th\hbox to
      14.6\P@{\lasyb\char'51\hskip-2.1\P@$\arrext$\hss
$\mathord\rightarrow$}}\limits} 
\def\leftmono{\mathrel{\m@th\hbox to
14.6\P@{$\mathord\leftarrow$\hss$\arrext$\hskip-2.1\P@\lasyb\char'50%
}}\limits} 
\mathchardef\arrext="0200       

\def\settypes(#1,#2,#3){\arrowtypea#1 \arrowtypeb#2 \arrowtypec#3}
\def\settoheight#1#2{\setbox\@tempboxa\hbox{#2}#1\ht\@tempboxa\relax}%
\def\settodepth#1#2{\setbox\@tempboxa\hbox{#2}#1\dp\@tempboxa\relax}%
\def\settokens`#1`#2`#3`#4`{%
     \def\tokena{#1}\def\tokenb{#2}\def\tokenc{#3}\def\tokend{#4}}
\def\setsqparms[#1`#2`#3`#4;#5`#6]{%
\arrowtypea #1
\arrowtypeb #2
\arrowtypec #3
\arrowtyped #4
\width #5
\height #6
}
\def\setpos(#1,#2){\xpos=#1 \ypos#2}

\def\settriparms[#1`#2`#3;#4]{\settripairparms[#1`#2`#3`1`1;#4]}%

\def\settripairparms[#1`#2`#3`#4`#5;#6]{%
\arrowtypea #1
\arrowtypeb #2
\arrowtypec #3
\arrowtyped #4
\arrowtypee #5
\width #6
\height #6
}

\def\resetparms{\settripairparms[1`1`1`1`1;500]\width 500}

\resetparms

\def\mvector(#1,#2)#3{
\put(0,0){\vector(#1,#2){#3}}%
\put(0,0){\vector(#1,#2){26}}%
}
\def\evector(#1,#2)#3{{
\arrowlength #3
\put(0,0){\vector(#1,#2){\arrowlength}}%
\advance \arrowlength by-30
\put(0,0){\vector(#1,#2){\arrowlength}}%
}}

\def\horsize#1#2{%
\settowidth{\tempdimen}{$#2$}%
#1=\tempdimen
\divide #1 by\unitlength
}

\def\vertsize#1#2{%
\settoheight{\tempdimen}{$#2$}%
#1=\tempdimen
\settodepth{\tempdimen}{$#2$}%
\advance #1 by\tempdimen
\divide #1 by\unitlength
}

\def\putvector(#1,#2)(#3,#4)#5#6{{%
\ifnum3<\arrowtype
\putdashvector(#1,#2)(#3,#4)#5\arrowtype
\else
\ifnum\arrowtype<-3
\putdashvector(#1,#2)(#3,#4)#5\arrowtype
\else
\xpos=#1
\ypos=#2
\run=#3
\rise=#4
\arrowlength=#5
\ifnum \arrowtype<0
    \ifnum \run=0
        \advance \ypos by-\arrowlength
    \else
        \tempcounta \arrowlength
        \multiply \tempcounta by\rise
        \divide \tempcounta by\run
        \ifnum\run>0
            \advance \xpos by\arrowlength
            \advance \ypos by\tempcounta
        \else
            \advance \xpos by-\arrowlength
            \advance \ypos by-\tempcounta
        \fi
    \fi
    \multiply \arrowtype by-1
    \multiply \rise by-1
    \multiply \run by-1
\fi
\ifcase \arrowtype
\or \put(\xpos,\ypos){\vector(\run,\rise){\arrowlength}}%
\or \put(\xpos,\ypos){\mvector(\run,\rise)\arrowlength}%
\or \put(\xpos,\ypos){\evector(\run,\rise){\arrowlength}}%
\fi\fi\fi
}}

\def\putsplitvector(#1,#2)#3#4{
\xpos #1
\ypos #2
\arrowtype #4
\halflength #3
\arrowlength #3
\gap 140
\advance \halflength by-\gap
\divide \halflength by2
\ifnum\arrowtype>0
   \ifcase \arrowtype
   \or \put(\xpos,\ypos){\line(0,-1){\halflength}}%
       \advance\ypos by-\halflength
       \advance\ypos by-\gap
       \put(\xpos,\ypos){\vector(0,-1){\halflength}}%
   \or \put(\xpos,\ypos){\line(0,-1)\halflength}%
       \put(\xpos,\ypos){\vector(0,-1)3}%
       \advance\ypos by-\halflength
       \advance\ypos by-\gap
       \put(\xpos,\ypos){\vector(0,-1){\halflength}}%
   \or \put(\xpos,\ypos){\line(0,-1)\halflength}%
       \advance\ypos by-\halflength
       \advance\ypos by-\gap
       \put(\xpos,\ypos){\evector(0,-1){\halflength}}%
   \fi
\else \arrowtype=-\arrowtype
   \ifcase\arrowtype
   \or \advance \ypos by-\arrowlength
       \put(\xpos,\ypos){\line(0,1){\halflength}}%
       \advance\ypos by\halflength
       \advance\ypos by\gap
       \put(\xpos,\ypos){\vector(0,1){\halflength}}%
   \or \advance \ypos by-\arrowlength
       \put(\xpos,\ypos){\line(0,1)\halflength}%
       \put(\xpos,\ypos){\vector(0,1)3}%
       \advance\ypos by\halflength
       \advance\ypos by\gap
       \put(\xpos,\ypos){\vector(0,1){\halflength}}%
   \or \advance \ypos by-\arrowlength
       \put(\xpos,\ypos){\line(0,1)\halflength}%
       \advance\ypos by\halflength
       \advance\ypos by\gap
       \put(\xpos,\ypos){\evector(0,1){\halflength}}%
   \fi
\fi
}

\def\putmorphism(#1)(#2,#3)[#4`#5`#6]#7#8#9{{%
\run #2
\rise #3
\ifnum\rise=0
  \puthmorphism(#1)[#4`#5`#6]{#7}{#8}#9%
\else\ifnum\run=0
  \putvmorphism(#1)[#4`#5`#6]{#7}{#8}#9%
\else
\setpos(#1)%
\arrowlength #7
\arrowtype #8
\ifnum\run=0
\else\ifnum\rise=0
\else
\ifnum\run>0
    \coefa=1
\else
   \coefa=-1
\fi
\ifnum\arrowtype>0
   \coefb=0
   \coefc=-1
\else
   \coefb=\coefa
   \coefc=1
   \arrowtype=-\arrowtype
\fi
\width=2
\multiply \width by\run
\divide \width by\rise
\ifnum \width<0  \width=-\width\fi
\advance\width by60
\if l#9 \width=-\width\fi
\putbox(\xpos,\ypos){#4}
{\multiply \coefa by\arrowlength
\advance\xpos by\coefa
\multiply \coefa by\rise
\divide \coefa by\run
\advance \ypos by\coefa
\putbox(\xpos,\ypos){#5} }%
{\multiply \coefa by\arrowlength
\divide \coefa by2
\advance \xpos by\coefa
\advance \xpos by\width
\multiply \coefa by\rise
\divide \coefa by\run
\advance \ypos by\coefa
\if l#9%
   \putrbox(\xpos,\ypos){#6}%
\else\if r#9%
   \putlbox(\xpos,\ypos){#6}%
\fi\fi }%
{\multiply \rise by-\coefc
\multiply \run by-\coefc
\multiply \coefb by\arrowlength
\advance \xpos by\coefb
\multiply \coefb by\rise
\divide \coefb by\run
\advance \ypos by\coefb
\multiply \coefc by70
\advance \ypos by\coefc
\multiply \coefc by\run
\divide \coefc by\rise
\advance \xpos by\coefc
\multiply \coefa by140
\multiply \coefa by\run
\divide \coefa by\rise
\advance \arrowlength by\coefa
\ifcase\arrowtype
\or \put(\xpos,\ypos){\vector(\run,\rise){\arrowlength}}%
\or \put(\xpos,\ypos){\mvector(\run,\rise){\arrowlength}}%
\or \put(\xpos,\ypos){\evector(\run,\rise){\arrowlength}}%
\fi}\fi\fi\fi\fi}}

\newcount\numbdashes \newcount\lengthdash \newcount\increment

\def\howmanydashes{
\numbdashes=\arrowlength \lengthdash=40
\divide\numbdashes by \lengthdash
\lengthdash=\arrowlength
\divide\lengthdash by \numbdashes
\increment=\lengthdash
\multiply\lengthdash by 3
\divide\lengthdash by 5
}

\def\putdashvector(#1)(#2,#3)#4#5{%
\ifnum#3=0 \putdashhvector(#1){#4}#5
\else
\ifnum#2=0
\putdashvvector(#1){#4}#5\fi\fi}

\def\putdashhvector(#1,#2)#3#4{{%
\arrowlength=#3 \howmanydashes
\multiput(#1,#2)(\increment,0){\numbdashes}%
{\vrule height .4pt width \lengthdash\unitlength}
\arrowtype=#4 \xpos=#1
\ifnum\arrowtype<0 \advance\arrowtype by 7 \fi
\ifcase\arrowtype
\or \advance\xpos by 10
    \put(\xpos,#2){\vector(-1,0){\lengthdash}}
    \advance\xpos by 40
    \put(\xpos,#2){\vector(-1,0){\lengthdash}}
\or \advance \xpos by 10
    \put(\xpos,#2){\vector(-1,0){\lengthdash}}
    \advance\xpos by  \arrowlength
    \advance\xpos by  -50
    \put(\xpos,#2){\vector(-1,0){\lengthdash}}
\or \advance\xpos by 10
    \put(\xpos,#2){\vector(-1,0){\lengthdash}}
\or \advance\xpos by \arrowlength
    \advance\xpos by -\lengthdash
    \put(\xpos,#2){\vector(1,0){\lengthdash}}
\or {\advance\xpos by 10
    \put(\xpos,#2){\vector(1,0){\lengthdash}}}
    \advance\xpos by \arrowlength
    \advance\xpos by -\lengthdash
    \put(\xpos,#2){\vector(1,0){\lengthdash}}
\or \advance\xpos by \arrowlength
    \advance\xpos by -\lengthdash
    \put(\xpos,#2){\vector(1,0){\lengthdash}}
    \advance\xpos by -40
    \put(\xpos,#2){\vector(1,0){\lengthdash}}
   \fi
}}

\def\putdashvvector(#1,#2)#3#4{{%
\arrowlength=#3 \howmanydashes
\ypos=#2 \advance\ypos by -\arrowlength
\multiput(#1,#2)(0,\increment){\numbdashes}%
    {\vrule width .4pt height \lengthdash\unitlength}
\arrowtype=#4 \ypos=#2
\ifnum\arrowtype<0 \advance\arrowtype by 7 \fi
\ifcase\arrowtype
\or \advance\ypos by \arrowlength \advance\ypos by -40
    \put(#1,\ypos){\vector(0,1){\lengthdash}}
    \advance\ypos by -40
    \put(#1,\ypos){\vector(0,1){\lengthdash}}
\or \advance\ypos by 10
    \put(#1,\ypos){\vector(0,1){\lengthdash}}
    \advance\ypos by \arrowlength \advance\ypos by -40
    \put(#1,\ypos){\vector(0,1){\lengthdash}}
\or \advance\ypos by \arrowlength \advance\ypos by -40
    \put(#1,\ypos){\vector(0,1){\lengthdash}}
\or \advance\ypos by 10
    \put(#1,\ypos){\vector(0,-1){\lengthdash}}
\or \advance\ypos by 10
    \put(#1,\ypos){\vector(0,-1){\lengthdash}}
    \advance\ypos by \arrowlength \advance\ypos by -40
    \put(#1,\ypos){\vector(0,-1){\lengthdash}}
\or \advance\ypos by 10
    \put(#1,\ypos){\vector(0,-1){\lengthdash}}
    \advance\ypos by 40
    \put(#1,\ypos){\vector(0,-1){\lengthdash}}
\fi
}}

\def\puthmorphism(#1,#2)[#3`#4`#5]#6#7#8{{%
\xpos #1
\ypos #2
\width #6
\arrowlength #6
\arrowtype=#7
\putbox(\xpos,\ypos){#3\vphantom{#4}}%
{\advance \xpos by\arrowlength
\putbox(\xpos,\ypos){\vphantom{#3}#4}}%
\horsize{\tempcounta}{#3}%
\horsize{\tempcountb}{#4}%
\divide \tempcounta by2
\divide \tempcountb by2
\advance \tempcounta by30
\advance \tempcountb by30
\advance \xpos by\tempcounta
\advance \arrowlength by-\tempcounta
\advance \arrowlength by-\tempcountb
\putvector(\xpos,\ypos)(1,0)\arrowlength\arrowtype
\divide \arrowlength by2
\advance \xpos by\arrowlength
\vertsize{\tempcounta}{#5}%
\divide\tempcounta by2
\advance \tempcounta by20
\if a#8 %
   \advance \ypos by\tempcounta
   \putbox(\xpos,\ypos){#5}%
\else
   \advance \ypos by-\tempcounta
   \putbox(\xpos,\ypos){#5}%
\fi}}

\def\putvmorphism(#1,#2)[#3`#4`#5]#6#7#8{{%
\xpos #1
\ypos #2
\arrowlength #6
\arrowtype #7
\settowidth{\xlen}{$#5$}%
\putbox(\xpos,\ypos){#3}%
{\advance \ypos by-\arrowlength
\putbox(\xpos,\ypos){#4}}%
{\advance\arrowlength by-140
\advance \ypos by-70
\ifdim\xlen>0pt
   \if m#8%
      \putsplitvector(\xpos,\ypos)\arrowlength\arrowtype
   \else
   \putvector(\xpos,\ypos)(0,-1)\arrowlength\arrowtype
   \fi
\else
   \putvector(\xpos,\ypos)(0,-1)\arrowlength\arrowtype
\fi}%
\ifdim\xlen>0pt
   \divide \arrowlength by2
   \advance\ypos by-\arrowlength
   \if l#8%
      \advance \xpos by-40
      \putrbox(\xpos,\ypos){#5}%
   \else\if r#8%
      \advance \xpos by40
      \putlbox(\xpos,\ypos){#5}%
   \else
      \putbox(\xpos,\ypos){#5}%
   \fi\fi
\fi
}}

\def\putsquarep<#1>(#2)[#3;#4`#5`#6`#7]{{%
\setsqparms[#1]%
\setpos(#2)%
\settokens`#3`%
\puthmorphism(\xpos,\ypos)[\tokenc`\tokend`{#7}]{\width}{\arrowtyped}b%
\advance\ypos by \height
\puthmorphism(\xpos,\ypos)[\tokena`\tokenb`{#4}]{\width}{\arrowtypea}a%
\putvmorphism(\xpos,\ypos)[``{#5}]{\height}{\arrowtypeb}l%
\advance\xpos by \width
\putvmorphism(\xpos,\ypos)[``{#6}]{\height}{\arrowtypec}r%
}}

\def\putsquare{\@ifnextchar <{\putsquarep}{\putsquarep%
   <\arrowtypea`\arrowtypeb`\arrowtypec`\arrowtyped;\width`\height>}}
\def\square{\@ifnextchar< {\squarep}{\squarep
   <\arrowtypea`\arrowtypeb`\arrowtypec`\arrowtyped;\width`\height>}}
\def\squarep<#1>[#2`#3`#4`#5;#6`#7`#8`#9]{{
\setsqparms[#1]
\diagram
\putsquarep<\arrowtypea`\arrowtypeb`\arrowtypec`
\arrowtyped;\width`\height>
(0,0)[#2`#3`#4`{#5};#6`#7`#8`{#9}]
\enddiagram
}}                                                 
\def\putptrianglep<#1>(#2,#3)[#4`#5`#6;#7`#8`#9]{{%
\settriparms[#1]%
\xpos=#2 \ypos=#3
\advance\ypos by \height
\puthmorphism(\xpos,\ypos)[#4`#5`{#7}]{\height}{\arrowtypea}a%
\putvmorphism(\xpos,\ypos)[`#6`{#8}]{\height}{\arrowtypeb}l%
\advance\xpos by\height
\putmorphism(\xpos,\ypos)(-1,-1)[``{#9}]{\height}{\arrowtypec}r%
}}

\def\putptriangle{\@ifnextchar <{\putptrianglep}{\putptrianglep
   <\arrowtypea`\arrowtypeb`\arrowtypec;\height>}}
\def\ptriangle{\@ifnextchar <{\ptrianglep}{\ptrianglep
   <\arrowtypea`\arrowtypeb`\arrowtypec;\height>}}
\def\ptrianglep<#1>[#2`#3`#4;#5`#6`#7]{{
\settriparms[#1]
\diagram
\putptrianglep<\arrowtypea`\arrowtypeb`
\arrowtypec;\height>
(0,0)[#2`#3`#4;#5`#6`{#7}]
\enddiagram
}}                                            

\def\putqtrianglep<#1>(#2,#3)[#4`#5`#6;#7`#8`#9]{{%
\settriparms[#1]%
\xpos=#2 \ypos=#3
\advance\ypos by\height
\puthmorphism(\xpos,\ypos)[#4`#5`{#7}]{\height}{\arrowtypea}a%
\putmorphism(\xpos,\ypos)(1,-1)[``{#8}]{\height}{\arrowtypeb}l%
\advance\xpos by\height
\putvmorphism(\xpos,\ypos)[`#6`{#9}]{\height}{\arrowtypec}r%
}}

\def\putqtriangle{\@ifnextchar <{\putqtrianglep}{\putqtrianglep
   <\arrowtypea`\arrowtypeb`\arrowtypec;\height>}}
\def\qtriangle{\@ifnextchar <{\qtrianglep}{\qtrianglep
   <\arrowtypea`\arrowtypeb`\arrowtypec;\height>}}
\def\qtrianglep<#1>[#2`#3`#4;#5`#6`#7]{{
\settriparms[#1]
\width=\height                                
\diagram
\putqtrianglep<\arrowtypea`\arrowtypeb`
\arrowtypec;\height>
(0,0)[#2`#3`#4;#5`#6`{#7}]
\enddiagram
}}

\def\putdtrianglep<#1>(#2,#3)[#4`#5`#6;#7`#8`#9]{{%
\settriparms[#1]%
\xpos=#2 \ypos=#3
\puthmorphism(\xpos,\ypos)[#5`#6`{#9}]{\height}{\arrowtypec}b%
\advance\xpos by \height \advance\ypos by\height
\putmorphism(\xpos,\ypos)(-1,-1)[``{#7}]{\height}{\arrowtypea}l%
\putvmorphism(\xpos,\ypos)[#4``{#8}]{\height}{\arrowtypeb}r%
}}

\def\putdtriangle{\@ifnextchar <{\putdtrianglep}{\putdtrianglep
   <\arrowtypea`\arrowtypeb`\arrowtypec;\height>}}
\def\dtriangle{\@ifnextchar <{\dtrianglep}{\dtrianglep
   <\arrowtypea`\arrowtypeb`\arrowtypec;\height>}}
\def\dtrianglep<#1>[#2`#3`#4;#5`#6`#7]{{
\settriparms[#1]
\width=\height                                
\diagram
\putdtrianglep<\arrowtypea`\arrowtypeb`
\arrowtypec;\height>
(0,0)[#2`#3`#4;#5`#6`{#7}]
\enddiagram
}}

\def\putbtrianglep<#1>(#2,#3)[#4`#5`#6;#7`#8`#9]{{%
\settriparms[#1]%
\xpos=#2 \ypos=#3
\puthmorphism(\xpos,\ypos)[#5`#6`{#9}]{\height}{\arrowtypec}b%
\advance\ypos by\height
\putmorphism(\xpos,\ypos)(1,-1)[``{#8}]{\height}{\arrowtypeb}r%
\putvmorphism(\xpos,\ypos)[#4``{#7}]{\height}{\arrowtypea}l%
}}

\def\putbtriangle{\@ifnextchar <{\putbtrianglep}{\putbtrianglep
   <\arrowtypea`\arrowtypeb`\arrowtypec;\height>}}
\def\btriangle{\@ifnextchar <{\btrianglep}{\btrianglep
   <\arrowtypea`\arrowtypeb`\arrowtypec;\height>}}
\def\btrianglep<#1>[#2`#3`#4;#5`#6`#7]{{
\settriparms[#1]
\width=\height                               
\diagram
\putbtrianglep<\arrowtypea`\arrowtypeb`
\arrowtypec;\height>
(0,0)[#2`#3`#4;#5`#6`{#7}]
\enddiagram
}}

\def\putAtrianglep<#1>(#2,#3)[#4`#5`#6;#7`#8`#9]{{%
\settriparms[#1]%
\xpos=#2 \ypos=#3
{\multiply \height by2
\puthmorphism(\xpos,\ypos)[#5`#6`{#9}]{\height}{\arrowtypec}b}%
\advance\xpos by\height \advance\ypos by\height
\putmorphism(\xpos,\ypos)(-1,-1)[#4``{#7}]{\height}{\arrowtypea}l%
\putmorphism(\xpos,\ypos)(1,-1)[``{#8}]{\height}{\arrowtypeb}r%
}}

\def\putAtriangle{\@ifnextchar <{\putAtrianglep}{\putAtrianglep
   <\arrowtypea`\arrowtypeb`\arrowtypec;\height>}}
\def\Atriangle{\@ifnextchar <{\Atrianglep}{\Atrianglep
   <\arrowtypea`\arrowtypeb`\arrowtypec;\height>}}
\def\Atrianglep<#1>[#2`#3`#4;#5`#6`#7]{{
\settriparms[#1]
\width=\height                                     
\diagram
\putAtrianglep<\arrowtypea`\arrowtypeb`
\arrowtypec;\height>
(0,0)[#2`#3`#4;#5`#6`{#7}]
\enddiagram
}}

\def\putAtrianglepairp<#1>(#2)[#3;#4`#5`#6`#7`#8]{{%
\settripairparms[#1]%
\setpos(#2)%
\settokens`#3`%
\puthmorphism(\xpos,\ypos)[\tokenb`\tokenc`{#7}]{\height}{\arrowtyped}b%
\advance\xpos by\height
\puthmorphism(\xpos,\ypos)[\phantom{\tokenc}`\tokend`{#8}]%
{\height}{\arrowtypee}b%
\advance\ypos by\height
\putmorphism(\xpos,\ypos)(-1,-1)[\tokena``{#4}]{\height}{\arrowtypea}l%
\putvmorphism(\xpos,\ypos)[``{#5}]{\height}{\arrowtypeb}m%
\putmorphism(\xpos,\ypos)(1,-1)[``{#6}]{\height}{\arrowtypec}r%
}}

\def\putAtrianglepair{\@ifnextchar <{\putAtrianglepairp}{\putAtrianglepairp%
   <\arrowtypea`\arrowtypeb`\arrowtypec`\arrowtyped`\arrowtypee;\height>}}
\def\Atrianglepair{\@ifnextchar <{\Atrianglepairp}{\Atrianglepairp%
   <\arrowtypea`\arrowtypeb`\arrowtypec`\arrowtyped`\arrowtypee;\height>}}

\def\Atrianglepairp<#1>[#2;#3`#4`#5`#6`#7]{{
\settripairparms[#1]
\settokens`#2`
\width=\height                                
\diagram
\putAtrianglepairp                            
<\arrowtypea`\arrowtypeb`\arrowtypec`
\arrowtyped`\arrowtypee;\height>
(0,0)[{#2};#3`#4`#5`#6`{#7}]
\enddiagram
}}

\def\putVtrianglep<#1>(#2,#3)[#4`#5`#6;#7`#8`#9]{{%
\settriparms[#1]%
\xpos=#2 \ypos=#3
\advance\ypos by\height
{\multiply\height by2
\puthmorphism(\xpos,\ypos)[#4`#5`{#7}]{\height}{\arrowtypea}a}%
\putmorphism(\xpos,\ypos)(1,-1)[`#6`{#8}]{\height}{\arrowtypeb}l%
\advance\xpos by\height
\advance\xpos by\height
\putmorphism(\xpos,\ypos)(-1,-1)[``{#9}]{\height}{\arrowtypec}r%
}}

\def\putVtriangle{\@ifnextchar <{\putVtrianglep}{\putVtrianglep
   <\arrowtypea`\arrowtypeb`\arrowtypec;\height>}}
\def\Vtriangle{\@ifnextchar <{\Vtrianglep}{\Vtrianglep
   <\arrowtypea`\arrowtypeb`\arrowtypec;\height>}}
\def\Vtrianglep<#1>[#2`#3`#4;#5`#6`#7]{{
\settriparms[#1]
\width=\height                                 
\diagram
\putVtrianglep<\arrowtypea`\arrowtypeb`
\arrowtypec;\height>
(0,0)[#2`#3`#4;#5`#6`{#7}]
\enddiagram
}}

\def\putVtrianglepairp<#1>(#2)[#3;#4`#5`#6`#7`#8]{{
\settripairparms[#1]%
\setpos(#2)%
\settokens`#3`%
\advance\ypos by\height
\putmorphism(\xpos,\ypos)(1,-1)[`\tokend`{#6}]{\height}{\arrowtypec}l%
\puthmorphism(\xpos,\ypos)[\tokena`\tokenb`{#4}]{\height}{\arrowtypea}a%
\advance\xpos by\height
\puthmorphism(\xpos,\ypos)[\phantom{\tokenb}`\tokenc`{#5}]%
{\height}{\arrowtypeb}a%
\putvmorphism(\xpos,\ypos)[``{#7}]{\height}{\arrowtyped}m%
\advance\xpos by\height
\putmorphism(\xpos,\ypos)(-1,-1)[``{#8}]{\height}{\arrowtypee}r%
}}

\def\putVtrianglepair{\@ifnextchar <{\putVtrianglepairp}{\putVtrianglepairp%
    <\arrowtypea`\arrowtypeb`\arrowtypec`\arrowtyped`\arrowtypee;\height>}}
\def\Vtrianglepair{\@ifnextchar <{\Vtrianglepairp}{\Vtrianglepairp%
    <\arrowtypea`\arrowtypeb`\arrowtypec`\arrowtyped`\arrowtypee;\height>}}
\def\Vtrianglepairp<#1>[#2;#3`#4`#5`#6`#7]{{
\settripairparms[#1]
\settokens`#2`
\diagram
\putVtrianglepairp                             
<\arrowtypea`\arrowtypeb`\arrowtypec`
\arrowtyped`\arrowtypee;\height>
(0,0)[{#2};#3`#4`#5`#6`{#7}]
\enddiagram
}}

\def\putCtrianglep<#1>(#2,#3)[#4`#5`#6;#7`#8`#9]{{%
\settriparms[#1]%
\xpos=#2 \ypos=#3
\advance\ypos by\height
\putmorphism(\xpos,\ypos)(1,-1)[``{#9}]{\height}{\arrowtypec}l%
\advance\xpos by\height
\advance\ypos by\height
\putmorphism(\xpos,\ypos)(-1,-1)[#4`#5`{#7}]{\height}{\arrowtypea}l%
{\multiply\height by 2
\putvmorphism(\xpos,\ypos)[`#6`{#8}]{\height}{\arrowtypeb}r}%
}}

\def\putCtriangle{\@ifnextchar <{\putCtrianglep}{\putCtrianglep
    <\arrowtypea`\arrowtypeb`\arrowtypec;\height>}}
\def\Ctriangle{\@ifnextchar <{\Ctrianglep}{\Ctrianglep
    <\arrowtypea`\arrowtypeb`\arrowtypec;\height>}}
\def\Ctrianglep<#1>[#2`#3`#4;#5`#6`#7]{{
\settriparms[#1]
\width=\height                               
\diagram
\putCtrianglep<\arrowtypea`\arrowtypeb`
\arrowtypec;\height>
(0,0)[#2`#3`#4;#5`#6`{#7}]
\enddiagram
}}                                           
\def\putDtrianglep<#1>(#2,#3)[#4`#5`#6;#7`#8`#9]{{%
\settriparms[#1]%
\xpos=#2 \ypos=#3
\advance\xpos by\height \advance\ypos by\height
\putmorphism(\xpos,\ypos)(-1,-1)[``{#9}]{\height}{\arrowtypec}r%
\advance\xpos by-\height \advance\ypos by\height
\putmorphism(\xpos,\ypos)(1,-1)[`#5`{#8}]{\height}{\arrowtypeb}r%
{\multiply\height by 2
\putvmorphism(\xpos,\ypos)[#4`#6`{#7}]{\height}{\arrowtypea}l}%
}}

\def\putDtriangle{\@ifnextchar <{\putDtrianglep}{\putDtrianglep
    <\arrowtypea`\arrowtypeb`\arrowtypec;\height>}}
\def\Dtriangle{\@ifnextchar <{\Dtrianglep}{\Dtrianglep
   <\arrowtypea`\arrowtypeb`\arrowtypec;\height>}}
\def\Dtrianglep<#1>[#2`#3`#4;#5`#6`#7]{{
\settriparms[#1]
\width=\height                              
\diagram
\putDtrianglep<\arrowtypea`\arrowtypeb`
\arrowtypec;\height>
(0,0)[#2`#3`#4;#5`#6`{#7}]
\enddiagram
}}                                          
\def\setrecparms[#1`#2]{\width=#1 \height=#2}%

\def\recursep<#1`#2>[#3;#4`#5`#6`#7`#8]{{\m@th
\width=#1 \height=#2
\settokens`#3`
\settowidth{\tempdimen}{$\tokena$}
\ifdim\tempdimen=0pt
  \savebox{\tempboxa}{\hbox{$\tokenb$}}%
  \savebox{\tempboxb}{\hbox{$\tokend$}}%
  \savebox{\tempboxc}{\hbox{$#6$}}%
\else
  \savebox{\tempboxa}{\hbox{$\hbox{$\tokena$}\times\hbox{$\tokenb$}$}}%
  \savebox{\tempboxb}{\hbox{$\hbox{$\tokena$}\times\hbox{$\tokend$}$}}%
  \savebox{\tempboxc}{\hbox{$\hbox{$\tokena$}\times\hbox{$#6$}$}}%
\fi
\ypos=\height
\divide\ypos by 2
\xpos=\ypos
\advance\xpos by \width
\bfig
\putCtrianglep<-1`1`1;\ypos>(0,0)[`\tokenc`;#5`#6`{#7}]%
\puthmorphism(\ypos,0)[\tokend`\usebox{\tempboxb}`{#8}]{\width}{-1}b%
\puthmorphism(\ypos,\height)[\tokenb`\usebox{\tempboxa}`{#4}]{\width}{-1}a%
\advance\ypos by \width
\putvmorphism(\ypos,\height)[``\usebox{\tempboxc}]{\height}1r%
\efig
}}

\def\recurse{\@ifnextchar <{\recursep}{\recursep<\width`\height>}}

\def\puttwohmorphisms(#1,#2)[#3`#4;#5`#6]#7#8#9{{%
\puthmorphism(#1,#2)[#3`#4`]{#7}0a
\ypos=#2
\advance\ypos by 20
\puthmorphism(#1,\ypos)[\phantom{#3}`\phantom{#4}`#5]{#7}{#8}a
\advance\ypos by -40
\puthmorphism(#1,\ypos)[\phantom{#3}`\phantom{#4}`#6]{#7}{#9}b
}}

\def\puttwovmorphisms(#1,#2)[#3`#4;#5`#6]#7#8#9{{%
\putvmorphism(#1,#2)[#3`#4`]{#7}0a
\xpos=#1
\advance\xpos by -20
\putvmorphism(\xpos,#2)[\phantom{#3}`\phantom{#4}`#5]{#7}{#8}l
\advance\xpos by 40
\putvmorphism(\xpos,#2)[\phantom{#3}`\phantom{#4}`#6]{#7}{#9}r
}}

\def\puthcoequalizer(#1)[#2`#3`#4;#5`#6`#7]#8#9{{%
\setpos(#1)%
\puttwohmorphisms(\xpos,\ypos)[#2`#3;#5`#6]{#8}11%
\advance\xpos by #8
\puthmorphism(\xpos,\ypos)[\phantom{#3}`#4`#7]{#8}1{#9}
}}

\def\putvcoequalizer(#1)[#2`#3`#4;#5`#6`#7]#8#9{{%
\setpos(#1)%
\puttwovmorphisms(\xpos,\ypos)[#2`#3;#5`#6]{#8}11%
\advance\ypos by -#8
\putvmorphism(\xpos,\ypos)[\phantom{#3}`#4`#7]{#8}1{#9}
}}

\def\putthreehmorphisms(#1)[#2`#3;#4`#5`#6]#7(#8)#9{{%
\setpos(#1) \settypes(#8)
\if a#9 %
     \vertsize{\tempcounta}{#5}%
     \vertsize{\tempcountb}{#6}%
     \ifnum \tempcounta<\tempcountb \tempcounta=\tempcountb \fi
\else
     \vertsize{\tempcounta}{#4}%
     \vertsize{\tempcountb}{#5}%
     \ifnum \tempcounta<\tempcountb \tempcounta=\tempcountb \fi
\fi
\advance \tempcounta by 60
\puthmorphism(\xpos,\ypos)[#2`#3`#5]{#7}{\arrowtypeb}{#9}
\advance\ypos by \tempcounta
\puthmorphism(\xpos,\ypos)[\phantom{#2}`\phantom{#3}`#4]{#7}{\arrowtypea}{#9}
\advance\ypos by -\tempcounta \advance\ypos by -\tempcounta
\puthmorphism(\xpos,\ypos)[\phantom{#2}`\phantom{#3}`#6]{#7}{\arrowtypec}{#9}
}}

\def\setarrowtoks[#1`#2`#3`#4`#5`#6]{%
\def\toka{#1}
\def\tokb{#2}
\def\tokc{#3}
\def\tokd{#4}
\def\toke{#5}
\def\tokf{#6}
}
\def\hex{\@ifnextchar <{\hexp}{\hexp<1000`400>}}
\def\hexp<#1`#2>[#3`#4`#5`#6`#7`#8;#9]{%
\setarrowtoks[#9]
\yext=#2 \advance \yext by #2
\xext=#1 \advance\xext by \yext
\bfig
\putCtriangle<-1`0`1;#2>(0,0)[`#5`;\tokb``\tokd]
\xext=#1 \yext=#2 \advance \yext by #2
\putsquare<1`0`0`1;\xext`\yext>(#2,0)[#3`#4`#7`#8;\toka```\tokf]
\advance \xext by #2
\putDtriangle<0`1`-1;#2>(\xext,0)[`#6`;`\tokc`\toke]
\efig
}
\makeatother

\newtheorem{lemma}{Lemma}[section]

\newtheorem{remark}[lemma]{Remark}

\newtheorem{theorem}[lemma]{Theorem}

\newtheorem{definition}[lemma]{Definition}
\newtheorem{corollary}[lemma]{Corollary}
\newtheorem{claim}[lemma]{Claim}
\newtheorem{conjecture}[lemma]{Conjecture}

\newcommand{\BProof}{{\it Proof.} \r}
\newcommand{\EProof}{\hfill \blacksquare}

\def\ome {\omega}

\def\XI{\xi}
\def\XII{\eta}
\def\CHI{\chi}

\def\lto{\longrightarrow}

\def\iso{\backsimeq}

\def\R{{\mathbb R}}
\def\Q{{\mathbb Q}}

\def\Z{{\mathbb Z}}
\def\C{{\mathbb C}}
\def\Fq {\mathbb{F}_{q}}   
\def\Fqm{\Fq^*}
\def\lgn {\sigma}  
\def\Fp{\mathbb{F}_{p}}

\def\FFp{\overline{\mathbb{F}}_{p}}
\def\Fpm{\Fp^*}

\def\Lm{\Lambda^*}

\def\GL{\mathrm{GL}}
\def\PGL{\mathrm{PGL}}
\def\GZ{\mathrm{SL}_2(\Z)}
\def\G{\mathrm{\Gamma}}     
\def\GG{\mathrm{SL}_2(\Fp)}
\def\ASL_2{\bf{SL_2}}

\def\AGG{{\bf SL_2}} 
\def\AT{{\bf T}}      
\def\GrA{\langle A \rangle}
\def\CA{{\mathrm T}_A}
\def\ACA{{\bf T}_A}
\def\CAD{{\mathrm T}_A^\vee} 
\def\Gf{\mathrm{\Gamma}_{p}}

\def\Sp {\mathrm{Sp}} 
\def\ASp {{\bf Sp}} 
\def\gH {\mathrm{H}}      
\def\Aheiz{{\bf H}}     
\def \semi{\mathrm{G}} 
\def \Asemi{{\bf G}} 

\def\Ga {\mathbb{G}_a} 
\def\Gm {\mathbb{G}_m} 

\def\Br {\mathrm{B}} 
\def\ABr {{\bf B}} 

\def\oB{\Br} 
\def\Ow{{\mathrm O}_w} 

\def\AoB{\ABr} 

\def\AUn{{{\bf U}^\circ}} 

\def\AoU{\bf U} 

\def\AO{{\bf O}}
\def\O{\mathrm O}

\def\AU{{\bf U}}

\def\ob{b} 
\def\ou{u}
\def\u{{u^\circ}}
\def\S0{S}

\def\FST{{\cal S}(\T)}
\def\F{C^{\infty}({\T})}
\def\T{{\mathbb T}}
\def\Td{\T^\vee} 

\def\Irr{\mathrm{Irr}({\cal A} _\hbar)}
\def\A{\cal A}
\def\Ad{ {\cal A} _\hbar}

\def\Adf{{\cal A}_{p}}
\def\hb{\hbar}
\def\h{ \hbar }

\def \Wigner {\mathcal{W}} 
\def \nWigner{\tilde{\mathcal{W}}} 

\def\H{{\cal H}}
\def\Hh{{\cal H} _\h}
\def\Hc {{\mathcal{H}_\V}} 
\def\V{\mathrm{V}} 
\def\AV{{\bf V}} 
\def\VI{\mathrm{V}_1} 
\def\VII{\mathrm{V}_2} 
\def\W{{\mathrm W}}

\def\rhoh{\rho _{_\h}}
\def\pih{\pi i {\h}}
\def\Pih{\pi_{_\hbar}}
\def\rhof {\rho_{_p}}

\def \bL {{\mathrm L}}
\def\bM{{\mathrm M}}
\def\Pif{\pi_{p}}

\def\FSVI{{\cal S}(\VI)}
\def\FSFq{{\cal S}(\Fq)}

\def\Y{Y_{0}}
\def\YY{{\mathrm Y}}
\def\AYY{{\bf Y}}

\def \Lmmf {\mathrm{V}}

\def \bA{\mathbb{A}}
\def\projI {\mathbb{P}^1}  
\def\P1{\mathbb{P}^1}

\def\X{\mathrm{X}}
\def\AX {{\bf X}} 

\def\Sw {\mathrm{O}_w}  
\def\ASw {{\bf O}_w}  
\def\O{{\bf O}} 
\def\calU{{\bf O}^{\times}} 

\def \AUnx {\AUn^{\times}}

\def\i_XI{i_{_{\XI}}}
\def\iXII{i_{_{\XII}}}
\def\p_XI{p_{_{\XI}}}

\def \mRSwE {\mathrm{R}}

\def\SE{\mathcal{E}}

\def\SF{\mathcal{F}}
\def\SG{\mathcal{G}}
\def\SL{\mathscr L}

\def \SI {\mathcal{I}}

\def\SK {\mathcal{K}} 

\def\SKoB {\SK_{\AoB}}  
\def\SKUx{\SK_{\AUn ^ \times}} 
\def\SKO{\SK_\O} 
\def\SKU{\SK_{\cal U}} 

\def\SKpi {\SK_{\pi}} 

\def\SKn {\tilde{\SK}}  
\def\SKnSw {\SKn_{\ASw}}
\def\SKnoB {\SKn_{\AoB}}
\def\SKnSwE {\SKn_\O}
\def\SKnoU {\SKn_{\AoU}}
\def\SKnUx {\SKn_{\AUn ^ \times}}

\def\SKnoUxw{\SKn_{\AoU^\times w}}

\def\SKnE{\SKn_{\Aheiz}}
\def\SKnw {\SKn_w}
\def\SKoU {\SK_{\AoU}} 
\def\SKUx {\SK_{\AUn ^ \times}} 
\def\SKoUxw {\SK_{{\AoU}^\times w}}

\def\SKob {\SK_{\ob}}

\def\SKw {\SK_w} 
\def\SKwI {\SK_{w^{-1}}}
\def\SAw  {\SA_w}

\def\SA {\mathcal{A}}  
\def\SASw {\SA_{\ASw}}
\def\SAoB {\SA_{\AoB}}
\def\SASwE  {\SA_{\bf O}}
\def\SAoU {\SA_{\AoU}}

\def \SAUx {\SA_{\AUn ^ \times}}
\def \SAoUxw {\SA_{\AoU^\times w}}

\def\Sartin{\SL_{\psi}} 
\def\Skummer {\SL_{\chi}} 

\def\Slegendre {\SL_{\lgn}}

\def\Tr{\mathrm{Tr}}
\def\Av{\mathrm{\bf Av}}
\def\ev{{\mathrm{e.v}}}

\def\End{\mathrm{End}}
\def\Hom{\mathrm{Hom}}
\def\dim{\mathrm{dim}}

\def\orn {\varrho}
\def\ornlag {\mathrm{Lag}^{\circ}} 
\def\lag {\mathrm{Lag}} 
\def\ornL {\mathrm{L^{^\circ}}} 
\def\ornLI {{\mathrm{L}^{^\circ}_1}} 
\def\ornLIm {{\mathrm{L}^{^{\overline{\circ}}}_1}} 
\def\ornLII {{\mathrm{L}^{^\circ}_2}} 
\def \ornLIII {{\mathrm{L}^{^\circ}_3}} 
\def\interLILII {{\mathrm{Int}}_{\ornLII,\ornLI}} 
\def\L {\mathrm{L}} 
\def\LI {{\mathrm{L}_1}} 
\def\LII {{\mathrm{L}_2}} 
\def \LIII {{\mathrm{L}_3}} 
\def\thetaLILII {\textsf{F}_{{\ornLII,\ornLI}}} 
\def \thetaLILIII  {\textsf{F}_{{\ornLIII,\ornLI}}} 
\def \thetaLIILIII  {\textsf{F}_{{\ornLIII,\ornLII}}} 
\def\thetagLIgLII {\textsf{F}_{{g\ornLII,g\ornLI}}} 
\def\ornofL {\orn_{_\L}} 
\def\ornVII  {{\mathrm{\VII^{^\circ}}}} 
\def\thetagVIIVII {\textsf{F}_{{\ornVII,g\ornVII}}} 
\def\thetaLIILI {\textsf{F}_{{\ornLI,\ornLII}}} 
\def\thetaLIILIm {\textsf{F}_{{\ornLIm,\ornLII}}} 

\def\thnLILII {\tilde{\textsf{F}}_{{\ornLII,\ornLI}}} 
\def \thnLILIII  {\tilde{\textsf{F}}_{{\ornLIII,\ornLI}}} 
\def \thnLIILIII  {\tilde{\textsf{F}}_{{\ornLIII,\ornLII}}} 
\def \thnLIILI  {\tilde{\textsf{F}}_{{\ornLI,\ornLII}}} 

\def\aLILII {\mathrm{a}_{\ornLII,\ornLI}} 
\def\aLILIII {\mathrm{a}_{\ornLIII,\ornLI}} 
\def\aLIILIII {\mathrm{a}_{\ornLIII,\ornLII}} 
\def\aLIILI {\mathrm{a}_{\ornLI,\ornLII}} 

\def\xiLI {\xi_{_\LI}} 
\def\xiLII {\xi_{_\LII}} 
\def\xiLIII {\xi_{_\LIII}} 

\def\rLIILIII {\mathrm{r}_{_{\LIII,\LII}}} 
\def\ri {\mathrm{r}_{_{\mathrm{L}_{i},\mathrm{L}_{k}}}} 

\def\Cconst{\mathrm{C}} 
\def\Dconst{\mathrm{D}} 

\def\cq {\mathrm{q}} 
\def\cp {\mathrm{p}} 

\def\Db{{\mathrm D}^b_{\mathrm{c}}}
\def\Perv{Perv}
\def\Fr{\mathrm{Fr}}
\def\chiFr{\chi_{_\Fr}}
\def\Qlb{\overline{\Q}_\ell}
\def\sint{\smallint}
\def\coH{{\mathrm H}}
\def\Swan{\mathrm{Swan}}

\def\rev{\quad}

\def\r{\;}
\def\above{\overset}

\def\half{\begin{smallmatrix} \frac{1}{2} \end{smallmatrix}}
\def\quarter{\begin{smallmatrix} \frac{1}{4} \end{smallmatrix}}

\begin{document}

\title{\texttt{PROOF OF THE KURLBERG-RUDNICK RATE CONJECTURE}}
\author{\textsf{SHAMGAR GUREVICH AND RONNY HADANI}}

\date{}

\maketitle

\bigskip

\centerline{\textit{To Joseph Bernstein for his 60's birthday with
admiration}} \vfill\null

\begin{abstract}
In this paper we present a proof of the {\it Hecke quantum unique
ergodicity rate conjecture} for the Hannay-Berry model. A model of
quantum mechanics on the two-dimensional torus. This conjecture
was stated in Z. Rudnick's lectures at MSRI, Berkeley 1999, and
ECM, Barcelona 2000.
\end{abstract}

\bigskip\numberwithin{equation}{subsection}

\setcounter{section}{-1}

\section{Introduction}
\subsection{Hannay-Berry model}
In the paper ``{\it Quantization of linear maps on the torus -
Fresnel diffraction by a periodic grating}'', published in 1980
\cite{HB}, the physicists J. Hannay and Sir M.V. Berry explore a
 model for quantum mechanics on the two-dimensional symplectic torus
$(\T,\ome)$. Hannay and Berry suggested to quantize simultaneously
the functions on the torus and the linear symplectic group $\G=
\GZ$.

\subsection{Quantum chaos}
One of their main motivations was to study the phenomenon of
quantum chaos \cite{B1, B2, R2, S} in this model. More precisely,
they considered an ergodic discrete dynamical system on the torus,
which is generated by a hyperbolic automorphism $A \in \GZ$.
Quantizing the system, we replace: the classical phase space
$(\T,\ome)$  by a \textit{finite dimensional} Hilbert space $\Hh$,
classical observables, i.e., functions $f \in \F$,  by operators
$\Pih(f) \in \End(\Hh)$, and classical symmetries by a unitary
representation  $\rhoh : \GZ \lto \mathrm{U}(\Hh)$. A fundamental
meta-question in the area of quantum chaos is to
\textit{understand} the ergodic properties of the quantum system
$\rhoh(A)$, at least in the semi-classical limit as $\h
\rightarrow 0$.

\subsection{Schnirelman's theorem}
analogous with the case of the Schr\"{o}dinger equation, consider
the following eigenstate problem:
\begin{equation}\label{eigenstate}
\rhoh(A)\Psi = \lambda\Psi, \rev \Psi \in \Hh.
\end{equation}
A fundamental result, valid for a wide class of quantum systems
which are associated to ergodic classical dynamics, is
Schnirelman's theorem \cite{Sc}, asserting that in the
semi-classical limit "almost all" eigenstates becomes
equidistributed in an appropriate sense. A variant of
Schnirelman's theorem also holds in our situation \cite{BD}. More
precisely, we have that in the semi-classical limit $\h \to 0 $,
for "almost all" eigenstates $\Psi$ of the operator $\rhoh(A)$,
the corresponding \textit{Wigner distribution} $\left <
\Psi|\Pih(\cdot) \Psi\right>:\F\lto \C$ approaches the phase space
average $\int_\T \cdot \,|\ome|$. In this respect, it seems
natural to ask whether there exists exceptional sequences of
eigenstates? Namely, eigenstates that do not obey the
Schnirelman's rule ("scarred" eigenstates). It was predicted by
Berry \cite{B1, B2}, that "scarring" phenomenon is not expected to
be seen for quantum systems associated with "generic" chaotic
classical dynamics. However, in our situation, the operator
$\rhoh(A)$ is not generic, and exceptional eigenstates were
constructed. Indeed, it was observed numerically, and then
confirmed mathematically in \cite{FND}, that certain
$\rhoh(A)$-eigenstates might localize. For example, in that paper,
a sequence of eigenstates $\Psi$ was constructed, for which the
corresponding Wigner distribution approaches the measure
$\half\delta_0 + \half|\ome|$ on $\T$.

\subsection{Hecke quantum unique ergodicity}
A quantum system that obeys the Schnirelman's rule is also called
quantum ergodic. Can one impose some natural conditions on the
eigenstates (\ref{eigenstate}) so that no exceptional eigenstates
will appear? Namely,  \textit{Quantum Unique Ergodicity} will
hold. This question was addressed in a paper by Kurlberg and
Rudnick \cite{KR1}. In this paper they formulated a rigorous
notion of Hecke quantum unique ergodicity for the case $\h =
\frac{1}{p}$. The following is a brief description of that work.
The basic observation is that the degeneracies of the operator
$\rhoh(A)$ are coupled with the existence of symmetries. There
exists a commutative group of operators that commutes with
$\rhoh(A)$. In more detail, the representation $\rhoh$ factors
through the quotient group $\Gf \iso \GG$. We denote by $\CA
\subset \Gf$ the centralizer of the element $A$, now considered as
an element of the quotient group $\Gf$. The group $\CA$ is called
(cf. \cite{KR1}) the \textit{Hecke torus} corresponding to the
element $A$. The Hecke torus acts semisimply on $\Hh$. Therefore,
we have a decomposition:
$$ \Hh = \bigoplus_{\chi : \CA \lto \C^*} \H_\chi,$$
where $\H_\chi$ is the Hecke eigenspace corresponding to the
character $\chi$. Consider a unit eigenstate $\Psi \in \H_\chi$
and the corresponding Wigner distribution $\Wigner_\chi : \F \lto
\C $, defined by the formula $\Wigner_\chi(f) = \left < \Psi|
\Pih(f) \Psi \right>$. The main statement in \cite{KR1} asserts
about an explicit bound of the semi-classical asymptotic of
$\Wigner_\chi(f)$:
$$ \left | \Wigner_\chi(f) - \int_\T f |\ome| \right | \leq \frac{C_f}{p^{1/4}},$$
where $C_f$ is a constant that depends only on the function $f$.
In Rudnick's lectures at MSRI, Berkeley 1999 \cite{R1}, and ECM,
Barcelona 2000 \cite{R2}, he conjectured that a stronger bound
should hold true, that is,:
\\
%
%
\begin{conjecture}[Rate Conjecture]\label{RC} The following bound holds:
\begin{eqnarray*}
\left | \Wigner_\chi(f) - \int_\T f |\ome| \right | \leq
\frac{C_f}{p^{1/2}}\,.
\end{eqnarray*}
\end{conjecture}
The basic \textit{clues} suggesting the validity of this stronger
bound come from two main sources. The first source is
\textit{computer} simulations \cite{Ku} accomplished over the
years to give extremely precise bounds for considerably large
values of $p$. A more mathematical argument is based on the fact
that for special values of $p$, in which the Hecke torus
\textit{splits}, namely, $\CA \simeq \Fpm$, one is able to compute
explicitly the eigenstate $\Psi \in \H_\chi$ and as a consequence
to give an explicit \textit{formula} for the Wigner distribution
\cite{KR2, DGI}. More precisely, in case $\xi \in \Td$, i.e., a
character, the distribution $\Wigner_\chi(\xi)$ turns out to be
equal to the exponential sum:
\begin{equation*}
\frac{1}{p} \sum_{a\in \Fp^*} \psi\left(\frac{a+1}{a-1}\right)
\lgn(a) \chi(a),
\end{equation*}
where $\lgn$ denotes the Legendre character, and $\psi$ is a
non-trivial additive character of $\Fp$. This sum is very much
similar to the Kloosterman sum and the classical Weil bound
\cite{W1} yields the result.

In this paper, a proof for the rate conjecture is presented,
treating both cases of split and inert (\textit{non-split}) tori
in a uniform manner. A fundamental idea in our approach concerns a
non-trivial relation between two seemingly different dynamical
systems. One which is attached to a split ("non-compact") torus
and the other which is attached to a non-split ("compact") torus.
This relation is geometric in nature and can be formally described
in the framework of algebraic geometry.
\subsection{Geometric approach}
The basic observation to be made is that the theory of quantum
mechanics on the torus, in case $\h = \frac{1}{p}$, can be
equivalently recast in the language of representation theory of
finite groups in characteristic $p$. We will endeavor to give a
more precise explanation of this matter. Consider the quotient
$\Fp$-vector space $\V = \Td / p \Td$, where $\Td$ is the lattice
of characters on $\T$. We denote by $\gH = \gH(\V)$ the Heisenberg
group. The group $\Gf \iso \GG$ is naturally identified with the
group of linear symplectomorphisms of $\V$. We have an action of
$\GG$ on $\gH$. The Stone-von Neumann theorem states that there
exists a unique irreducible representation $\pi : \gH \lto
\GL(\H)$, with the non-trivial central character $\psi$, for which
its isomorphism class is fixed by $\GG$. This is equivalent to
saying that $\H$ is equipped with a compatible projective
representation $\rho : \GG \lto \PGL(\H)$. Noting that $\gH$ and
$\GG$ are the sets of rational points of corresponding algebraic
groups, it is natural to \textit{ask} whether there exists an
algebra-geometric object that underlies the pair $(\pi,\rho)$?.
The answer to this question is \textit{positive}. The construction
is proposed in an unpublished letter of Deligne to Kazhdan
\cite{D1}. In one sentence, the content of this letter is a
construction of \textit{Representation Sheaves} $\SK_\pi$ and
$\SK_\rho$ on the algebraic varieties $\Aheiz$ and $\AGG$
respectively. One obtains, as a consequence, the following general
principle:
\\
\\
\textbf{(*) Motivic principle}: \textit{All quantum mechanical
quantities in the Hannay-Berry model are motivic in nature.}
\\
\\
By this we mean that every quantum-mechanical quantity
$\mathcal{Q}$, is associated with a vector space $\V_\mathcal{Q}$
endowed with a Frobenius action $\Fr:\V_\mathcal{Q} \lto
\V_\mathcal{Q}$, so that:
$$ \mathcal{Q} = \Tr (\Fr_{|_{\V_\mathcal{Q}}}).$$
The \textit{main contribution} of this paper is to implement this
principle. In particular, we show that there exists a
two-dimensional vector space $\V_\chi$, endowed with an action
$\Fr : \V_\chi \lto \V_\chi$, so that:
$$ \Wigner_\chi(\xi) = \Tr (\Fr_{|_{\V_\chi}}).$$
This, combined with a bound on the modulus of the eigenvalues of
Frobenius, i.e., $$\left|\ev(\Fr_{|_{\V_\chi}})\right| \leq
\frac{1}{p^{1/2}},$$ completes the proof of the rate conjecture.
\\
\\
\subsection{Remarks}
There are several remarks that we would like to make at this
point:
\\
\\
\textbf{Remark 1: Discreteness principle.} ``Every'' quantity
$\mathcal{Q}$ that appears in the Hannay-Berry model admits
discrete spectrum in the following arithmetic sense: $\mathcal{Q}$
can take only values which are finite linear combinations of terms
with absolute value of the form $p^{i/2}$ for $i \in \Z$. This is
a consequence of the motivic principle (*) and Deligne's weight
theory \cite{D2}. This puts some restrictions on the possible
values of the modulus $|\mathcal{Q}|$. We believe that this
principle can be effectively used in various situations in order
to derive strong bounds out of weaker bounds. A \textit{striking
example} would be a possible alternative trivial "proof" for the
bound $|\Wigner_\chi(\xi)| \leq \frac{C_\xi}{p^{1/2}}$:

$$\left|\Wigner_\chi(\xi)\right| \leq \frac{C_\xi}{p^{1/4}} \r  \Rightarrow \r \left|\Wigner_\chi(\xi)\right|
\leq \frac{C_\xi}{p^{1/2}}\,.$$ Kurlberg and Rudnick proved in
their paper \cite{KR1} the weak bound $|\Wigner_\chi(\xi)| \leq
\frac{C_\xi}{p^{1/4}}$. This strongly indicates that the stronger
bound $ |\Wigner_\chi(\xi)| \leq \frac{C_\xi}{{p}^{1/2}}$ is
valid.
\\
\\
\textbf{Remark 2: Higher dimensional exponential sums.} Proving
the bound $| \Wigner_\chi(f) | \leq \frac{C_f}{\sqrt{p}}$ can be
equivalently stated as bounding by $\frac{C_f}{\sqrt{p}}$ the
spectral radius of the operator $\Av_{\CA} (f) = \frac{1}{|\CA|}
\sum\limits_{B \in \CA} \rhoh(B) \Pih(f) \rhoh(B^{-1})$. This
implies a bound on the $L_N$ norms, for every $N \in \Z^+$:
\begin{equation} \label{Lnorm}
\left \| \Av_{\CA} (f) \right \|_{N} \leq \frac{C_f}{p^N}.
\end{equation}
In particular, for $0 \neq f = \xi \in \Td$ one can compute
explicitly the left hand side of (\ref{Lnorm}) and obtain:
$$\left \| \Av_{\CA} (\xi) \right \|_{N} = \Tr ( |\Av_{\CA} (\xi)|^N )
=  \frac{1}{|\CA|^{2N}}\sum_{(x_1,\ldots,x_{2N}) \in
\X}\psi(\sum_{i < j} \ome(x_i,x_j)),$$
where $\X = \{(x_1,\ldots,x_{2N})|\r x_i\in {\cal O}_\xi,\r\sum
x_i = 0 \}$ and ${\cal O}_\xi = \CA \cdot \xi \subset \V$ denotes
the orbit of $\xi$ under the action of $\CA$. Therefore, referring
to (\ref{Lnorm}) we obtained a \textit{non-trivial} bound for a
higher dimensional exponential sum. It would be interesting to
know whether there exists an independent proof for this bound and
whether this representation theoretic approach can be used to
prove optimal bounds for other interesting higher dimensional
exponential sums.
\subsection{Sato-Tate conjecture}
The next level of the theory is to understand the \textit{complete
statistics} of the Hecke-Wigner distributions for different Hecke
states. More precisely, let us fix a character $\xi \in \Td$. For
every character $\chi :\CA \lto \C^*$ we consider the normalized
value $ \nWigner_\chi(\xi) = \frac{1}{\sqrt{p}}
\Wigner_\chi(\xi)$, which lies in the interval $[-2,2]$. Now,
running over all multiplicative characters we define the following
atomic measure on the interval $[-2,2]$:
%
%
$$ \mu_p = \frac{1}{|\CA|} \sum_{\chi} \delta_{\nWigner_\chi(\xi)}. $$
One would like to describe the limit measure (if it exists!). This
is the content of another conjecture of Kurlberg and Rudnick
\cite{KR2}:
\\
\\
\textbf{Conjecture (Sato-Tate Conjecture).} The following limit
exists:
$$ \lim_{p \rightarrow \infty} \mu_p = \mu_{ST},$$
where $\mu_{ST}$ is the push-forward of the Haar measure on
$\mathrm{SU}(2)$ to the interval $[-2,2]$ through the map $g
\mapsto\Tr(g)$.
\\
\\
We hope that using the methodology described in this paper one
will be able to gain some progress in proving this conjecture.
\\
\\
\textbf{Remark.} Note that the family $\{ \nWigner_\chi(\xi)
\}_{\chi \in \CAD}$ runs over a non-algebraic space of parameters.
Hence Deligne's equidistribution theory (cf. Weil II \cite{D2})
can not be applied directly in order to solve the Sato-Tate
Conjecture.

\subsection{Results} \label{results}
\begin{enumerate}
\item \textbf{Kurlberg-Rudnick conjecture.} The main result of this paper is Theorem \ref{GH3}, which
is the \textit{proof} of the Kurlberg-Rudnick rate conjecture
(Conjecture \ref{RC}) on the asymptotic behavior of the
Hecke-Wigner distributions.
\item \textbf{Weil representation.} We introduce two new constructions of the Weil
  representation over finite fields.
  \begin{enumerate}
  \item The \textit{first} construction is stated in Theorem \ref{GH2},
  and is based on the Rieffel \textit{quantum torus} $\Ad$, for $\h = \frac{1}{p}$. This approach is essentially
  equivalent to the classical approach \cite{Ge, H, Kl,W2} that uses the representation
  theory of the Heisenberg group in characteristic $p$. The fundamental difference is that the quantum torus is well
  defined for every value of the parameter $\h$.
  \item \textbf{Canonical Hilbert space (Kazhdan's question).} The \textit{second} construction uses
  the \textit{"method of Canonical Hilbert
  Space"} (see Appendix \ref{metaplectique}). This approach is based on
  the following statement:
\\
\\
\textbf{Proposition (Canonical Hilbert Space).} Let $(\V,\ome)$ be
a two-dimensional symplectic vector space over the finite field
$\Fq$. There exists a canonical Hilbert space $\Hc$ attached to
$(\V,\ome)$.
\\
\\
An immediate consequence of this proposition is that all
symmetries of $(\V,\ome)$ automatically act on $\H_\V$. In
particular, we obtain a \textit{linear} representation of the
group $\Sp = \Sp(\V,\ome)$ on $\H_\V$. This approach has higher
dimensional generalization, for the case where $\V$ is of
dimension $2n$. This generalization will be published by the
authors elsewhere.
\\
\\
\textbf{Remark.} Note the main difference of our construction from
the classical approach due to Weil (cf. \cite{W2}). The
\textit{classical construction} proceeds in two stages. Firstly,
one obtains a projective representation of $\Sp$, and secondly,
using general arguments about the group $\Sp$ one proves the
existence of a linearization. A consequence of \textit{our
approach} is that there exists a \textit{distinguished} linear
representation, and its existence is not related to any group
theoretic property of $\Sp$. We would like to mention that this
approach answers, in the case of the two-dimensional Heisenberg
group, \textit{a question} of David Kazhdan \cite{Ka} dealing with
the existence of Canonical Hilbert Spaces for co-adjoint orbits of
general unipotent groups. The main motive behind our construction
is the notion of \textit{oriented Lagrangian subspace}. This idea
was suggested to us by Joseph Bernstein \cite{B}.
\end{enumerate}
\item \textbf{Deligne's Weil representation sheaf.} We include for the sake of completeness (see Appendix
  \ref{delignes_letter}) a formal presentation of
  \textit{Deligne's letter to Kazhdan} \cite{D1}, that places the Weil representation
  on a complete algebro-geometric ground. As far as we know, the content of
  this letter was never published. This construction plays a \textit{central
  rule} in the proof of the Kurlberg-Rudnick conjecture.
\end{enumerate}
\subsection{Structure of the paper}
The paper is naturally separated into four parts:
\\
\\
\textbf{Part I.} Consists of sections \ref{classicaltorus},
\ref{quantization} and \ref{QHUE}. In section \ref{classicaltorus}
we discuss classical mechanics on the torus. In section
\ref{quantization} we discuss quantum mechanics \'{a}-la Hannay
and Berry, using the Rieffel quantum torus model. In section
\ref{QHUE} we \textit{formulate} the quantum unique ergodicity
theorem, i.e., Theorem \ref{GH3}. This part of the paper is
self-contained and consists of mainly linear algebraic
considerations.
\\
\\
\textbf{Part II.} Section \ref{Proof}. This is the main part of
the paper, consisting of the \textit{proof} of the
Kurlberg-Rudnick conjecture. The proof is given in two stages. The
first stage consists of mainly linear algebraic manipulations to
obtain a more transparent formulation of the statement, resulting
in Theorem \ref{GH4}. In the second stage we venture into
algebraic geometry. All linear algebraic constructions are
replaced by sheaf theoretic objects, concluding with the
\textit{Geometrization Theorem}, i.e., Theorem \ref{deligne}.
Next, the statement of Theorem \ref{GH4} is reduced to a geometric
statement, the \textit{Vanishing Lemma}, i.e., Lemma
\ref{vanishing}. The remainder of the section is devoted to the
proof of Lemma \ref{vanishing}. For the convenience of the reader
we include a large body of intuitive explanations for all the
constructions involved. In particular, we devote some space
explaining the Grothendieck {\it Sheaf to Function Correspondence}
procedure which is the basic bridge connecting Parts I and II.
\\
\\
\textbf{Part III}: Appendix \ref{metaplectique}. In section
\ref{canonical_hilbert} we describe the \textit{method of
canonical Hilbert space}. In section \ref{weilrep} we describe the
Weil representation in this manifestation. In section
\ref{realization} we relate the invariant construction to the more
classical constructions, supplying explicit formulas that will be
used later. In section \ref{delignes_letter} we give a formal
presentation of \textit{Deligne's letter to Kazhdan} \cite{D1}.
The main statement of this section is Theorem \ref{main_thm}, in
which the \textit{Weil representation sheaf} $\SK$ is introduced.
We include in our presentation only the parts of that letter which
are most relevant to our needs. In particular, we consider only
the two-dimensional case of this letter. In section
\ref{app_proofs} we supply proofs for all technical lemmas and
propositions appearing in the previous sections of the Appendix.
\\
\\
\textbf{Part IV}: Appendix \ref{proofs}. In this Appendix we
supply the proofs for all statements appearing in Part I and Part
II. In particular, we give the proof of Theorem \ref{deligne}
which essentially consists of taking the \textit{Trace} of
Deligne's {\it Weil representation sheaf} $\SK$.
\subsection*{Acknowledgments}
It is a pleasure to thank our Ph.D. adviser J. Bernstein for his
interest and guidance in this project. We appreciate the
contribution of P. Kurlberg and Z. Rudnick who discussed with us
their papers and explained their results. We thank P. Sarnak for
his invitation to visit the Courant Institute in September 2004 to
discuss our work. We would like to thank David Kazhdan for sharing
his thoughts about the possible existence of canonical Hilbert
spaces. We thank our friend D. Gaitsgory for the discussions on
$\ell$-adic cohomologies and Deligne's letter. We thank Y. Flicker
for helpful remarks. We thank B. Lewis for the editorial work. We
thank M. Baruch, the Technion, and Mina Teicher, the Emmy Noether
institute, for all the support. We would like to acknowledge the
referee for very important remarks. Finally, we would like to
thank Prof. P. Deligne for letting us publish his ideas about the
geometrization of the Weil representation which appeared in a
letter he wrote to David Kazhdan in 1982.
\section{Classical Torus} \label{classicaltorus}
Let $(\T,\ome)$ be the two-dimensional symplectic torus. Together
with its linear symplectomorphisms $\G \iso \GZ$ it serves as a
simple model of classical mechanics (a compact version of the
phase space of the harmonic oscillator). More precisely, let $\T=
\W/ \Lambda$ where $\W$ is a two-dimensional real vector space,
i.e.,  $\W \simeq \R^2$ and $\Lambda$ is a rank two lattice in
$\W$, i.e., $\Lambda \simeq \Z^2$. We obtain the symplectic form
on $\T$ by taking a non-degenerate symplectic form on $\W$:
$$\ome: \W \times \W \lto \R.$$
We require $\ome$ to be integral, namely, $\ome : \Lambda \times
\Lambda \lto
\Z$ and normalized, i.e., Vol$(\T) =1$.\\\\
Let ${\mathrm{Sp}}(\W, \ome)$ be the group of linear
symplectomorphisms, i.e., $\mathrm{Sp}(\W,\ome) \simeq
\mathrm{SL}_2(\R)$. Consider the subgroup $\G \subset
\mathrm{Sp}(\W,\ome)$ of elements that preserve the lattice
$\Lambda$, i.e., $\G (\Lambda) \subseteq \Lambda $. Then $\G
\simeq \GZ$. The
subgroup $\G$ is the group of linear symplectomorphisms of $\T$.\\
We denote by $\Lm \subseteq \W^*$ the dual lattice $\Lm = \{ \xi
\in \W^* | \r\r \xi (\Lambda) \subset \Z \} $. The lattice $\Lm$
is identified with the lattice of characters of $\T$ by the
following map:
\begin{equation*}
  \xi \in \Lm \longmapsto e^{2 \pi i <\xi, \cdot>} \in \r \Td,
\end{equation*}
where $\Td = \Hom(\T,\C^*)$.
\subsection{Classical mechanical system}
We consider a very simple discrete mechanical system. An
hyperbolic element $A \in \G$, i.e., $|\Tr(A)| > 2$, generates an
ergodic discrete dynamical system. The {\it Birkhoff's Ergodic
Theorem} states that:
\begin{equation*} 
  \lim_{N \rightarrow \infty} \frac{1}{N}\sum_{k=1}^{N} f(A^k
  x)= \int_{\T}f |\ome|,
\end{equation*}
for every $f \in \FST$ and for almost every point $x \in \T$. Here
$\FST$ stands for a good class of functions, for example
trigonometric polynomials or smooth functions.\\\\
We fix an hyperbolic element $A \in \G$ for the remainder of the paper.
\section{Quantization of the Torus} \label{quantization}
Quantization is one of the big mysteries of modern mathematics,
indeed it is not clear at all what is the precise structure which
underlies quantization in general. Although physicists have been
using quantization for almost a century, for mathematicians the
concept remains all-together unclear. Yet, in specific cases,
there are certain formal models for quantization that are well
justified mathematically. The case of the symplectic torus is one
of these cases. Before we employ the formal model, it is
worthwhile to discuss the general phenomenological
principles of quantization which are surely common for all models.\\\\
Let us start with a model of classical mechanics, namely, a
symplectic manifold, serving as a classical phase space. In our
case this manifold is the symplectic torus $\T$. Principally,
quantization is a protocol by which one associates a quantum
"phase" space $\H$ to the classical phase space $\T$, where $\H$
is a Hilbert space. In addition, the protocol gives a rule by
which one associates to every classical observable, namely a
function $f \in \FST$, a quantum observable ${\mathrm{Op}}(f) : \H
\lto \H$, an operator on the Hilbert space. This rule should send
a
real function into a self adjoint operator.\\\\
To be more precise, quantization should be considered not as a
single protocol, but as a one parameter family of protocols,
parameterized by $\hb$, the Planck constant. For every fixed value
of the parameter $\hb$ there is a protocol which associates to
$\T$ a Hilbert space $\H_{\hb}$ and for every function $f \in
\FST$ an operator ${\mathrm{Op}_{\hb}}(f) : \H_{\hb} \lto \H_{\hb}
$. Again the association rule should send real
functions to self adjoint operators.\\\\
Accepting the general principles of quantization, one searches for
a formal model by which to quantize, that is a mathematical model
which will manufacture a family of Hilbert spaces $\H_{\hb}$ and
association rules $\FST \leadsto \End(\H_{\hb})$. In this work we
employ a model of quantization called the {\it Weyl quantization
model}.
\subsection {The Weyl quantization model}
The Weyl quantization model works as follows. Let $\Ad$ be a one
parameter deformation of the algebra $\A$ of trigonometric
polynomials on the torus. This algebra is known in the literature
as the Rieffel torus \cite{Ri}. The algebra $\Ad $ is constructed
by taking the free algebra over $\C$ generated by the symbols
$\{s(\xi) \r | \r \xi \in \Lm \}$ and quotient out by the relation
$ s(\xi + \eta) = e^{\pih \ome(\xi,\eta)}s(\xi)s(\eta)$. Here
$\ome$ is the form on $\W^*$ induced by the original form $\ome$
on $\W$. We point out two facts about the algebra $\Ad$. First,
when substituting $\hb = 0$ one gets the group algebra of $\Lm$,
which is exactly equal to the algebra of trigonometric polynomials
on the torus. Second, the algebra $\Ad$ contains as a standard
basis the lattice $\Lm$:
\begin{equation*}                
s: \Lm \lto \Ad.
\end{equation*}
Therefore, one can identify the algebras $\Ad \simeq \A$ as vector
spaces. Therefore, every function $f \in \A$ can be viewed as an
element of
$\Ad$.\\\\
For a fixed $\hbar$ a representation $\Pih:\Ad \lto \End(\Hh)$
serves as a quantization protocol, that is, for every function $f
\in \A$ one has:
\begin{equation*}\label{quantprotocol}
 f \in {\A}  \simeq {\Ad}  \longmapsto \Pih (f) \in \End(\Hh)\def\ornlag
 {\cal{L}ag^{\circ}}.
\end{equation*}
An equivalent way of saying this is:
\begin{equation*}\label{quantprot2}
  f \longmapsto \sum_{\xi \in \Lm} a_{\xi} \Pih (\xi),
\end{equation*}
where $f = \sum\limits_{\xi \in \Lm} a_{\xi} \cdot \xi $ is the Fourier expansion of $f$.\\\\
To summarize: every family of representations $\Pih : \Ad \lto
\End(\Hh)$ gives us a complete quantization protocol. Yet, a
serious question now arises, namely what representations to
choose? Is there a correct choice of representations, both
mathematically, but  also perhaps physically? A possible
restriction on the choice is to choose an irreducible
representation. Yet, some ambiguity still remains because there
are several irreducible classes for specific values of $\hbar$.\\\\
We present here a partial solution to this problem in the case
where the parameter $\hbar$ is restricted to take only rational
values \cite{GH1}. Even more particularly, we take $\hbar$ to be
of the form $\hbar = \frac{1}{p}$ where $p$ is an odd prime
number. Before any formal discussion one should recall that our
classical object is the symplectic torus $\T$ \textit{together}
with its linear symplectomorphisms $\G$. We would like to quantize
not only the observables $\A$, but also the symmetries $\G$. Next,
we shall construct an equivariant quantization of $\T$.
\subsection {Equivariant Weyl quantization of the torus} \label{weilquant}
Let $\hbar = \frac{1}{p}$ and consider the additive character
$\psi :\Fp \lto \C^*,\r \psi(t) = e^{\frac{2 \pi i t}{p}}$. We
give here a slightly different presentation of the algebra $\Ad$.
Let $\Ad$ be the free $\C$-algebra generated by the symbols
$\{s(\xi) \;|\; \xi \in \Lm \}$ and the relations $s(\xi + \eta) =
\psi( \half \ome(\xi,\eta) ) s(\xi)s(\eta)$. Here we consider
$\ome$ as a map $\ome : \Lm \times \Lm \lto \Fp$. The lattice
$\Lm$ serves as a standard basis for $\Ad$:
\begin{equation*}                      
s: \Lm \lto \Ad.
\end{equation*}
The group $\G$ acts on the lattice $\Lm$. Therefore, it acts on
$\Ad$. It is easy to see that $\G$ acts on $\Ad$ by homomorphisms
of algebras. For an element $B \in \G$, we denote by $ f
\longmapsto f^B$ the action of $B$ on an element $f \in \Ad$.\\\\
An equivariant quantization of the torus is a pair:
\begin{eqnarray*}          
\Pih   :  {\Ad} & \lto & \End(\Hh),
\\
\rhoh  :  \G & \lto & \PGL(\Hh),
\end{eqnarray*}
where $\Pih$ is a representation of $\Ad$ and $\rhoh$ is a
projective representation of $\G$. These two should be compatible
in the following manner:
\begin{equation}\label{eqvquant3}
  \rhoh(B) \Pih(f) \rhoh(B)^{-1} = \Pih(f^B),
\end{equation}
for every $B \in \G$ and $f \in \Ad$. Equation (\ref{eqvquant3})  is called the {\it Egorov identity}.\\\\
Let us suggest now a construction of an equivariant quantization
of
the torus.\\\\
Given a representation $\pi: \Ad \lto \End(\H)$ and an element $B
\in \G$, we construct a new representation $\pi^B:\Ad \lto
\End(\H)$:
\begin{equation}\label{actcat}
  \pi^B(f) = \pi(f^B).
\end{equation}
This gives an action of $\G$ on the set $\Irr$ of classes of irreducible representations. The set $\Irr$ has a
very regular structure as a principal homogeneous space over $\T$. Moreover, every irreducible representation
of $\Ad$ is finite dimensional and of dimension $p$. The following theorem plays a central role in the
construction.
\begin{theorem}[Canonical invariant representation \cite{GH1}]\label{GH}
Let $\hbar = \frac{1}{p},$ where $p$ is a prime\footnote{The
Theorem holds more generally, i.e., for all Planck constants of
the form $\h = M/N$ where $M,N$ are co-prime integers. However, in
this paper we will not need to consider this generality}. There
exists a \textit{unique} $($up to isomorphism$)$ irreducible
representation $(\Pih,\Hh)$ of $\Ad$ for which its equivalence
class is fixed by $\G$.
\end{theorem}
Let $(\Pih,\Hh)$ be a representative of the fixed irreducible
equivalence class. Then for every $B \in \G$ we have:
\begin{equation}\label{iso}
\Pih^{B} \simeq \Pih.
\end{equation}
This means that for every element $B \in \G$ there exists an
operator $\rhoh(B)$ acting on $\Hh$ which realizes the isomorphism
(\ref{iso}). The collection $\{\rhoh(B) : B \in \G \}$ constitutes
a projective representation:
\begin{equation}\label{projrep}
  \rhoh:\G \lto \PGL(\Hh).
\end{equation}
Equations (\ref{actcat}) and (\ref{iso}) also imply the  Egorov
identity (\ref{eqvquant3}).\\\\
The group $\G \simeq \GZ$ is almost a free group and it is
finitely presented. A brief analysis \cite{GH1} shows that every
projective representation of $\G$ can be lifted (linearized) into
a true representation. More precisely, it can be linearized in 12
different ways, where 12 is the number of characters of $\G$. In
particular, the projective representation (\ref{projrep}) can be
linearized (\textit{not} uniquely) into an honest representation.
The next theorem (to be proved in Appendix \ref {proofs}) proposes
the existence of a canonical linearization. Let $\Gf \iso \GG$
denotes the quotient group of $\G$ modulo $p$.
\begin{theorem}[Canonical linearization]\label{GH2} Let $\hbar =
\frac{1}{p}$, where $p \neq 2,\,3$. There exists a \textit{unique}
linearization:
\begin{equation*}
\rhoh:\G \lto \GL(\Hh),
\end{equation*}
characterized by the property that it factors through the quotient
group $\Gf$:
\[
\qtriangle[\G`\Gf`\GL(\Hh);`\rhoh`\bar{\rho}_{_\h}]
\]
\end{theorem}
From now on $\rhoh$ means the linearization of Theorem \ref{GH2}.
\\
\\
\textbf{Summary.} Theorem \ref{GH} confirms the existence of a
unique invariant representation of $\Ad$, for every $\hbar =
\frac{1}{p}$. This gives a canonical equivariant quantization
$(\Pih,\rhoh,\Hh)$. Moreover, for $p \neq 2,\,3$ by Theorem
\ref{GH2}, the projective representation $\rhoh$ can be linearized
in a canonical way to give an honest representation of $\G$ which
factors through $\Gf$\footnote{This is the famous Weil
representation of $\GG$.}. Altogether this gives a pair:
\begin{eqnarray*}
\Pih  :  {\Ad} & \lto & \End(\Hh),
\\
\rhoh  :  \Gf & \lto & \GL(\Hh)
\end{eqnarray*}
satisfying the following compatibility condition (Egorov identity):
\begin{equation*}
  \rhoh(B) \Pih(f) \rhoh(B)^{-1} = \Pih(f^B),
\end{equation*}
for every $B \in \Gf$, $f \in \Ad$. The notation $\Pih(f^B)$ means
that we take any pre-image $\bar{B} \in \G$ of $B \in \Gf$ and act
by it on $f$. Note, that the operator $\Pih(f^{\bar{B}})$ does not
depend on the choice of $\bar{B}$. In the  following, we denote
the Weil representation $\bar{\rho}_{_\h}$ by $\rhoh$ and consider
$\Gf$ to be the default domain.
\subsection{Quantum mechanical system}
Let $(\Pih,\rhoh,\Hh)$ be the canonical equivariant quantization.
Let $A$ be our fixed hyperbolic element, considered as an element
of $\Gf$ . The element $A$ generates a quantum dynamical system as
follows. Take  a (pure) quantum state $\Psi \in S(\Hh) = \{ \Psi
\in \Hh :\|\Psi\|=1\}$ and act on it with $A$:
\begin{equation}\label{Qsystem}
  \Psi \longmapsto \Psi^A = \rhoh(A)\Psi.
\end{equation}
\section{Hecke Quantum Unique Ergodicity} \label{QHUE}
The main silent question of this paper is whether the system
(\ref{Qsystem}) is quantum ergodic. Before discussing this
question, one is obliged to define a notion of quantum ergodicity.
As a first approximation, follow the classical definition, but
replace each classical notion by its quantum counterpart. That is,
for every  $f \in \Ad$ and almost every quantum state $\Psi \in
S(\Hh)$, the following holds:
\begin{eqnarray}\label{quantumergodicity}
 \lim_{N \rightarrow \infty} \frac{1}{N}\sum_{k=1}^{N}
  < \Psi|\Pih(f^{A^k})\Psi > \above{?}= \int_{\T}f |\ome|.
\end{eqnarray}
Unfortunately (\ref{quantumergodicity}) is literally not true. The
limit is never exactly equal to the integral for a fixed $\hbar$.
Let us now give a true statement which is a slight modification of
(\ref{quantumergodicity}), called the {\it Hecke Quantum Unique
Ergodicity}. First, rewrite (\ref{quantumergodicity}) in an
equivalent form. We have:
\begin{equation}\label{rw1}
  <\Psi|\Pih(f^{A^k})\Psi> = <\Psi|\rhoh(A^k) \Pih(f) \rhoh(A^k)^{-1}
  \Psi>,
\end{equation}
using the Egorov identity (\ref{eqvquant3}).\\\\
Now, note that the elements $A^k$ run inside the finite group
$\Gf$. Denote by $\GrA \subseteq \Gf$ the cyclic subgroup
generated by $A$. It is easy to see, using (\ref{rw1}), that:
\begin{equation*}
\lim_{N \rightarrow \infty} \frac{1}{N}\sum_{k=1}^{N}
<\Psi|\Pih(f^{A^k})\Psi>= \frac{1}{|\GrA|}\sum_{B \in \GrA}
<\Psi|\rhoh(B) \Pih(f) \rhoh(B)^{-1} \Psi>.
\end{equation*}
Altogether (\ref{quantumergodicity}) can be written in the form:
\begin{equation}\label{rw3}
\Av_{_{\GrA}} (<\Psi|\Pih(f)\Psi>) \above{?}= \int_{\T}f |\ome|,
\end{equation}
where $\Av_{_{\GrA}}$ denotes the average of the \textit{Wigner
distribution} $ <\Psi|\Pih(f)\Psi>$ with respect to the group
$\GrA$.
\subsection{Hecke theory}
Denote by $\CA$ the centralizer of $A$ in $\Gf \iso \GG$. The
finite group $\CA$ consists of the rational points of an algebraic
group $\ACA$. Moreover, in the case where the characteristic of
the field does not divide $\Tr(A)^2-4$ the group $\ACA$ is an
algebraic torus. We call $\CA$ the {\it Hecke torus} (cf.
\cite{KR1}). One has, $\GrA \subseteq \CA \subseteq \Gf$. Now, in
(\ref{rw3}) take the average with respect to the group $\CA$
instead of the group $\GrA$. The statement of the
\textbf{Kurlberg-Rudnick rate conjecture} (cf. \cite{KR1, R1, R2})
is given\footnote{Our conjecture is equivalent to the original
statement \cite{KR1}, which treats only the case of common
eigenstates of the Hecke torus.} in the following theorem:
\begin{theorem}[Hecke Quantum Unique Ergodicity]\label{GH3} Let $\hbar =
\frac{1}{p}$, where $p$ is a sufficiently large
prime\footnote{What one really needs here is that any non-trivial
fixed element $\xi \in \Lm$ will not be an eigenvector for the
action of $A$ on the quotient $\Fp$-vector space $\Lm/p\Lm$ for
sufficiently large $p$. Hence, Theorem \ref{GH3} holds true for
every regular element in $\Gamma$ that has no eigenvectors in the
integral lattice $\Lm$. The last property holds not only for
hyperbolic elements! For example, Theorem \ref{GH3} holds for the
Weyl element $w = \left (\begin{smallmatrix} 0 & -1
\\ 1 & 0
\end{smallmatrix} \right )$.}. For every $f \in \Ad$ and $\Psi \in
S(\Hh)$, we have:
\begin{equation}\label{qheckerg}
\left| \Av_{_{\CA}} ( <\Psi| \Pih(f) \Psi> ) - \int_{\T}f |\ome|
\right| \leq \frac{C_{f}}{\sqrt{p}},
\end{equation}
where $C_{f}$ is an explicit constant depending only on $f$.
\end{theorem}
Section \ref{Proof} is devoted to proving Theorem \ref{GH3}.
\section{Proof of the Hecke Quantum Unique Ergodicity Conjecture}\label{Proof}
The proof is given in two stages. The first stage is a preparation
stage and consists mainly of linear algebra considerations. We
reduce statement (\ref{qheckerg}) in several steps into an
equivalent statement which will be better suited to our needs. In
the second stage we introduce the main part of the proof, invoking
tools from algebraic geometry in the framework of $\ell$-adic
sheaves and $\ell$-adic cohomology (cf. \cite{M, BBD}).
\subsection{Preparation stage}
\textbf{Step 1.} It is enough to prove Theorem \ref{GH3} for the
case when $f$ is a non-trivial character $\xi \in \Lm$. Because
$\int_{\T}\xi |\ome| = 0$, statement (\ref{qheckerg}) becomes :
\begin{equation}\label{qheckerg2}
\left| \Av_{_{\CA}} ( <\Psi|\Pih(\xi) \Psi> )\right| \leq
\frac{C_\xi}{\sqrt{p}}.
\end{equation}
The statement for general $f \in \Ad$ follows directly from the
triangle inequality.
\\\\
\textbf{Step 2.} It is enough to prove (\ref{qheckerg2}) in case
$\Psi \in S(\Hh)$ is a {\it Hecke} eigenstate. To be more precise,
the {\it Hecke} torus $\CA$ acts semisimply on $\Hh$ via the
representation $\rhoh$, thus  $\Hh$ decomposes to a direct sum of
character spaces:
\begin{equation}\label{decom}
\Hh = \bigoplus_{\chi:\CA \lto \C^*}\H_{\chi}.
\end{equation}
The sum in (\ref{decom}) is over multiplicative characters of the
torus $\CA$. For every $\Psi \in \H_{\chi}$ and $B \in \CA$, we
have:
\begin{equation*}
\rhoh(B)\Psi = \chi(B)\Psi.
\end{equation*}
Taking $\Psi \in \H_{\chi}$, statement (\ref{qheckerg2}) becomes:
\begin{equation}\label{qheckerg3}
\left|<\Psi|\Pih(\xi)\Psi> \right| \leq \frac{C_\xi}{\sqrt{p}}.
\end{equation}
Here $C_\xi = 2+o(1)$, where we use here the standard $o$
notation.\footnote{In the case where $\CA$ is non-split we have
$C_\xi = 2$.}
\\\\
The  averaged operator:
\begin{equation*}
\Av_{_{\CA}} (\Pih(\xi)) = \frac{1}{|\CA|}\sum_{B \in \CA}
\rhoh(B) \Pih(\xi) \rhoh(B)^{-1},
\end{equation*}
is essentially\footnote{This follows from Remark \ref{lemma1}. If
$\CA$ does not split over $\Fp$ then $\Av_{_{\CA}}(\Pih(\xi))$ is
diagonal in the Hecke basis. In case $\CA$ splits then for the
Legendre character $\sigma$ we have that dim $\H_\sigma = 2$.
However, in the later case one can prove (\ref{qheckerg2}) for
$v\in \H_\sigma$ by a computation of explicit eigenstates (cf.
\cite{KR2}).} diagonal in the Hecke base. Knowing this, statement
(\ref{qheckerg2}) follows from (\ref{qheckerg3}) by invoking the
triangle inequality.\\\\
\textbf{Step 3.} Let $P_{\chi}:\Hh \lto \Hh$ be the orthogonal
projector on the eigenspace $\H_{\chi}$.
\begin{remark}\label{lemma1}
For $\chi$ other then the quadratic character of $\;\CA$ we have
$\dim\;\H_{\chi} = 1.$\footnote{This fact, which is needed if we
want to stick with the matrix coefficient formulation of the
conjecture, can be proven by algebro-geometric techniques or
alternatively by a direct computation (cf. \cite{Ge}).}
\end{remark}
Using Remark \ref{lemma1} we can rewrite (\ref{qheckerg3}) in the
form:
\begin{equation}\label{statment}
\left|\Tr(P_{\chi} \Pih(\xi)) \right| \leq \frac{C_\xi}{\sqrt{p}}.
\end{equation}
The projector $P_{\chi}$ can be defined in terms of the representation
$\rhoh$:
\begin{equation*}               
P_{\chi} = \frac{1}{|\CA|}\sum_{B \in \CA} \chi^{-1}(B) \rhoh(B).
\end{equation*}
Now write (\ref{qheckerg3}) in the form:
\begin{equation}\label{qheckerg4}
\frac{1}{|\CA|} \left| \sum_{B \in \CA} \Tr( \rhoh(B) \Pih(\xi))
\chi^{-1}(B) \right| \leq \frac{C_\xi}{\sqrt{p}}.
\end{equation}
On noting that $|\CA| = p \pm1$ and multiplying both sides of
(\ref{qheckerg4}) by $|\CA|$ we obtain that it is enough to prove the following statement:\\
\begin{theorem}[Hecke Quantum Unique Ergodicity (Restated)]\label{GH4}
Fix a non-trivial $\xi \in \Lm$. Let $\hbar = \frac{1}{p}$, where
$p$ is a sufficiently large prime. For every character $\chi$ the
following holds:
\begin{equation*}
\left| \sum_{B \in \CA} \Tr( \rhoh(B) \Pih(\xi)) \chi(B) \right|
\leq 2 \sqrt{p}.
\end{equation*}
\end{theorem}
\subsection {The trace function}
We prove Theorem \ref{GH4} using sheaf theoretic techniques.
Before diving into geometric considerations, we investigate
further the ingredients appearing in Theorem \ref{GH4}. Denote by
$F$ the function $F : \Gf \times \Lm \lto \C$ defined by:

\begin{equation}\label{functionF}
F(B,\xi) = \Tr(\rho(B) \Pih(\xi)).
\end{equation}

We denote by $\V = \Lm / p \Lm$ the quotient vector space, i.e.,
$\V \simeq \Fp^2$. The symplectic form $\ome$ specializes to give
a symplectic form on $\V$. The group $\Gf$ is the group of linear
symplectomorphisms of $\V$, i.e., $\Gf = \Sp(\V)$. Set $\Y = \Gf
\times \Lm$ and $\YY = \Gf \times \V$. One has (for a proof, see
Section \ref{proofFACT}) the quotient map:
\begin{equation*}
\Y \lto \YY.
\end{equation*}
\begin{lemma}\label{factorization}
The function $F : \Y \lto \C$ factors through the quotient $\YY$.
\[
\qtriangle[\Y`\YY`\C;`F`\overline{F}]
\]
\end{lemma}
Denote the function $\overline{F}$ also by $F$ and from now on
$\YY$ will be considered as the default domain. The function
$F:\YY \lto \C$ is invariant under a certain group action of
$\Gf$. To be more precise, let $S \in \Gf$. Then:
\begin{equation*}
\Tr(\rhoh(B) \Pih(\xi)) = \Tr(\rhoh(S) \rhoh(B) \rhoh(S)^{-1}  \rhoh(S) \Pih(\xi) \rhoh(S)^{-1}).
\end{equation*}
Applying the Egorov identity (\ref{eqvquant3}) and using the fact
that $\rhoh$ is a representation we get:
\begin{equation*}
\Tr( \rhoh(S) \rhoh(B) \rhoh(S)^{-1}  \rhoh(S) \Pih(\xi) \rhoh(S)^{-1}
)= \Tr(\Pih(S \xi) \rhoh(S B S^{-1})).
\end{equation*}
Altogether we have:
\begin{equation}\label{invar3}
F(B, \xi) = F(SBS^{-1} , S \xi).
\end{equation}
%
%
Putting (\ref{invar3}) in a more diagrammatic form: there is an
action of $\Gf$ on $\YY$ given by the following formula:
\begin{equation}\label{actionset}
  \begin{CD}
   \Gf \times \YY    @>\alpha>> \YY,\\
   (S,(B, \xi))      @>>>  (SBS^{-1} , S \xi).
   \end{CD}
\end{equation}
Consider the following diagram:
\begin{equation*}
  \begin{CD}
   \YY    @<pr<<    \Gf\times \YY    @>\alpha>> \YY,\\
  \end{CD}
\end{equation*}
where $pr$ is the projection on the $\YY$ variable. Formula
(\ref{invar3}) can be stated equivalently as:
\begin{equation*}
  \alpha^{*}(F) = pr^{*}(F),
\end{equation*}
where $\alpha^{*}(F)$ and $pr^{*}(F)$ are the pullbacks of the function $F$ on $\YY$
via the maps $\alpha$ and $pr$ respectively.
\subsection{Geometrization (Sheafification)}
Our next goal is to reduce Theorem \ref{GH4} to a geometric
statement, i.e., Lemma \ref{vanishing}. The main tool which we
invoke is called the "Geometrization" procedure. In this procedure
one replaces sets by algebraic varieties and functions by sheaf
theoretic objects (which are quite similar to vector bundles). The
main statement of this section will be presented in the
"Geometrization Theorem", i.e., Theorem \ref{deligne}.

\subsubsection{Algebraic geometry}
First, we have to devote some space recalling notions and
notations from algebraic geometry over finite fields and the
theory of $\ell$-adic sheaves.

\textbf{Varieties.} In the sequel, we shall translate back and
forth between algebraic varieties defined over the finite field
$\Fp$, and their corresponding sets of rational points. In order
to prevent confusion between the two, we use bold-face letters for
denoting a variety $\AX$, and normal letters for denoting its
corresponding set of rational points $\X$. An algebraic variety,
which is defined over the finite field $\Fp$, is an algebraic
variety equipped with an endomorphism called Frobenius:

\begin{equation*}   
  \Fr:\AX \lto \AX.
\end{equation*}

This is also sometimes called {\it rational} structure. We denote
by $\X$ the set of points fixed by the Frobenius, that is,:

\begin{equation*}
  \X = \AX^{\Fr} = \{ x \in \AX : \Fr(x) = x \}.
\end{equation*}

Another common notation for this set is $\X = \AX(\Fp)$.

In more detail, in this paper we decided not to use scheme
theoretic language. Hence, all spaces considered are plain
algebraic varieties defined over the algebraically closed field
$\FFp$, and points are morphisms $Spec(\FFp) \rightarrow \AX$.
Given an algebraic variety $\AX$, there exists the following
Cartesian square:

\begin{equation*}
\begin{CD}
\AX_0         @>\Fr>>           \AX \\
      @VVV                               @VVV \\
 Spec(\FFp) @>\Fr>>                      Spec(\FFp),
\end{CD}
\end{equation*}

where $\Fr:Spec(\FFp) \rightarrow Spec(\FFp)$, corresponds by
duality to the Frobenius endomorphism of the field $\FFp$. A
variety $\AX$ is said to be defined over the finite field $\Fp$,
if it is equipped with an isomorphism $\nu:\AX \rightarrow \AX_0$.
We denote the composition $ \Fr \circ \nu : \AX \rightarrow \AX $,
also by $\Fr$.
\\
\\
\textbf{Sheaves.} Let $\Db ({\AX} )$  denote the "triangulated"
category of constructible $\ell$-adic sheaves on $\AX $ (cf.
\cite{M, BBD}, in addition see \cite{BL} for equivariant sheaves
theory). We denote by $\Perv (\AX)$ the Abelian category of
perverse sheaves on $\AX$, that is the heart with respect to the
autodual perverse t-structure in $\Db(\AX)$ (cf. \cite{BBD}). We
will use also the notion of $N$-perversity: an object $\SF$ in
$\Db (\AX)$ is called $N$-perverse if $\SF [-N] \in \Perv (\AX)$.

Finally, we define the notion of a Weil structure (or Frobenius
structure). By a {\it Weil structure} on an object $\SF \in \Db
(\AX)$ we mean an isomorphism:

$$ \theta : \Fr^* \SF \simeq \SF.$$

The pair $(\SF, \theta)$ above is called a {\it Weil object}. By
an abuse of notation we often denote $\theta$ also by $\Fr$. We
fix once an identification $\Qlb \simeq \C$. Therefore, all
sheaves are considered to be with coefficients over the complex
numbers.

In the sequel we will use the following two standard sheaves. We
denote by $\Sartin$ the Artin-Schreier sheaf on the group $\Ga$
that corresponds to the additive character $\psi$ on the group
$\Fp = \Ga(\Fp)$, and by $\SL_{\lgn}$ the Kummer sheaf on the
multiplicative group $\Gm$ that corresponds to the Legendre
quadratic character $\lgn$ on the group $\Fp^\ast = \Gm(\Fp)$.

\subsubsection{The Geometrization Theorem}
Having said that, we can begin replacing all our sets with their
corresponding algebraic varieties. The symplectic vector space
$(\V,\omega)$ is identified as the set of rational points of an
algebraic variety $\AV$. The variety $\AV$ is equipped with a
morphism $\omega:\AV \times \AV \rightarrow \FFp$ respecting the
Frobenius structure on both sides. The group $\Gf$ is identified
as the set of rational points of the algebraic group $\ASp$.
Finally, the set $\YY$ is identified as the set of rational points
of the algebraic variety $\AYY$. More precisely, $\AYY \simeq \ASp
\times \AV$.  We denote by $\alpha$ the action of $\ASp$ on the
variety $\AYY$ (cf. (\ref{actionset})). We choose once an
identification of the symplectic vector space with the standard
symplectic plane:

\begin{equation} \label{identification1}
  (\AV,\omega) \simeq (\bA^2, \omega_{\mathrm{std}}),
\end{equation}

where $\omega_{\mathrm{std}}$ is the standard symplectic form
defined by the condition $\omega_{\mathrm{std}}((1,0),(0,1)) = 1$.
This induces an identification:

\begin{equation} \label{identification2}
  \ASp \simeq \AGG.
\end{equation}

Consider the standard torus $\AT \subset \AGG$, i.e., $\AT$
consists of diagonal matrices of the form $\left
(\begin{smallmatrix} a & 0
\\ 0 & a^{-1} \end{smallmatrix} \right )$. We use the notation
$\AT^{\times}$ for denoting the punctured torus $\AT
\smallsetminus \mathrm{I}$.
\\
\\
Our next goal is to replace functions by appropriate sheaf
theoretic objects. The following theorem (for a proof, see Section
\ref{proofdeligne}) proposes an appropriate sheaf theoretic object
standing in place of the function $F : \YY \lto \C$
(\ref{functionF}).
\begin{theorem}[Geometrization Theorem]\label{deligne}
There exists a Weil object $\SF \in \Db(\AYY) $ satisfying the
following properties:
\end{theorem}
%
%
\begin{enumerate}

\item \label{prop_del1} (Perversity) The object $\SF$ is geometrically irreducible
$\dim(\AYY)$-perverse of pure weight $w(\SF) = 0$.\footnote{We
thank the referee for bringing to our attention a deep observation
regarding these striking properties.}

\item \label{prop_del2} (Function) The function $F$ is associated to $\SF$ via \textit{sheaf-to-function correspondence}:
\begin{equation*}
  f^{\SF} = F.
\end{equation*}
\item \label{prop_del3} (Equivariance)  For every element $S \in \ASp$ there exists an
isomorphism:

  $$\alpha_S ^* \SF \simeq \SF.$$

\item \label{prop_del4} (Formula) Restricting the sheaf $\SF$ on the subvariety $\AT^{\times} \times
\AV$. Using the identifications (\ref {identification1}),
(\ref{identification2}). We have an explicit formula:
\begin{equation}\label{dependentformula}
  \SF_{|_{\AT^{\times} \times \AV}} ( \left (\begin{smallmatrix} a & 0
\\ 0 & a^{-1} \end{smallmatrix} \right ) , \lambda, \mu)  \simeq \SL_{\lgn(a)} \otimes
  \SL_{\psi(\half \frac{a+1}{a-1} \lambda \cdot \mu )}.
\end{equation}
\end{enumerate}

\textbf{Comments.}
\begin{enumerate}
\item In Property \ref{prop_del3}, in fact a finer statement is true. In the case that $S \in
\ASp(\Fq)$, $q=p^n$, one can show that the isomorphism $\alpha^*_S
\SF \simeq \SF$ is an isomorphism of Weil sheaves on $\AYY$,
considered as an algebraic variety defined over $\Fq$.

\item In Property $\ref{prop_del4}$, one can produce an invariant formula, without
using any identification. Moreover, this formula is valid on an
open subvariety. We shall now explain this further. Let $j: \AU
\hookrightarrow \ASp$ denote the open subvariety consisting of
elements $g \in \ASp$ such that $g-\mathrm{I}$ is invertible.
Restricting the sheaf $\SF$ to the open subvariety $\AU \times
\AV$, we have the following isomorphism:
\begin{equation}\label{invariantformula}
\SF_{|_{\AU \times \AV}} \simeq \SL_{\psi(\quarter \omega ( \frac
{g+\mathrm{I}}{g-\mathrm{I}} v,v))} \otimes
\SL_{\lgn(\Tr(g)-2)}.\footnote{In particular, this implies that
$\SF$ coincides with the middle extension $j_{!*} (
\SL_{\psi(\quarter \omega (
\frac{g+\mathrm{I}}{g-\mathrm{I}}v,v))} \otimes
\SL_{\lgn(\Tr(g)-2)})$. This means that one obtains a complete
description of the sheaf $\SF$.}
\end{equation}
It is a direct calculation to verify that the invariant formula
(\ref{invariantformula}) coincides with the coordinate dependent
formula (\ref{dependentformula}) when one restricts to the
standard (punctured) torus $\AT^{\times}$.
\end{enumerate}

%
%
\underline{Explanation}
\\
\\
We give here an \textit{intuitive} explanation of Theorem
{\ref{deligne}, part by part, as it was stated. An object $\SF \in
\Db(\AYY)$ can be considered as a vector bundle $\SF$ over $\AYY$:
\begin{equation*}
  \begin{CD}
  \SF \\
  @VVV \\
  \AYY.
\end{CD}
\end{equation*}
Saying that $\SF$ is a \textit{Weil sheaf} means that it is
equipped with a lifting of the Frobenius, that is,:
\begin{equation*}
  \begin{CD}
   \SF  @<\Fr<<  \SF \\
   @VVV        @VVV \\
   \AYY @>\Fr>>  \AYY
  \end{CD}
\end{equation*}

\textbf{Remark.} We deliberately choose the lifting above in the
opposite direction in order to make our intuitive explanation
consistent with the formal definitions.
\\
\\
To be even more precise, think of $\SF$ not as a single vector
bundle, but as a complex $\SF = \SF^{\bullet}$ of vector bundles
over $\AYY$:
\begin{equation*}
  \begin{CD}
   ...  @>d>> \SF^{-1}  @>d>> \SF^{0}  @>d>> \SF^{1}  @>d>> ...
   \end{CD}
\end{equation*}
The complex $\SF^{\bullet}$ is equipped with a lifting of Frobenius:
\begin{equation*}
  \begin{CD}
   ...  @>d>> \SF^{-1}  @>d>> \SF^{0}  @>d>> \SF^{1}  @>d>> ... \\
    &    &      @V\Fr VV         @V\Fr VV           @V\Fr VV            \\
   ...  @>d>> \SF^{-1}  @>d>> \SF^{0}  @>d>> \SF^{1}  @>d>> ...
   \end{CD}
\end{equation*}
Here the Frobenius commutes with the differentials.
Next, we explain the meaning of Property \ref{prop_del1}. We will
not try to explain here the notion of perversity, neither the
notion of purity (cf. \cite{BBD, D2}). Going into these matters
will take us to far a part. It is enough for the purpose of this
paper to explain the meaning of  the sheaf $\SF$ being of a mixed
weight $w(\SF) \leq 0$. This condition is implied by the condition
of $\SF$ being of pure weight $w(\SF) = 0$. Let $y \in \YY$ be a
point fixed by Frobenius. Denote by $\SF_{y}$ the fiber of $\SF$
at the point $y$. Thinking of $\SF$ as a complex of vector
bundles, it is clear what one means by taking the fiber at a
point. The fiber $\SF_{y}$ is just a complex of vector spaces.
Because the point $y$ is fixed by the Frobenius, it induces an
endomorphism of $\SF_{y}$:
\begin{equation}\label{liftingcompy}
  \begin{CD}
   ...  @>d>> \SF^{-1}_{y}  @>d>> \SF^{0}_{y}  @>d>> \SF^{1}_{y}  @>d>> ... \\
    &    &      @V\Fr VV         @V\Fr VV           @V\Fr VV            \\
   ...  @>d>> \SF^{-1}_{y}  @>d>> \SF^{0}_{y}  @>d>> \SF^{1}_{y}  @>d>> ...
   \end{CD}
\end{equation}
The Frobenius acting as in (\ref{liftingcompy}) commutes with the
differentials. Hence, it induces an action on cohomologies. For
every $i \in \Z $ we have an endomorphism:
\begin{equation}\label{actcohom}
  \Fr :\coH^{i}(\SF_{y}) \lto \coH^{i}(\SF_{y}).
\end{equation}
Saying that an object $\SF$ has mixed weight $w(\SF) \leq w$ means
that for every point $y \in \YY$ and for every $i \in \Z$ the
absolute value of the eigenvalues of $\Fr$ acting on the $i$'th
cohomology (\ref{actcohom}) satisfy:
\begin{equation*}
 \left |\ev(\Fr \big{|}_{\coH^{i}(\SF_{y})}) \right | \leq
 \sqrt{p}^{w+i}.
\end{equation*}
In our case $w = 0$ and therefore:
\begin{equation}\label{weightdefinition}
 \left |\ev(\Fr \big{|}_{\coH^{i}(\SF_{y})}) \right | \leq
 \sqrt{p}^{\;i}.
\end{equation}
Property \ref{prop_del2} of Theorem \ref{deligne} concerns a
function $f^{\SF}: \YY \lto \C$ associated to the sheaf $\SF$. To
define $f^{\SF}$, we only have to describe its value at every
point $y \in \YY$. For a given $y \in \YY$ the Frobenius acts on
the cohomologies of the fiber $\SF_{y}$  (cf. (\ref{actcohom}) ).
Now put:
\begin{equation*}
  f^{\SF}(y) = \sum_{i \in \Z} (-1)^i \Tr(\Fr
  \big{|}_{\coH^i(\SF_y)}).
\end{equation*}
In other words, the value $f^{\SF}(y)$ is the alternating sum of
traces of the operator $\Fr$ acting on the cohomologies of the
fiber $\SF_y$. This alternating sum is called the {\it Euler
characteristic} of Frobenius and it is denoted by:
\begin{equation*}
  f^{\SF}(y) = \chi_{_{\Fr}}(\SF_y).
\end{equation*}
Theorem \ref{deligne} confirms that $f^{\SF}$ is the trace
function $F$ defined earlier by formula (\ref{functionF}).
Associating the function $f^{\SF}$ on the set $\AYY^{\Fr}$ to the
sheaf $\SF$ on $\AYY$ is a particular case of a general procedure
called {\it Sheaf-to-Function Correspondence} \cite{G}. As this
procedure will be used later, next we spend some space explaining
it in greater details (cf.
\cite{Ga}).\\\\
\underline{\textbf{Grothendieck's Sheaf-to-Function
Correspondence}}
\\
\\
Let $\AX$ be an algebraic variety defined over $\Fq$. Let ${\cal
L} \in \Db(\AX)$ be a Weil sheaf. One can associate to ${\cal L}$
a function $f^{\cal L}$ on the set $\X$ by the following formula:
\begin{equation*}
  f^{\cal L}(x) = \sum_{i \in \Z} (-1)^i \Tr(\Fr \big{|}_{\coH^i({\cal
  L}_x)}).
\end{equation*}
This procedure is called {\it Sheaf-To-Function correspondence}.
Next, we describe some important functorial
properties of this procedure.\\\\
Let $\AX_1$, $\AX_2$ be algebraic varieties defined over $\Fq$.
Let $\X_1 = \AX^{\Fr}_1$ and $\X_2 = \AX^{\Fr}_2$ be the
corresponding sets of rational points. Let $\pi : \AX_1 \lto
\AX_2$ be a morphism of algebraic varieties. Denote also by
$ \pi : \X_1 \lto \X_2$ the induced map on the level of sets.\\\\
%
%
%
\textbf{First statement}}. Suppose we have a Weil sheaf ${\cal L}
\in \Db(\AX_2)$. The following holds:
\begin{equation}\label{sfc1}
 f^{\pi^{*}({\cal L})} = \pi^{*}(f^{\cal L}),
\end{equation}
where on the function level $\pi^{*}$ is just the pull back of
functions. On the sheaf theoretic level $\pi^{*}$ is the pull-back
functor of sheaves (think of pulling back a vector bundle).
Equation (\ref{sfc1}) states that the
{\it Sheaf-to-Function Correspondence} commutes with the operation of pull back.\\\\
%
%
\textbf{Second statement}. Suppose we have a Weil sheaf ${\cal L}
\in \Db(\AX_1)$.  The following holds:
\begin{equation}\label{sfc2}
 f^{\pi_{!}({\cal L})} = \pi_{!}(f^{\cal L}),
\end{equation}
where on the function level, $\pi_{!}$ means to sum up the values
of the function along the fibers of the map $\pi$. On the sheaf
theoretic level, $\pi_{!}$ is a compact integration of sheaves
(here we have no analogue under the vector bundle interpretation).
Equation (\ref{sfc2}) states that the {\it Sheaf-to-Function
Correspondence} commutes with integration.\\\\
%
%
\textbf{Third statement}. Suppose we have two Weil sheaves ${\cal
L}_{1}, {\cal L}_{2} \in \Db(\AX_1)$. The following holds:
\begin{equation}\label{sfc3}
f^{{\cal L}_{1} \otimes {\cal L}_{2}} = f^{{\cal L}_{1}} \cdot
f^{{\cal L}_{2}}.
\end{equation}
In other words, {\it Sheaf-to-Function Correspondence} takes
tensor product of sheaves to multiplication of the corresponding
functions.
\subsection{Geometric statement}
Fix an element $\xi \in \Lm$ and a sufficiently large prime $p$ so
that $\xi$ (mod p)
 is not a $\CA$-eigenvector\footnote{Note that
this can be done for every $\xi\in \Lm$ due the fact that $A \in
\GZ$ is an hyperbolic element and does not have eigenvectors in
$\Lm$.}. We denote by $\i_XI$ the inclusion map $\i_XI : \CA
\times \xi \lto \YY $.
%
%
Returning to Theorem \ref{GH4} and putting its content in a
functorial notation, we write the following inequality:
\begin{equation*}
  \left| pr_! ( \i_XI ^* (F) \cdot \CHI) \right| \leq 2\sqrt {p}.
\end{equation*}
In other words, taking the function $F : \YY \lto \C$ and:
\begin{itemize}
  \item Restrict $F$ to $\CA \times \xi $ and get $\i_XI^ * (F)$.
  \item Multiply $\i_XI^ * F $ by the character $\CHI$ to get
  $\i_XI^ * (F) \cdot \CHI$.
  \item Integrate $\i_XI^ * (F)  \cdot \CHI$ to the point, this means to sum up all
  its values, and get a scalar $a_{\CHI} = pr_! ( \i_XI^ * (F)  \cdot
\CHI)$. Here $pr$ stands for
  the projection $pr : \CA \times \xi \lto pt$.
\end{itemize}
Then Theorem \ref{GH4} asserts that the scalar $a_{\CHI}$ is of  an absolute value less than $2\sqrt{p}$.\\\\
%
%
Repeat the same steps in the geometric setting. We denote again by
$\i_XI$ the closed imbedding  $  \i_XI: \ACA \times \xi \lto \AYY
$. Take the sheaf $\SF$ on $\AYY$ and apply the following sequence
of operations:

\begin{itemize}
  \item Pull-back $\SF$ to the closed subvariety $ \ACA \times \xi $ and  get the sheaf $\i_XI^ * (\SF)$.

  \item Take the tensor product of $\i_XI^ * (\SF)$ with the Kummer
  sheaf ${\SL}_{\CHI}$ and get $\i_XI^
* (\SF) \otimes
  {\SL}_\CHI$.

  \item Integrate $\i_XI^ * (\SF) \otimes {\SL}_\CHI$ to the point
    and get the sheaf $pr_! (\i_XI^ * (\SF) \otimes {\SL}_\CHI)$ on the point.
\end{itemize}
The operation of {\it Sheaf-to-Function Correspondence} commutes
both with pullback (\ref{sfc1}), with integration (\ref{sfc2}) and
takes the tensor product of sheaves to the multiplication of
functions (\ref{sfc3}). This means that it intertwines the
operations carried out on the level of sheaves with those carried
out on the level of functions. The following diagram describes
pictorially what has been said so far:
\begin{equation*}
\begin{CD}
   \SF                                        @>\chiFr >>              & F               \\
   @A\i_XIAA            &                                            @A \i_XI AA            \\
  \i_XI^{*}(\SF) \otimes {\SL}_{\CHI}      @>\chiFr >>    & \i_XI^{*}(F) \cdot \CHI    \\
   @V pr  VV           &                                             @V pr VV        \\
 {pr}_{!}(\i_XI^{*}(\SF) \otimes {\SL}_{\CHI})  @>\chiFr >>     &  {pr}_{!}(\i_XI^{*}(F)\cdot
 \CHI).
\end{CD}
\end{equation*}
%
%
Recall the weight property $w(\SF) \leq 0$. Now, the effect of
functors $\i_XI^*$, $pr_!$ and tensor product $\otimes$ on the property of weight should be examined.\\\\
%
%
The functor $\i_XI^{*}$ does not increase weight. Observing the
definition of weight this claim is immediate. Therefore, we get:
\begin{equation*}
  w(\i_XI^ * (\SF)) \leq 0.
\end{equation*}
%
%
Assume we have two sheaves  ${\cal L}_1$ and  ${\cal L}_2$ of
mixed weights $w({\cal L}_1) \leq w_{_1}$ and $w({\cal L}_{2})
\leq w_{_2}$. Then the weight of the tensor product ${\cal L}_{1}
\otimes {\cal L}_{2}$ is of mixed weight $w({\cal L}_{1} \otimes
{\cal L}_{2}) \leq w_{_1} + w_{_2}$. This is again immediate from
the definition of weight.
\\\\
%
%
Knowing that the Kummer sheaf has pure weight $w(\SL_{\CHI}) = 0$,
we deduce:
\begin{equation*}
  w(\i_XI^*(\SF) \otimes \SL_{\CHI}) \leq 0.
\end{equation*}
%
%
Finally, one has to understand the effect of the functor $pr_!$.
The following theorem, proposed by  Deligne \cite{D2}, is a very
deep and important result in the theory of weights. Briefly
speaking, the theorem states that compact integration of sheaves
does not increase weight. Here is the precise statement:
\begin{theorem}[Deligne, Weil II \cite{D2}]\label{deligne2}
Let $\pi :\AX_1 \lto \AX_2$ be a morphism of algebraic varieties.
Let ${\cal L} \in \Db(\AX_1)$ be a sheaf of mixed weight $w({\cal
L}) \leq w$ then the sheaf $\pi_! ({\cal L})$ is of mixed weight
$w(\pi_! ({\cal L})) \leq w$.
\end{theorem}
Using Theorem \ref{deligne2} we get:
\begin{equation*}
 w( pr_! (\i_XI^*(\SF) \otimes \SL_{\CHI})) \leq 0.
\end{equation*}
Consider the sheaf $\SG = pr_! (\i_XI^*(\SF) \otimes \SL_{\CHI})$.
It is an object in $\Db(pt)$. This means it is merely a complex of
vector spaces, $\SG = \SG^{\bullet}$, together with an action of
Frobenius:
\begin{equation*}
  \begin{CD}
   ...  @>d>> \SG^{-1}  @>d>> \SG^{0}  @>d>> \SG^{1}  @>d>> ... \\
    &    &      @V\Fr VV         @V\Fr VV           @V\Fr VV            \\
   ...  @>d>> \SG^{-1}  @>d>> \SG^{0}  @>d>> \SG^{1}  @>d>> ...
   \end{CD}
\end{equation*}
The complex $\SG^{\bullet}$ is associated by {\it Sheaf-To-Function correspondence} to the scalar $a_{\CHI}$:
\begin{equation}\label{geulerchar}
  a_{\CHI} = \sum_{i \in \Z} (-1)^i \Tr(\Fr \big{|}_{\coH^i(\SG)}).
\end{equation}
Finally, we can give the geometric statement  about $\SG$,  which will imply Theorem \ref{GH4}.
\begin{lemma}[Vanishing Lemma]\label{vanishing}
Let $\SG = pr_! (\i_XI^*(\SF) \otimes \SL_{\CHI})$, where $\xi$ is
not a $\CA$-eigenvector. All cohomologies $\coH^{i}(\SG)$ vanish
except for $i=1$. Moreover, $\coH^1(\SG)$ is a two-dimensional
vector space.
\end{lemma}
Theorem \ref{GH4} now follows easily. By Lemma \ref{vanishing}
only the first cohomology $\coH^1(\SG)$ does not vanish and it is
two-dimensional. Having $w(\SG) \leq 0$ implies (using
(\ref{weightdefinition})) that the eigenvalues of Frobenius acting
on $\coH^1(\SG)$ are of absolute value $ \leq \sqrt{p}$. Hence,
using formula (\ref{geulerchar}) we get:
%
\begin{equation*}
  | a_{\CHI} | \leq 2 \sqrt{p}.
\end{equation*}
The remainder of this section is devoted to the proof of Lemma
\ref{vanishing}.
\subsection{Proof of the Vanishing Lemma}
The proof will be given in several steps.\\\\
\textbf{Step 1.} We use the identifications
(\ref{identification1}), and (\ref{identification2}). Note that
all tori in $\AGG$ are conjugated. Therefore, there exists an
element $\S0 \in \AGG$ conjugating the {\it Hecke} torus $\ACA
\subset \AGG$ with the standard torus $\AT$:
\begin{equation*}
\S0 \ACA \S0^{-1} = \AT.
\end{equation*}
%
%
The situation is displayed in the following diagram:
\begin{equation*}
  \begin{CD}
    \AGG \times \bA^2                        @>\alpha_{_{\S0}}>>                          \AGG \times \bA^2 \\
      @A\i_XIAA                                   @                                             A \iXII AA \\
    \ACA \times \xi                        @>\alpha_{_{\S0}}>>                          \AT \times \XII \\
     @V pr VV                                                                              @V pr VV     \\
      pt                                  @=                                                   pt
 \end{CD}
\end{equation*}
where $\XII = \S0 \cdot \XI$ and  $\alpha_{_{\S0}}$ is the
restriction of the action $\alpha$ to the element $\S0$.
\\
\\
%
%
\textbf{Step 2.} Using the equivariance property of the sheaf
$\SF$ (see Theorem \ref{deligne}, Property  \ref{prop_del3}) we
will show that it is \textit{sufficient} to prove the Vanishing
Lemma for the sheaf $\SG_{st} = pr_! (\iXII ^* \SF \otimes
{\alpha_{_\S0}}_! \Skummer )$.
\\
\\
Indeed, we have:
\begin{equation}\label{seq1}
  \SG = pr_! ( \i_XI^* \SF  \otimes \Skummer ) \iso pr_! {\alpha_{_\S0}}_{_{!}} ( \i_XI^* \SF  \otimes
  \Skummer).
\end{equation}
The morphism $\alpha_{_\S0}$ is an isomorphism. Therefore,
${\alpha_{_\S0}}_!$ commutes with taking $\otimes$, hence we
obtain:
\begin{equation}\label{seq2}
 pr_! {\alpha_{_{\S0}}}_{_{!}} (\i_XI^*(\SF) \otimes \Skummer ) \iso pr_! ({\alpha_{_{\S0}}}_! (\i_XI^* \SF) \otimes
{\alpha_{_{\S0}}}_! (\Skummer)).
\end{equation}
Applying base change we obtain:
\begin{equation}\label{seq3}
 {\alpha_{_{\S0}}}_! \i_XI^* \SF  \iso \iXII^* {\alpha_{_{\S0}}}_!
 \SF.
\end{equation}
Now using the equivariance property of the sheaf $\SF$ we have the isomorphism:
\begin{equation}\label{seq4}
   {\alpha_{_{\S0}}}_! \SF  \simeq \SF.
\end{equation}
Combining (\ref{seq1}), (\ref{seq2}), (\ref{seq3}) and (\ref{seq4})  we get:
\begin{equation}\label{seq5}
   pr_!( \i_XI^* \SF  \otimes \Skummer ) \iso  pr_! (\iXII^* \SF  \otimes {\alpha_{_{\S0}}}_! \Skummer
   ).
\end{equation}
Therefore, we see from (\ref{seq5}) that it is sufficient to prove
vanishing of cohomologies for:
\begin{equation}\label{seq6}
  \SG_{st} = pr_! (\iXII^* \SF  \otimes {\alpha_{_{\S0}}}_!
  \Skummer).
\end{equation}
However, this is  a situation over the standard torus and we can
compute explicitly all the sheaves involved!
\\
\\
\textbf{Step 3.} The Vanishing Lemma holds for the sheaf $\SG_{st}$.\\\\
What remains is to compute (\ref{seq6}). We write $ \XII =
(\lambda,\mu)$. By Theorem \ref{deligne} Property \ref{prop_del4},
we have $\iXII^* \SF \simeq \SL_{\psi( \half \frac{a+1}{a-1}
\lambda \cdot \mu)} \otimes \SL_{\lgn(a)}$, where $a$ is the
coordinate of the standard torus $\AT$ and $\lambda \cdot \mu \neq
0$\footnote{Recall that $\eta$ is not a
$\mathrm{T}$-eigenvector.}. The sheaf $ {\alpha_{_{\S0}}}_!
\Skummer $
  is a character sheaf on the torus $\AT$. Hence we get that
(\ref{seq6}) is a kind of a Kloosterman-sum sheaf. A direct
computation (Appendix \ref{proofs}  section \ref{compvanishing})
proves the Vanishing Lemma for this sheaf. This completes the
proof of the Hecke quantum unique ergodicity conjecture. $\EProof$

\textbf{Comment.} We mention that, in this paper we obtained also
an alternative proof of the Vanishing Lemma, i.e., using the
invariant formula for the sheaf $\SF$ (see Formula
(\ref{invariantformula})). Indeed, noting that the invariant
formula of the sheaf $\SF$ is valid on an open subvariety $\AU
\subset \ASp \times \AV$, which contains $\ACA^{\times} \times
\AV$, we proceed and prove the statement directly, without the
need to use the equivariance property, and with essentially the
same cohomological computations.
%
%
%
%
%
\section*{\center Appendix}
\begin{appendix}
\section{Metaplectique}\label{metaplectique}
In the first part of this Appendix we give new construction of the
{\it Weil} ({\it metaplectic}) {\it representation}
$(\rho,\mathrm{Sp}(\V),\Hc)$, attached to a two-dimensional
symplectic vector space $(\V,\ome)$ over $\Fq$, which appears in
the body of the paper. The difference is that here the
construction is slightly more general. Even more importantly, it
is obtained in completely natural geometric terms. The focal step
in our approach is the introduction of a \textit{canonical Hilbert
space} on which the Weil representation is naturally manifested.
The motivation to look for this space was initiated by a question
of David Kazhdan \cite{Ka}. The key idea behind this construction
was suggested to us by Joseph Bernstein \cite{B}. The upshot is to
replace the notion of a Lagrangian subspace by a more refined
notion of an {\it oriented Lagrangian subspace}\footnote{We thank
A. Polishchuk for pointing out to us that this is an
$\Fq$-analogue of well known considerations, due to Lion and
Vergne \cite{LV}, with usual oriented Lagrangians giving
explicitly the metaplectic covering of $\mathrm{Sp}(2n,\R)$.}.
\\
\\
In the second part of this Appendix we apply a
\textit{geometrization} procedure to the construction given in the
first part, meaning that all sets are replaced by algebraic
varieties and all functions are replaced by $\ell$-adic sheaves.
This part is based on a letter of Deligne to Kazhdan from 1982
\cite{D1}. We extract from that work only the part that is most
relevant to this paper. Although all basic ideas appear already in
the letter, we tried to give here a slightly more general and
detailed account of the construction. As far as we know, the
contents of this mathematical work has never been published. This
might be a good enough reason for writing this part.
\\
\\
The following is a description of the Appendix. In section
\ref{canonical_hilbert} we introduce the notion of oriented
Lagrangian subspace and the construction of the canonical Hilbert
space. In section  \ref{weilrep}  we obtain a natural realization
of the Weil representation. In section \ref{realization} we give
the standard Schr\"{o}dinger realization (cf. \cite{Ge, H, W2}).
We also include several formulas for the kernels of basic
operators. These formulas will be used in section
\ref{delignes_letter} where the geometrization procedure is
described. In section \ref{app_proofs} we give proofs of all
lemmas and propositions which appear in previous sections.
\\
\\
For the remainder of the Appendix we fix the following notations.
Let $\Fq$ denotes the finite field of characteristic $p \neq 2$
and $q$ elements. Fix $\psi : \Fq \lto \C^*$ a non-trivial
additive character. Denote by $\lgn : \Fqm \lto \C^*$ the {\it
Legendre} multiplicative quadratic character.
\subsection{Canonical Hilbert space} \label{canonical_hilbert}
\subsubsection{Oriented Lagrangian subspace}
Let $(\V,\ome)$ be a $2$-dimensional symplectic vector space over
$\Fq$.
%
%
\begin{definition} \label{oriented_lagrangian_def}
An oriented Lagrangian subspace is a pair $(\L,\orn_{_\L})$, where
$\L$ is a Lagrangian subspace of $\V$ and $\orn_{_\L}: \L
\smallsetminus \{0\} \to \{\pm 1\}$ is a function which satisfies
the following equivariant property:
%
%
\begin{equation*}
\orn_{_\L}(t \cdot l) = \lgn(t) \orn_{_\L}(l),
\end{equation*}
where $t \in \Fqm$ and $\lgn$ the Legendre character of $\Fqm$.
\end{definition}
We denote by $\ornlag$ the space of oriented Lagrangians
subspaces. There is a forgetful map $\ornlag \lto \lag$, where $\lag$
is the space of Lagrangian subspaces, $\lag \simeq \projI(\Fq)$. In the
sequel we use the notation $\L^{\circ}$ to specify that $\L$ is
equipped with an orientation.
\subsubsection{The Heisenberg group}
Let $\gH$ be the Heisenberg group. As a set we have $\gH = \V
\times \Fq$. The multiplication is defined by the following
formula:
%
%
\begin{equation*}
(v,\lambda) \cdot (v',\lambda ') = (v+v',\lambda + \lambda ' +
\half \ome(v,v')).
\end{equation*}
We have a projection $\pi: \gH \lto \V$. We fix a section of this
projection:
\begin{equation}\label{section}
 s:\V \dashrightarrow \gH, \rev s(v) = (v,0).
\end{equation}
\subsubsection{Models of irreducible representation}
Given $\ornL = (\L,\orn_{_\L}) \in \ornlag$, we construct the
Hilbert space $\H_{\ornL} = {\mathrm{Ind}}_{\tilde{\L}}^{\gH} \;
\C_{\tilde{\psi}}$, where $\tilde{\L} = \pi^{-1}(\L)$ and
$\tilde{\psi}$ is the extension of the additive character $\psi$
to $\tilde{\L}$ using the section $s$, i.e., $\tilde{\psi} :
\tilde{\L} = \L  \times \Fq   \lto \C^*$ is given by the formula:
%
%
\begin{equation*}
\tilde{\psi}(l,\lambda) = \psi(\lambda).
\end{equation*}
More concretely: $\H_{\ornL} = \{ f:\gH \lto \C \; | \; f(\lambda
l h) = \psi(\lambda) f(h) \}$. The group $\gH$ acts on
$\H_{\ornL}$ by multiplication from the right. It is well known
(and easy to prove) that the representations $\H_{\ornL}$ of $\gH$
are irreducible and for different $\ornL$'s they are all
isomorphic. These are different models of the same irreducible
representation. This is  stated in the following theorem:
%
%
\begin{theorem}[Stone-von Neumann]\label{isomorphic_models}
For an oriented Lagrangian subspace $\ornL$, the representation
$\H_{\ornL}$ of $\gH$ is irreducible. Moreover, for any two
oriented Lagrangians $\ornLI,\ornLII \in \ornlag$ one has
$\H_{\ornLI} \simeq \H_{\ornLII}$ as representations of $\gH$.
$\EProof$
\end{theorem}
\subsubsection{Canonical intertwining operators}
Let $\ornLI,\ornLII \in \ornlag$ be two oriented Lagrangians. Let
$\H_\ornLI, \H_\ornLII$ be the corresponding representations of
$\gH$. We denote by $\interLILII = \Hom_{_\gH}
(\H_\ornLI,\H_\ornLII)$ the space of intertwining operators
between the two representations. Because all representations are
irreducible and isomorphic to each other we have
$\dim(\interLILII) = 1$. Next, we construct a canonical element in
$\interLILII$.
\\
\\
Let $\ornLI = (\LI,\orn_{_\LI})$, $\ornLII = (\LII, \orn_{_\LII})$
be two oriented Lagrangian subspaces. Assume they are in general
position, i.e., $\LI \neq \LII$. We define the following specific
element $\thetaLILII \in \interLILII, \;\; \thetaLILII :
\H_{\ornLI} \lto \H_{\ornLII}$. It is defined by the following
formula:
\begin{equation}\label{different_presentation}
\thetaLILII = \aLILII \cdot \thnLILII,
\end{equation}
where $\thnLILII : \H_\ornLI \lto \H_\ornLII$ denotes the standard
averaging operator and $\aLILII$ denotes the normalization factor.
The formulas are:
%
%
\begin{equation*}
\thnLILII(f)(h) = \sum_{l_2 \in \LII} f(l_2 h),
\end{equation*}
where $f \in \H_\ornLI$.
%
%
\begin{equation*}
\aLILII = \frac{1}{q} \sum_{l_1 \in \LI} \psi( \half \ome(l_1,
\xiLII)) \orn_{_\LI}(l_1) \orn_{_\LII}(\xiLII),
\end{equation*}
where $\xiLII$ is a fixed non-zero vector in $\LII$. Note that
$\aLILII$ does not depend on $\xiLII$.
\\
\\
Now we extend the definition of $\thetaLILII$ to
the case where $\LI =\LII$. Define:
%
%
\begin{equation*}
\thetaLILII = \left \{
\begin{matrix}
\;\;\mathrm{I}, & & \orn_{_\LI} & = & \orn_{_\LII} \\
    \mathrm{-I},& & \orn_{_\LI} & = &  -\orn_{_\LII} \\
\end{matrix} \right .
\end{equation*}
The \textit{main claim} is that the collection $\{ \thetaLILII
\}_{_{\ornLI,\ornLII \in\ornlag} }$ is associative. This is
formulated in the following theorem:
%
%
\begin{theorem}[Associativity]\label{associativity_thm}
Let $\ornLI,\ornLII,\ornLIII \in \ornlag$ be a triple of oriented
Lagrangian subspaces. The following associativity condition holds:
$$\thetaLIILIII \circ \thetaLILII = \thetaLILIII.$$
\end{theorem}

For a proof, see Section \ref{app_proofs}.
\subsubsection{Canonical Hilbert space}
Define the \textit{canonical Hilbert space} $\Hc \subset
\bigoplus\limits_{\ornL \in \ornlag} \H_{\ornL}$ as the subspace
of compatible systems of vectors, that is,:
$$\Hc = \{(f_{_\ornL})_{_{\ornL \in \ornlag}} ;\rev
\thetaLILII(f_{_\ornLI}) = f_{_\ornLII}\}.$$
%
%
%
\subsection{The Weil representation} \label{weilrep}
In this section we construct the Weil representation using the
Hilbert space $\Hc$. We denote by $\Sp = \mathrm{Sp}(\V,\ome)$ the
group of linear symplectomorphisms of $\V$. Before giving any
formulas, note that the space $\Hc$ was constructed out of the
symplectic space $(\V,\ome)$ in a complete canonical way. This
immediately implies that all the symmetries of $(\V,\ome)$
automatically acts on $\Hc$. In particular, we obtain a
\textit{linear} representation of the group $\Sp$ in the space
$\Hc$. This is the famous Weil representation of $\Sp$ and we
denote it by $\rho : \Sp \lto \GL (\Hc)$. It is given by the
following formula:
\\
\begin{equation}\label{Wrf} 
\rho(g)[(f_{_\ornL})] = (f_{_\ornL}^g).
\end{equation}
\\
Let us elaborate on this formula. The group $\Sp$ acts on the
space $\ornlag$. Any element $g \in \Sp$ induces an automorphism $g
: \ornlag \lto \ornlag$ defined by:
%
%
\begin{equation*}
(\L,\ornofL) \longmapsto (g \L,\ornofL^g),
\end{equation*}
where $\ornofL^g(l) = \ornofL(g^{-1} l)$. Moreover, $g$ induces
an isomorphism of vector spaces $g:\H_{\ornL} \lto \H_{g\ornL}$
defined by the following formula:
%
%
\begin{equation}\label{actionfiber}
f_{_\ornL}  \longmapsto f_{_\ornL}^g, \;\; f_{_\ornL}^g(h) =
f_{_\ornL}(g^{-1} h),
\end{equation}
where the action of $ g \in \Sp$ on $h = (v,\lambda) \in \gH$ is
given by $g(v,\lambda) = (gv,\lambda)$. It is easy to verify that
the action (\ref{actionfiber}) of $\Sp$ commutes with the
canonical intertwining operators, that is, for any two
$\ornLI,\ornLII \in \ornlag$ and any element $g \in \Sp$ the
following diagram is commutative:
\begin{equation*}  
\begin{CD}
\H_{\ornLI}   @>\thetaLILII>>     \H_{\ornLII} \\
  @VgVV                             @VgVV       \\
\H_{g\ornLI}   @>\thetagLIgLII>>   \H_{g\ornLII}.
\end{CD}
\end{equation*}
From this we deduce that formula (\ref{Wrf}) indeed gives the
action of $\Sp$ on $\Hc$.
\subsection{Realization and formulas}\label{realization}
In this section we give the standard Schr\"{o}dinger realization
of the Weil representation. Several formulas for the kernels of
basic operators are also included.
\subsubsection{Schr\"{o}dinger realization}
Fix  $\V = \VI \oplus \VII$ to be a Lagrangian decomposition of
$\V$. Fix $\orn_{_{\VII}}$ to be an orientation on $\VII$. Denote
by $\ornVII = (\VII,\orn_{_{\VII}})$ the oriented space. Using the
system of canonical intertwining operators we identify $\Hc$ with
a specific representative $\H_{\ornVII}$. Using the section $s:\V
\dashrightarrow \gH$ (cf. \ref{section}) we further make the
identification $s: \H_{\ornVII} \simeq \FSVI$, where $\FSVI$ is
the space of complex valued functions on $\VI$. We denote $\H =
\FSVI$. In this realization the Weil representation, $\rho : \Sp
\lto \GL (\H)$, is given by the following formula:
\begin{equation*}  
\rho(g)(f) = \thetagVIIVII(f^g),
\end{equation*}
where $f \in \H \simeq \H_{\ornVII}$ and $ g \in \Sp$.
\subsubsection{Formulas for the Weil representation}
First we introduce a basis $e \in \VI$ and the dual basis $e^* \in
\VII$ normalized so that $\ome(e,e^*) = 1$. In terms of this basis
we have the following identifications: $\V \simeq \Fq^2$,
$\VI,\VII \simeq \Fq$, $\Sp \simeq \mathrm{SL}_2(\Fq)$ and $\gH
\simeq \Fq^2 \times \Fq$ (as sets). We also have $\H \simeq
\FSFq$.
\\
\\
For every element $g \in \Sp$ the operator $\rho(g) : \H \lto \H$
is represented by a kernel $K_g:\Fq^2 \lto \C$. The multiplication
of operators becomes convolution of kernels.  The collection $\{
K_g \}_{g \in \Sp}$ gives a single function of ``kernels'' which
we denote by $K_\rho : \Sp \times \Fq^2 \lto \C$. For every
element $g \in \Sp$ the kernel $K_\rho(g)$ is of the form:
%
%
\begin{equation*}  
K_\rho(g,x,y) = a_g \cdot \psi(R_g(x,y)),
\end{equation*}
where $a_g$ is a certain normalizing coefficient and $R_g : \Fq^2 \lto \Fq$ is a quadratic function supported
on some linear subspace of $\Fq^2$. Next, we give an explicit description of the kernels $K_\rho(g)$.
\\
\\
Consider the (opposite) Bruhat decomposition $\Sp = \oB w \oB \cup
\oB$ where:
%
%
\begin{equation*}
\oB = \left (
\begin{matrix}
*  &   \\
*  &  * \\
\end{matrix}
\right),
\end{equation*}
and $w = \left(\begin{smallmatrix} 0 & 1 \\ -1 & 0
\end{smallmatrix} \right)$ is the standard Weyl element.
\\
\\
If $g \in \oB w \oB$ then:
%
%
\begin{equation*}  
g = \left (
\begin{matrix}
a & b \\
c & d
\end{matrix}
\right ),
\end{equation*}
where $ b \neq 0$. In this case we have:
%
%
\begin{eqnarray*}  
a_g & = & \frac{1}{q} \sum_{t \in \Fq} \psi(\frac{b}{2}t)
\lgn(t),\\
%
%
R_g(x,y) & = & \frac{-b^{-1}d}{2} x^2 + \frac{b^{-1} -c +
ab^{-1}d}{2} xy - \frac{ab^{-1}}{2} y^2.
\end{eqnarray*}
Altogether we have:
%
%
\begin{equation*}  
K_\rho (g,x,y) = a_g \cdot \psi(\frac{-b^{-1}d}{2} x^2 +
\frac{b^{-1} -c + ab^{-1}d}{2} xy - \frac{ab^{-1}}{2} y^2).
\end{equation*}
If $g \in \oB$ then:
%
%
\begin{equation*}  
g = \left (
\begin{matrix}
a & 0 \\
r & a^{-1}
\end{matrix}
\right ).
\end{equation*}
In this case we have:
%
%
\begin{eqnarray*}
a_g & = & \lgn(a),\\  
%
%
R_g(x,y) & = & \frac{-r a^{-1}}{2} x^2 \cdot \delta_{y=a^{-1}x}.  
\end{eqnarray*}
Altogether we have:
%
%
\begin{equation} \label{K_borel}
K_\rho (g,x,y) = a_g \cdot \psi(\frac{-r a^{-1}}{2} x^2)
\delta_{y=a^{-1}x}.
\end{equation}
\subsubsection{Formulas for the Heisenberg representation}
On $\H$ we also have a representation of the Heisenberg group
$\gH$. We denote it by $\pi : \gH \lto \GL(\H)$. For every element
$h \in \gH$ we have a kernel $K_h :\Fq^2 \lto \C$. We denote by
$K_\pi : \gH \times \Fq^2 \lto \C$ the function of kernels. For an
element $h \in \gH$ the kernel $K_\pi (h)$ has the form
$\psi(R_h(x,y))$ where $R_h$ is an affine function which is
supported on a certain one dimensional subspace of $\Fq^2$. Here
are the exact formulas:
\\
\\
For an element $h=(\cq,\cp,\lambda)$ we have:
%
%
$$ R_h(x,y) =  (\frac{\cp \cq}{2} + \cp x  + \lambda) \delta_{y=x+\cq},$$
%
%
\begin{eqnarray}\label{K_heizenberg}
K_\pi(h,x,y) & = & \psi(\frac{\cp \cq}{2} + \cp x  + \lambda)
\delta_{y=x+\cq}.
\end{eqnarray}
\subsubsection{Formulas for the representation of the Jacobi group}
The representations $\rho : \Sp \lto \GL(\H)$ and $\pi : \gH \lto
\GL(\H)$ combine together to give a representation of the
semi-direct product $\semi = \Sp \ltimes \gH$. The group $\semi$
is sometimes referred in literature as the {\it Jacobi group}.  We
denote the total representation by $ \rho \ltimes \pi : \semi \lto
\GL(\H)$, $\rho \ltimes \pi (g,h) = \rho(g) \cdot \pi(h)$. The
representation $ \rho \ltimes \pi$ is given by a kernel $K_{\rho
\ltimes \pi} : \semi  \times \Fq^2 \lto \C$. We denote this kernel
simply by $K$.
\\
\\
We give an explicit formula for the kernel $K$ only in the case
where $(g,h) \in \oB w \oB \times \gH$, i.e.,:
%
%
\begin{equation*}  
g = \left (
\begin{matrix}
a & b \\
c & d
\end{matrix}
\right),
\end{equation*}
where $ b \neq 0$ and $ h = (\cq,\cp,\lambda)$. In this case:
%
%
\begin{eqnarray}
\mRSwE(g,h,x,y) & = & R_g(x,y-\cq) + R_h(y-\cq,y) \label{R_map}, \\
%
%
K(g,h,x,y) & = & a_g \cdot \psi(R_g(x,y-\cq) + R_h(y-\cq,y)).
\label{K_Delta}
\end{eqnarray}
\subsection{Deligne's letter} \label{delignes_letter}
In this section we \textit{geometrize} (Theorem \ref{main_thm})
the total representation $\rho \ltimes \pi : \semi \lto \H$.
First, we realize all finite \textit{sets} as rational points of
certain algebraic \textit{varieties}. Beginning with the vector
space, we take $\V = \AV(\Fq)$. Next we replace all groups. We
take, $\gH = \Aheiz(\Fq)$, where $\Aheiz = \AV \times \Ga$, $\Sp =
\ASp(\Fq)$ and finally $\semi = \Asemi(\Fq)$, where $\Asemi = \ASp
\times \Aheiz$. The second step is to replace the kernel $K =
K_{\rho \ltimes \pi} : \semi \times \Fq^2 \lto \C$ (see
(\ref{K_Delta})) by a \textit{ sheaf} theoretic object. Recall
that $K$ is a kernel of a representation and hence satisfies
multiplication property:

\begin{equation} \label{mm}
(m,\mathrm{Id})^*K = p_1^*K * p_2^*K,
\end{equation}

where $m$ denotes the multiplication map, and $ m \times
\mathrm{Id} : \semi \times \semi \times \Fq^2 \rightarrow \semi
\times \Fq^2$ is given by $(g_1,g_2,x,y) \mapsto (g_1·g_2,x,y)$.
Finally, $p_i(g_1,g_2,x,y)=(g_i,x,y)$ are the projections on the
first and second $\semi$-coordinate respectively. The r.h.s of
(\ref{mm}) is the convolution:
$$p_1^*K * p_2^* K(g_1,g_2,x,y) =
\sum_{z \in \Fq} K(g_1,x,z)\cdot K(g_2,z,y).
$$
In the sequel, we will usually suppress the $\Fq^2$ coordinates,
writing $m : \semi \times \semi \rightarrow \semi$, and $p_i :
\semi \times \semi \rightarrow \semi$. Moreover, to enhance the
clarity of the notation, we will also suppress the projections
$p_i$ in (\ref{mm}), yielding a much cleaner statement:

\begin{equation} \label{mm1}
 m^*K=K*K,
\end{equation}

These conventions will continue to hold also in the geometric
setting, which is exactly where we will proceed to. We replace the
kernel $K$ by Deligne's \textit{Weil representation sheaf}
\cite{D1}. This is an object $\SK \in \Db(\Asemi \times \bA^2)$
that satisfies\footnote{However, in this paper we will prove a
weaker property (see Theorem \ref{main_thm}) which is sufficient
for our purposes.} the analogue (to (\ref{mm1})) multiplication
property:
\begin{equation*}
m^*\SK \iso \SK*\SK,
\end{equation*}
and its function is:
$$
f^{\SK} = K.
$$
\subsubsection{Existence of the Weil representation sheaf}
\textbf{The strategy.} The method of constructing the Weil
representation sheaf $\SK$ is reminiscent to some extent to the
construction of an analytic function via an analytic continuation.
In the realm of perverse sheaves one uses the operation of
\textit{perverse extension} (cf. \cite{BBD}), or, maybe,
preferably called in our context {\it middle
extension}\footnote{We use this as a unified terminology for
taking a perverse extension with respect to any chosen perverse
t-structure}. The main idea is to construct, using formulas, an
explicit irreducible (shifted) perverse sheaf $\SKO$ on a "good"
open subvariety $\O \subset \Asemi \times \bA^2$ and then we
obtain the sheaf $\SK$ by perverse extension of $\SKO$ to the
whole variety $\Asemi \times \bA^2$.
\\
\\
\textbf{Preliminaries.} We use in our construction the
identifications $(\AV,\omega) \simeq
(\bA^2,\omega_{\mathrm{std}})$, and $\ASp \simeq \AGG$. We denote
by $\O$ the open subvariety:
$$\O = \ASw \times \Aheiz \times \bA^2,$$
where $\ASw$ denotes the (opposite) big Bruhat cell $\AoB w \AoB
\subset \AGG$.
\\
\\
In the sequel we will frequently make use of the
\textit{character} property (cf. \cite{Ga}) of the sheaves
$\Sartin$ and $\SL_{\lgn}$, that is,:
%
%
\begin{eqnarray}
s^*\Sartin & \simeq & \Sartin \boxtimes \Sartin, \label{cspSa}
\\
m^*\SL_{\lgn} & \simeq & \SL_{\lgn} \boxtimes \SL_{\lgn},
\label{cspSl}
\end{eqnarray}
where $s: \Ga \times \Ga \to \Ga$ and $m: \Gm \times \Gm \to \Gm$
denotes the addition and multiplication morphisms correspondingly,
and $\boxtimes$ means exterior tensor product of sheaves.
\\
\\
Finally, given a sheaf ${\cal L}$ we use the notation ${\cal L}
[i]$ for the translation functors and the notation ${\cal L} (i)$
for the i'th Tate twist.
\\
\\
\textbf{Construction of the sheaf $\SK$.} In the first step we
sheafify the kernel $K_{\rho \ltimes \pi}$ of the total
representation, when restricted to the set $\mathrm{O} = \Sw
\times \gH \times \Fq^2$, using the formula (\ref{K_Delta}). We
obtain a sheaf $\SKO$ on the open subvariety $\AO = \ASw \times
\Aheiz \times \bA^2$:
\begin{equation*}
\SKO = \SASwE \otimes \SKnSwE,
\end{equation*}
where $\SKnSwE$ is the sheaf of the non-normalized kernels and
$\SASwE$ is the sheaf of the normalization coefficients. The
sheaves $\SKnSwE$ and $\SASwE$ are constructed as follows. Define
the morphism $\mRSwE : \ASw \times \Aheiz \times \bA^2 \lto \bA^1$
by formula (\ref{R_map}) and let $pr: \ASw \times \Aheiz \times
\bA^2 \lto \ASw$ be the projection morphism on the $\ASw$
coordinate. Now take:
\begin{eqnarray*}
\SKnSwE =  {\mRSwE}^* \Sartin, \\
\SASwE =  pr ^* \SASw,
\end{eqnarray*}
where:
\begin{equation*}
\SASw(g) = \int_{ x \in \bA^1} \SL_{\psi(\half bx)} \otimes
\SL_{\lgn(x)}[2](1).
\end{equation*}

Here $g = \left(\begin{smallmatrix} a & b \\ c & d
\end{smallmatrix}\right )$, and the notation $\int_{x \in \bA^1}$ means integration
with compact support along the $\bA^1$ fiber. It is a direct
consequence from the construction:

\begin{corollary}\label{propofSKO}
The sheaf $\SKO$  is geometrically irreducible
$[\dim(\Asemi)+1]$-perverse of pure weight zero and its function
agrees with K on $O$, that is, $f^{\SKO} = K_{|_O}$.
\end{corollary}

Let $j$ denote the open imbedding $ j: \AO \rightarrow \Asemi
\times \bA^2$.  We define the Weil representation sheaf $\SK$ as
the middle extension:

\begin{equation} \label{construction}
\SK = j_{!*} \SKO.
\end{equation}

We are now ready to state the main theorem of this section:
%
%
\begin{theorem}[Weil representation sheaf \cite{D1}]\label{main_thm}
The sheaf $\SK$ is geometrically irreducible $[\dim(\Asemi) +
1]$-perverse Weil sheaf on  $ \Asemi \times \bA^2$, of pure weight
$w(\SK) = 0$. The sheaf $\SK$ satisfies the following properties:
\end{theorem}
\begin{enumerate}
\item \label{prop1} (Function) The function associated to $\SK$, by sheaf to function correspondence,
is the Weil representation kernel $f^\SK = K$.
\item \label{prop2}(Multiplication) For every element $g \in \Asemi$ there exists
isomorphisms:
$$\SK_{|_g} * \SK \simeq L_g^*( \SK ),$$
and:
$$\SK * \SK_{|_g} \simeq R_g^*( \SK ),$$

where $\SK_{|_g}$ denotes the restriction of $\SK$ to the closed
subvariety $g \times \bA^2$ and $R_g,\; L_g :\Asemi \lto \Asemi$
are the morphisms of right and left multiplication by $g$
respectively.
\end{enumerate}
For a proof, see Section \ref{app_proofs}.

\textbf{Remark.} In Property \ref{prop2}. the notation $ g \in
\Asemi$ means that $g$ is a morphism $g : Spec(\FFp) \rightarrow
\Asemi$. In fact a finer statement is true, namely, if $g \in
\Asemi(\Fq),\, q=p^n$, then:

$$\SK_{|_g} * \SK \simeq L_g^*( \SK ),$$
and:
$$\SK * \SK_{|_g} \simeq R_g^*( \SK ),$$
are isomorphisms of Weil sheaves on $\Asemi \times \bA^2$,
considered as an algebraic variety over $\Fq$.

\subsection{Proofs}\label{app_proofs}
In this section we give the proofs for all technical facts that
appeared in Part \ref{metaplectique} of the Appendix.
\\
\\
\textbf{Proof of Theorem \ref{associativity_thm}}. Before giving
the proof, we introduce a structure which is inherent to
configurations of triple Lagrangian subspaces. Let $\LI,\LII,\LIII
\subset \V$ be three Lagrangian subspaces which are in a general
position. In our case, these are just three different lines in the
plane. Then the space $\mathrm{L}_j$ induces an isomorphism $\ri :
\mathrm{L}_{k} \lto \mathrm{L}_{i}$, $i\neq j \neq k$, which is
given by the rule $\ri(l_{k}) = l_{i}$, where $l_{k} + l_{i} \in
\mathrm{L}_j$.
\\
\\
The actual proof of the theorem will be given in two parts. In the
first part we deal with the case where the three lines
$\LI,\LII,\LII \in \lag$ are in a general position. In the second
part we deal with the case when two of the  three lines are equal
to each other.
\\
\\
\textbf{Part 1.} (General position) Let $\ornLI,\ornLII,\ornLIII
\in \ornlag$ be three oriented lines in a general position. Using
the presentation (\ref{different_presentation}) we can write:
%
%
\begin{eqnarray*}
\thetaLIILIII \circ \thetaLILII & = & \aLIILIII \cdot  \aLILII \cdot \thnLIILIII \circ \thnLILII, \\
\thetaLILIII & = & \aLILIII \cdot \thnLILIII.
\end{eqnarray*}
The result for Part 1 is a consequence of the following three
simple lemmas (to be proved below):
%
%
\begin{lemma} \label{ass_cocycle_lemma}
The following equality holds:
\begin{equation*}
\thnLIILIII \circ \thnLILII = \Cconst \cdot \thnLILIII,
\end{equation*}
where $\Cconst = \sum\limits_{l_2 \in \LII} \psi(\half
\ome(l_2,\rLIILIII(l_2)))$.
\end{lemma}
%
%
\begin{lemma} \label{ass_factors_lemma}
The following equality holds:
\begin{equation*}
\aLIILIII \cdot \aLILII = \Dconst \cdot \aLILIII,
\end{equation*}
where $\Dconst = \frac{1}{q} \sum\limits_{l_2 \in \LII}
\psi(-\half \ome(l_2, \rLIILIII(\xiLII))) \orn_{_\LII}(l_2)
\orn_{_\LII}(\xiLII)$
\end{lemma}
%
%
\begin{lemma}\label{DCequal_lm}
The following equality holds:
\begin{equation*}
\Dconst \cdot \Cconst = 1.
\end{equation*}
\end{lemma}
\textbf{Part 2.} (Non-general position) It is enough to check the
following equalities:
%
%
\begin{eqnarray}
\thetaLIILI \circ \thetaLILII & = & \;\;\mathrm{I} \label{equall1_eq}, \\
%
%
\thetaLIILIm \circ \thetaLILII & = & -\mathrm{I},
\label{equall2_eq}
\end{eqnarray}
where $\ornLIm$ has the opposite orientation to $\ornLI$. We
verify equation (\ref{equall1_eq}). The verification of
(\ref{equall2_eq}) is done in the same way, therefore we omit it.
\\
\\
Write:
\begin{equation} \label{part2_eq1}
\thnLIILI(\thnLILII(f))(h) = \sum_{l_1 \in \LI} \sum_{l_2 \in
\LII} f(l_2 l_1 h),
\end{equation}
where $ f \in \H_\LI$ and $h \in \gH$. Both sides of
(\ref{equall1_eq}) are self intertwining operators of $\H_\LI$ and
therefore they are proportional. Hence it is sufficient to compute
(\ref{part2_eq1}) for a specific function $f$ and specific element
$ h\in \gH$. We take $h=0$ and $ f = \delta_0$, where
$\delta_0(\lambda l h) = \psi(\lambda)$ if $h=0$  and equals 0
otherwise. We get:
\begin{equation*}
 \sum_{l_1 \in \LI} \sum_{l_2 \in \LII} f(l_2 l_1 h) = q.
\end{equation*}
Now write:
\begin{equation*}
\aLIILI \cdot \aLILII = \frac{1}{q^2} \sum_{l_1 \in \LI,\; l_2 \in
\LII} \psi(\half \ome (l_2,\xiLI) + \half \ome(l_1,\xiLII))
\orn_{_\LII}(l_2) \orn_{_\LI}(\xiLI) \orn_{_\LI}(l_1)
\orn_{_\LII}(\xiLII).
\end{equation*}
We identify $\LII$ and $\LI$  with the field $\Fq$ by the rules $
s \cdot 1 \longmapsto s \cdot \xiLII$ and $t \cdot 1 \longmapsto t
\cdot \xiLI$ correspondingly. In terms of these identifications we
get:
%
%
\begin{equation} \label{part2_eq4}
\aLIILI \cdot \aLILII = \frac{1}{q^2} \sum_{ t,\; s \in \Fq}
\psi(\half s \ome(\xiLII,\xiLI) + \half t \ome(\xiLI,\xiLII))
\lgn(t) \lgn(s).
\end{equation}
Denote by $a = \ome(\xiLII,\xiLI)$. The right-hand side of (\ref{part2_eq4}) is equal to:
\begin{equation*}
\frac{1}{q^2} \sum_{s \in \Fq} \psi( \half a s) \lgn(s) \cdot
\sum_{ t \in \Fq} \psi( - \half a t)\lgn(t) =\frac{q}{q^2} =
\frac{1}{q}.
\end{equation*}
All together we get:
\begin{equation*}
\thetaLIILI \circ \thetaLILII = \mathrm{I}.
\end{equation*}
\\
\\
\textbf{Proof of Lemma \ref{ass_cocycle_lemma}.} The proof is by
direct computation. Write:
%
%
\begin{eqnarray}
\thnLILIII(f)(h) & = & \sum_{l_3 \in \LIII} f(l_3 h),  \label{L1_to_L3_eq} \\
\thnLIILIII( \; \thnLILII(f) \;)(h) & = & \sum_{l_3 \in \LIII}
\sum_{l_2 \in \LII} f(l_2 l_3 h), \label{L1_to_L2_to_L3_eq}
\end{eqnarray}
where $f \in \H_\LI$ and $h \in \gH$. Both (\ref{L1_to_L3_eq}) and
(\ref{L1_to_L2_to_L3_eq}) are intertwining operators from $\H_\LI$
to $\H_\LIII$, therefore they are proportional. In order to
compute the proportionality coefficient $\Cconst$ it is enough to
compute (\ref{L1_to_L3_eq}) and (\ref{L1_to_L2_to_L3_eq}) for
specific $f$ and specific $e$. We take $h = 0$ and $f = \delta_0$
where  $\delta_0(\cq,\cp,\lambda) = \psi(\lambda)$. We get:
%
%
\begin{equation*}
\thnLILIII(\delta_0) = 1,
\end{equation*}
\begin{equation} \label{specific_L1_to_L2_to_L3_eq}
\thnLIILIII( \; \thnLILII(\delta_0) \; )(0) = \sum_{l_2 + l_3 \in
\LI} \psi( \half \ome(l_2,l_3)).
\end{equation}
But the right-hand side of (\ref{specific_L1_to_L2_to_L3_eq}) is equal to:
\begin{equation*}  
\sum_{l_2 \in \LII} \psi( \; \half \ome(l_2,\rLIILIII(l_2))).
\end{equation*}
$\EProof$
\\
\\
%
%
%
%
%
{\bf Proof of Lemma \ref{ass_factors_lemma}.} The proof is by
direct computation. Write:
%
%
\begin{equation*}
\aLILIII = \frac{1}{q} \sum_{l_1 \in \LI} \psi( \half \ome(l_1,
\xiLIII)) \orn_{_\LI}(l_1) \orn_{_\LII}(\xiLIII),
\end{equation*}
%
%
\begin{equation} \label{ass_factor_eq2}
\aLIILIII \cdot \aLILII = \frac{1}{q^2} \sum_{ l_1 \in \LI,\; l_2
\in \LII} \psi( \half \ome(l_1,\xiLII) + \half \ome(l_2,\xiLIII))
\orn_{_\LI}(l_1) \orn_{_\LII}(\xiLII) \orn_{_\LII}(l_2)
\orn_{_\LIII}(\xiLIII).
\end{equation}
The term $\psi( \half \ome(l_1,\xiLII) + \half \ome(l_2,\xiLIII))$ is equal to:
%
%
\begin{equation} \label{ass_factor_eq3}
\psi(\half \ome(l_1,\xiLII - \xiLIII) + \half \ome(l_2,\xiLIII))
\cdot \psi( \half \ome(l_1,\xiLIII)).
\end{equation}
We are free to choose $\xiLIII$ such that $\xiLII-\xiLIII \in
\LI$. Therefore, using (\ref{ass_factor_eq3}) we get that the
right-hand side of (\ref{ass_factor_eq2}) is equal to:
%
%
\begin{equation} \label{ass_factor_eq4}
\frac{1}{q} \sum_{l_2 \in \LII}\psi(\half \ome(l_2,\xiLIII))
\orn_{_\LII}(\xiLII) \orn_{_\LII}(l_2) \cdot \aLILIII.
\end{equation}
Now, substituting $\xiLIII = -\rLIILIII(\xiLII)$ in
(\ref{ass_factor_eq4}) we obtain:
%
%
\begin{equation*} 
\frac{1}{q} \sum_{l_2 \in \LII}  \psi(-\half
\ome(l_2,\rLIILIII(\xiLII))) \orn_{_\LII}(l_2)
\orn_{_\LII}(\xiLII) \cdot \aLILIII.
\end{equation*}
$\EProof$
\\
\\
%
%
%
%
%
%
\textbf{ Proof of Lemma \ref{DCequal_lm}.} Identify $\LII$ with
$\Fq$ by the rule $ t \cdot 1 \longmapsto  t \cdot \xiLII$. In
terms of this identification we get:
%
%
\begin{eqnarray*}
\Dconst & = & \frac{1}{q} \sum_{t \in \Fq} \psi(- \half \ome(t
\xiLII , \rLIILIII(\xiLII))) \orn(t),
\\
\Cconst & = & \sum_{t \in \Fq} \psi(\half \ome(t \xiLII ,
\rLIILIII(t\xiLII))).
\end{eqnarray*}
Denote by $a = \ome(\xiLII , \rLIILIII(\xiLII))$. Then:
\begin{eqnarray*}
\Dconst & = & \frac{1}{q} \sum_{t \in \Fq} \psi(-\half a t)
\orn(t),
\\
%
%
\Cconst & = & \sum_{ t \in \Fq} \psi (\half a t^2).
\end{eqnarray*}
Now, we have the following remarkable equality:
%
%
\begin{equation*}  
\sum_{t \in \Fq} \psi(\half a t) \orn(t) = \sum_{ t \in \Fq} \psi
(\half a t^2).
\end{equation*}
This, combined with $\Cconst \cdot \overline{\Cconst} = q$, gives
the result. $\EProof$
\\
\\
This completes the proof of Part 2 and of Theorem
\ref{associativity_thm} $\EProof$
\\
\\
\textbf{Proof of Theorem \ref{main_thm}.} The sheaf $\SK$ is
obtained by middle extension from the open subvariety $\AO$ of the
sheaf $\SKO$. The sheaf $\SKO$ is clearly geometrically
irreducible $[\dim(\Asemi)+1]$-perverse of pure weight 0. This
implies that the sheaf $\SK$ is also geometrically irreducible
$[\dim(\Asemi) + 1]$-perverse of pure weight $w(\SK)=0$.
\\
\\
\textbf{Proof of Property \ref{prop1}.} Assuming the validity of
Property \ref{prop2}, we prove Property \ref{prop1}. Restricting
the sheaf $\SK$ to the open subvariety $\AO$, and taking sheaf to
function correspondence, one obtains $f^\SK = f^{\SKO} = K_{|\O}$.
Applying sheaf to function correspondence to the multiplication
isomorphism, we get that $f^\SK$ is multiplicative. Finally, we
use the fact that the set $\Ow \times \gH$ is a generating set of
$\semi$. Therefore, we have two functions $K$, and $f^\SK$ that
coincide on a generating set and satisfy multiplication property.
This implies that they must coincide on the whole domain.
\\
\\
\textbf{Proof of Property \ref{prop2}.} What remains is to prove
the multiplication property, namely, Property \ref{prop2}. We will
prove the left multiplication isomorphism. The proof of the right
multiplication isomorphism follows the same lines, therefore we
omit it. In the course of the proof we shall use the following
auxiliary sheaves:
\\
\\
$\bullet$ We sheafify the kernel $K_\pi$ using the formula
(\ref{K_heizenberg}) and obtain a sheaf on $\Aheiz \times \bA^2$:
\begin{equation*}
\SKpi(h,x,y) = \SL_{\psi(\half \cp\cq+\cp x + \lambda)} \otimes
\delta_{y=x+\cq}.
\end{equation*}
Where $h = (\cq,\cp,\lambda)$, and we use the notation $\delta$
for the skyscraper sheaf.
\\
\\
$\bullet$ We sheafify the kernel $K_\rho$ when restricted to the
set $\oB \times \Fq^2$ using the formula (\ref{K_borel}) and
obtain a sheaf on the variety $\AoB \times \bA^2$, which we denote
by $\SKoB$,
\begin{equation*}
\SKoB(b,x,y) = \SL_{\lgn (a)} \otimes \SL_{\psi(\half r a^{-1}
x^2)} \otimes \delta_{y=a^{-1}x}.
\end{equation*}
Where $b = \left ( \begin{smallmatrix} a & 0 \\ r & a^{-1}
\end{smallmatrix} \right )$.
\\
\\
$\bullet$ We will frequently make use of several other sheaves
obtained by restrictions from $\SKO, \r {\mathcal A}_{\O}, \r
\SKoB$ and $\SAoB$. Suppose ${\bf X} \subset \O_w \times \Aheiz$
is a subvariety. Then we define ${\SK}_{\bf X} = {\SKO}_{|_{{\bf
X} \times \bA^2}}$ and ${\mathcal A}_{\bf X} = {{\mathcal
A}_{\O}}_{|_{{\bf X} \times \bA^2}}$. The same when ${\bf X}
\subset \AoB$. Finally, we denote by $\delta_0$ the {\it
sky-scraper} sheaf on $\bA^1$ which corresponds to the delta
function at zero.
\\
\\
The proof of Property \ref{prop2} will proceed in several steps:
\\
\\
\textbf{Step 1.} It is sufficient to prove the multiplication
property separately for the Weyl element $w$, an element $\ob \in
\AoB$ and an element $h \in \Aheiz$. This follows from the Bruhat
decomposition, Corollary \ref{corr1} below and the following
decomposition lemma:
%
%
\begin{lemma}\label{DL}
There exists isomorphisms:
\begin{equation*}
\SKO \simeq \SKoB \ast \SKw \ast  \SKoU \ast \SKpi,
\end{equation*}
where $\AoU$ denote the unipotent radical of $\AoB$ and $w$ is the
Weyl element.
\end{lemma}
%
%
\textbf{Step 2.} We prove Property \ref{prop2} for the Weyl
element, $g = w$. We want to construct an isomorphism:
%
%
\begin{equation} \label{existence_eq1}
\SK_{|_w} * \SK \simeq L_w ^* \SK.
\end{equation}
Both sides of (\ref{existence_eq1}) are irreducible $[\dim(\Asemi)
+ 1]$-perverse, therefore it is sufficient to construct an
isomorphism on the open subvariety ${\calU} = \O \cap w\O$. This
has the advantage that over $\calU$ we have formulas for $\SK$,
and moreover, $L_w$ maps $\calU$ into itself. We consider two
decompositions:

\begin{eqnarray}
\calU & \simeq & {\AUn}^\times \times \AoB \times \Aheiz, \label{cs1}\\
\calU & \simeq  & {\AoU^\times} w \times \AoB \times \Aheiz,
\nonumber
\end{eqnarray}

where ${\AoU}^{\times} =  {\AoU} \smallsetminus \{{\mathrm I}\}$
and $\AUn$ denotes the standard unipotent radical (Recall that in
our convention $\AoU$ denotes the group of unipotent lower
triangular matrices) . In terms of the above decompositions we
have the following isomorphisms:

\begin{claim}\label{ics}
There exist isomorphisms:
\end{claim}
\begin{enumerate}
\item $\SKU (\u \ob h) \simeq \SKUx (\u) \ast \SKoB (\ob) \ast \SKpi(h) \label{ics1}$.
\item $ \SKU (\ou w \ob h)  \simeq   \SKoUxw(\ou w) \ast \SKoB (\ob) \ast \SKpi(h)$. \label{ics2}
\end{enumerate}
Now, restricting to $\calU$ and using the decomposition
(\ref{cs1}) our \textit{main statement} is the existence of an
isomorphism:
\begin{equation}\label{main}
\SK_w \ast \SKU(\u \ob h)  \simeq  \SKU(w \u \ob h).
\end{equation}
Indeed, on developing the right-hand side of (\ref{main}) we obtain:
\begin{eqnarray*}
\SKU(w \u \ob h) & = & \SKU({\u}^w w \ob h) \\
& \simeq & \SKoUxw({\u}^w w) \ast \SKoB (\ob) \ast \SKpi(h) \\
& \simeq & \SK_w \ast (\SKUx (\u) \ast \SKoB (\ob) \ast \SKpi(h))\\
& \simeq & \SK_w \ast \SKU (\u \ob h),
\end{eqnarray*}
where ${\u}^w = w \u w^{-1}$. The first and third isomorphisms are
applications of Claim \ref{ics} parts \ref{ics2} and \ref{ics1}
respectively. The second isomorphism is a result of associativity
of convolution and the following \textit{central lemma}:
%
%
\begin{lemma} \label{centralemma}
There exists an isomorphism:
\begin{equation} \label{main_iso}
\SKoUxw (\u^w w) \simeq  \SKw * \SKUx (\u).
\end{equation}
\end{lemma}
The following is a consequence of (\ref{existence_eq1}).
%
%
\begin{corollary}\label{corr1}
There exists an isomorphism:
\begin{equation} \label{corr1_eq}
\SK_{|_{\AoB \times \Aheiz}} \simeq \SKoB \ast \SKpi.
\end{equation}
\end{corollary}
$\BProof$ On developing the left-hand side of (\ref{corr1_eq}) we
obtain:
\begin{eqnarray*}
\SK_{|_{\AoB \times \Aheiz}} & \simeq & \SKw * \SK_{|{w^{-1} \AoB
\times \Aheiz}}
\\
& \simeq & \SKw * (\SKwI * \SKoB \ast \SKpi)
\\
& \simeq & (\SKw * \SKwI ) * \SKoB \ast \SKpi
\\
& \simeq & \SKoB \ast \SKpi.
\end{eqnarray*}
The first isomorphism is a consequence  of (\ref{existence_eq1}).
The second isomorphism is a consequence of Lemma \ref{DL}. The
third isomorphism is the associativity property of convolution.
The last isomorphism is a property of the Fourier transform
\cite{KL}, that is, $\SKw * \SKwI \simeq \SI$, where $\SI$ is the
kernel of the identity operator. $\EProof$
\\
\\
\textbf{Step 3.} We prove Property \ref{prop2} for element $\ob
\in \AoB$. Using corollary \ref{corr1} we have $\SK_{|_{\ob}}
\simeq {\SKoB}_{|_{\ob}} = \SKob$. We want to construct an
isomorphism:
\begin{equation}\label{existence_eq3}
\SKob * \SK \simeq L_{\ob} ^* \SK.
\end{equation}
Since both sides of (\ref{existence_eq3}) are irreducible
(shifted) perverse sheaves, it is enough to construct an
isomorphism on the open set $\O = \ASw \times \Aheiz \times
\bA^2$. Write:
$$ \SKob * \SK_{|_{\O}} \simeq \SKob * \SKO \simeq \SKob * (\SKoB * \SKw *
\SKoU \ast \SKpi ) \simeq (\SKob * \SKoB) * (\SKw * \SKoU \ast
\SKpi).$$
The first isomorphism is by construction. The second isomorphism
is an application of Lemma \ref{DL}. The third isomorphism is the
associativity property of the convolution operation between
sheaves. From the last isomorphism we see that it is enough to
construct an isomorphism $\SKob * \SKoB \simeq L_{\ob}^*(\SKoB)$,
where $\L_{\ob} : \AoB \lto \AoB$. The construction is an easy
consequence of formula (\ref{K_borel}) and the character sheaf
property (\ref{cspSa}) of $\Sartin$.
\\
\\
\textbf{Step 4.} We prove Property \ref{prop2} for an element $ h
\in \Aheiz$. We want to construct an isomorphism:
\begin{equation} \label{existence_eq4}
\SK_{|_h} * \SK \simeq L_h ^* \SK.
\end{equation}
Both sides of (\ref{existence_eq4}) are irreducible $[\dim(\Asemi)
+ 1]$-perverse. Therefore, it is sufficient to construct an
isomorphism on the open set $\O$. This is done by a direct
computation, very similar to what has been done before, hence we
omit it. This concludes the proof of Theorem \ref{main_thm}.
$\EProof$
\\
\\
%
%
%
%
%
%
%
%
%
%
%
%
%
%
%
%
\textbf{Proof of Lemma \ref{DL}}. We will prove the lemma in two
steps.
\\
\\
%
%
%
%
\textbf{ Step 1.} We prove that $\SKO  \simeq \SK_{\O_w} \ast
\SKpi$. In a more explicit form we want to show:
%
%
\begin{equation} \label{lm3_explicit}
\SA_\O \otimes \SKnSwE \simeq \SASw \otimes \SKnSw * \SKnE.
\end{equation}
It is sufficient to show the existence of an isomorphism $\SKnSwE
\simeq \SKnSw * \SKnE$. On developing the left-hand side of
(\ref{lm3_explicit}) we obtain:
$$ \SKnSwE(g,h,x,y) = \SL_{\psi(R(g,h,x,y))}.$$
On developing the right-hand side we obtain:
\begin{eqnarray*}
\SKnSw * \SKnE (\; (g,h) \; ,x,y) & = & \sint_{z \in \bA^1} \SKnSw(g,x,z) \otimes \SKnE(h,z,y) \\
& = & \sint_{\bA^1} \SL_{\psi (R_g(x,z))} \otimes \SL_{\psi(R_h(z,y))} \otimes \delta_{y=z-\cq} \\
&  \simeq & \SL_{\psi(R_g(x,y-\cq))} \otimes \SL_{\psi(R_h(y-\cq,y))} \\
&  \simeq & \SL_{\psi(R_g(x,y-\cq) + R_h(y-\cq,y))} \\
& = & \SL_{\psi(R(g,h,x,y))}.
\end{eqnarray*}
The only non-trivial isomorphism is the last one and  it is a
consequence of the Artin-Schreier sheaf being a character sheaf on
the additive group $\Ga$.
\\
\\
\textbf{Step 2.} We prove that $\SK_{\O_w} \simeq  \SKoB \ast \SKw
\ast \SKoU$. In a more explicit form we want to show:
%
%
\begin{equation*} 
\SASw \otimes \SKnSw \simeq  \SAoB \otimes \SAw \otimes  \SAoU
\otimes  \SKnoB * \SKnw * \SKnoU.
\end{equation*}
We will separately show the existence of two isomorphisms:
%
%
\begin{eqnarray}
\SKnSw & \simeq & \SKnoB  * \SKnw *  \SKnoU, \label{lm1_iso1}\\
%
%
 \SASw & \simeq & \SAoB  \otimes \SAw \otimes \SAoU. \label{lm1_iso2}
\end{eqnarray}
\textbf{ Isomorphism (\ref{lm1_iso1}).} We have the decomposition,
$\ASw \iso \AoB \times w \times \AoU$. Let $ \ob = \left (
\begin{smallmatrix} a & 0
  \\ r & a^{-1} \end{smallmatrix} \right )$ and  $\ou = \left (
\begin{smallmatrix} 1 & 0 \\ s & 1 \end{smallmatrix} \right )$  be
general elements in the groups $\AoU$ and $\AoB$ respectively. In
terms of the coordinates $ (a,r,s) $ a general element in $\ASw$
is of the form $ g = \left ( \begin{smallmatrix} as & a \\ rs -
a^{-1} & r
\end{smallmatrix} \right )$. Developing the left-hand side of
(\ref{lm1_iso1}) in terms of the coordinates $ (a,r,s)$ we obtain:
$$ \SKnSw(\ob w \ou ,x,y) = \SL_{\psi(-\half a^{-1} r x^2 + a^{-1}xy
-\half s y^2)}.$$
On developing the right-hand side of (\ref{lm1_iso1}) we obtain:
\begin{eqnarray*}
\SKnoB * \SKnw * \SKnoU (\ob w \ou,x,y) & = &  \sint\limits_{z,z'
\in \bA^1} \SKnoB(\ob,x,z) \otimes \SKnw(w,z,z') \otimes
\SKnoU(\ou,z',y)
\\
& = & \sint\limits_{z,z' \in \bA^1} \SL_{\psi(-\half r a^{-1}
x^2)} \otimes \delta_{x = a z} \otimes \SL_{\psi(zz')} \otimes
\SL_{\psi(-\half s z'^2)} \otimes \delta_{y=z'}
\\
& \simeq & \SL_{\psi(-\half r a^{-1} x^2)} \otimes \SL_{\psi(a^{-1} x y)} \otimes
\SL_{\psi(-\half s y^2)}
\\
& \simeq & \SL_{\psi(-\half ra^{-1} x^2 + a^{-1}xy -\half s y^2)}.
\end{eqnarray*}
The last isomorphism is a consequence of the fact that the
Artin-Schreier sheaf is a character sheaf (\ref{cspSa}).
Altogether we obtained isomorphism (\ref{lm1_iso1}).
\\
\\
\textbf{Isomorphism (\ref{lm1_iso2}).} On developing the left-hand
side of (\ref{lm1_iso2}) in terms of the coordinates $(a,r,s)$ we
obtain:
\begin{eqnarray*}
\SASw( \ob w \ou ) & = & \SG(\psi_a,\lgn)[2](-1)\\
& \simeq & \SG(\psi,\lgn_{a^{-1}}) [2](-1)\\
& \simeq & \SL_{\lgn(a^{-1})} \otimes \SG(\psi,\lgn) [2](-1)\\
%
& \simeq & \SL_{\lgn(a)} \otimes  \SG(\psi,\lgn) [2](-1) \\
& = & \SAoB \otimes \SAw \otimes \SAoU( \ob w \ou),
\end{eqnarray*}
where $\SG(\psi_s,\lgn_a) = \int_{\bA^1} \SL_{\psi(\half sz)}
\otimes \SL_{\lgn(az)}$ denotes the quadratic {\it Gauss-sum}
sheaf. The second isomorphism is a change of coordinates $z
\mapsto az$ under the integration. The third isomorphism is a
consequence of the Kummer sheaf $\SL_\lgn$ being a character sheaf
on the multiplicative group $\Gm$ (\ref{cspSl}). The fourth
isomorphism is a specific property of the Kummer sheaf which is
associated to the quadratic character $\lgn$. This completes the
construction of isomorphism (\ref{lm1_iso2}). $\EProof$
\\
\\
%
%
%
\textbf{Proof of claim \ref{ics}.} Carried out in  exactly the
same way as the proof of the decomposition Lemma \ref{DL}. Namely,
using the explicit formulas of the sheaves $\SKO, \SKoB, \SKpi$
and the character sheaf property of the sheaves $\Sartin$ and
$\Slegendre$. $\EProof$
\\
\\
\textbf{Proof of Lemma \ref{centralemma}.} First, we write
isomorphism (\ref{main_iso}) in a more explicit form:
%
%
\begin{equation} \label{main_explicit}
\SAoUxw \otimes \SKnoUxw (\u^w w) \simeq  \SAw \otimes \SAUx
\otimes \SKnw * \SKnUx (\u).
\end{equation}
Let $ \u = \left ( \begin{smallmatrix} 1 & s \\ 0  & 1
\end{smallmatrix} \right ) \in \AUnx$ be a non-trivial unipotent.
Then $\u^w w = w \u = \left (\begin{smallmatrix} 0 & 1 \\ -1 & -s \end{smallmatrix} \right)$.
On developing the left-hand side of (\ref{main_explicit}) we obtain:
%
%
\begin{eqnarray*}
\SKnoUxw (\u^w w,x,y) & = & \SL_{\psi(\half s x^2 +xy)}, \\
\SAoUxw (\u^w w) & = & \SG(\psi,\lgn) [2](1),
\end{eqnarray*}
where $\SG(\psi_s,\lgn_a) = \int_{\bA^1} \SL_{\psi(\half s z)}
\otimes \SL_{\lgn(az)}$.
\\
\\
On developing the right-hand side of (\ref{main_explicit}) we obtain:
\begin{eqnarray*}
\SKnw * \SKnUx (\u,x,y) & = &  \sint_{z \in \bA^1} \SKnw(x,z) \otimes \SKnUx(u,z,y) \\
& = & \sint_{\bA^1}\SL_{\psi(xz)} \otimes \SL_{\psi(-\half s^{-1} z^2 + s^{-1}zy -\half s^{-1} y^2)} \\
& \simeq & \sint_{\bA^1} \SL_{\psi(xz -\half s^{-1} z^2 + s^{-1} zy -\half s^{-1} y^2)} \\
& \simeq & \sint_{\bA^1} \SL_{\psi(-\half s^{-1} ( z-sx-y)^2)} \otimes \SL_{\psi(\half s x^2 +xy)} \\
& \simeq &  \sint_{\bA^1} \SL_{\psi(-\half s^{-1} z^2)} \otimes
\SL_{\psi(\half s x^2 +xy)}.
\end{eqnarray*}
By applying change of coordinates $z \mapsto sz $ under the last integration we obtain:
\begin{equation}\label{main_eq1}
\sint_{\bA^1} \SL_{\psi(-\half s^{-1} z^2)} \otimes
\SL_{\psi(\half s x^2 +xy)} \simeq  \sint_{z \in \bA^1}
\SL_{\psi(-\half s z^2)} \otimes \SL_{\psi( \half s x^2 +xy)}.
\end{equation}
Now write:
\begin{equation} \label{main_eq2}
\SAw \otimes \SAUx (\u) = \SG(\psi,\lgn) [2](1) \otimes
\SG(\psi_s,\lgn)[2](1).
\end{equation}
Combining (\ref{main_eq1}) and (\ref{main_eq2}) we obtain that the
right-hand side of (\ref{main_explicit}) is isomorphic to:
\begin{equation*} 
\left (\SG(\psi_s,\lgn) [2](1) \otimes \sint_{\bA^1}
\SL_{\psi(-\half s z^2)} \right ) \otimes \left ( \SG(\psi,\lgn)
[2](1) \otimes \SL_{\psi( \half s x^2 +xy)} \right).
\end{equation*}
The main argument is the existence of the following isomorphism:
%
%
\begin{equation*} 
\SG(\psi_s,\lgn) [2](1) \otimes  \sint_{\bA^1} \SL_{\psi(- \half s
z^2)} \simeq \Qlb.
\end{equation*}
It is a direct consequence of the following lemma:
\begin{lemma}[Main lemma] \label{mainlemma}
There exists a canonical isomorphism of sheaves on $\Gm$:
\begin{equation*} 
\sint_{\bA^1} \SL_{\psi(\half s z)} \otimes \SL_{\lgn(z)} \simeq
\sint_{\bA^1} \SL_{\psi(\half s z^2)},
\end{equation*}
where $s \in \Gm$.
\end{lemma}
$\BProof$  The parameter $s$ does not play any essential role in
the argument. Therefore, it is sufficient to prove:
%
%
\begin{equation} \label{mainlemma_eq1}
\sint_{\bA^1} \SL_{\psi(z)} \otimes \SL_{\lgn(z)} \simeq
\sint_{\bA^1} \SL_{\psi(z^2)}.
\end{equation}
Define the morphism $ p:\Gm \lto \Gm$, $ p(x) = x^2 $. The
morphism $p$ is an \`{e}tale double cover. We have $ p_* \Qlb
\simeq \SL_{\lgn} \oplus \Qlb$. Now on developing the
left-hand-side of (\ref{mainlemma_eq1}) we obtain:
$$ \sint_{\bA^1} \SL_{\psi(z)} \otimes \SL_{\lgn(z)}  = \pi_! (\Sartin
\otimes \SL_{\lgn}) \simeq \pi_!(\Sartin \otimes ( \SL_\lgn \oplus
\Qlb)).$$
The first step is just a translation to conventional notations,
where $\pi$ stands for the projection $\pi :\Gm \lto pt$. The second
isomorphism uses the fact that  $\pi_! \Sartin \simeq 0 $. Next:
$$ \pi_!(\Sartin \otimes ( \SL_\lgn \oplus
\Qlb)) \simeq \pi_! (\Sartin \otimes p_* \Qlb) \simeq \pi_! p^*
\Sartin = \sint_{\bA^1} \SL_{\psi(z^2)}. \rev \EProof$$
This completes the proof of proposition \ref{centralemma}.
$\EProof$
\\
\\
\section{Proofs Section}\label{proofs}
For the remainder of this section we fix the following notations.
Let $\hb = \frac{1}{p}$, where $p$ is a fixed prime $\neq 2, \,
3$. Consider the lattice $\Lm$ of characters of the torus $\T$ and
the quotient vector space $ \V = \Lm / p\Lm $. The integral
symplectic form on $\Lm$ is specialized to give a symplectic form
on $\V$, i.e., $\ome : \V \times \V \longrightarrow \Fp$. Fix
$\psi: \Fp \longrightarrow \C^*$ a non-trivial additive character.
Let $\Ad$ be "the algebra of functions on the quantum torus"
and $\G \simeq \GZ$ its group of symmetries.\\
\subsection{Proof of Theorem \ref{GH2}}\label{proofGH}
{\it Basic set-up}: let $ (\Pih,\Hh)$ be a representation of
$\Ad$, which is a representative of the unique irreducible class
which is fixed by $\G$ (cf. Theorem \ref{GH}). Let $\rhoh : \G
\longrightarrow \PGL(\Hh)$ be the associated projective
representation. Here we give a proof that $\rhoh$ can be
linearized in a unique way which factors through the quotient
group $\Gf \simeq \GG$:
\[
\qtriangle[\G`\Gf`\GL(\Hh);`\rhoh`\bar{\rho}_{_\h}]
\]
The proof will be given in several steps.\\\\
\textbf{ Step 1.} {\it Uniqueness}. The uniqueness of the
linearization follows directly  from the fact that the group
$\GG$, $p \neq 2, \, 3$, has no characters.\\\\
\textbf{ Step 2.} {\it Existence.}\footnote{This statement and
proof work more generally for all Planck constants of the form $\h
= M/N,$ with $M, N$ co-prime integers, and N odd. One should
replace then the finite Field $\Fp$ with the finite ring
$\Z/N\Z$.} The main technical tool in the proof of existence is a
construction of a finite dimensional quotient of the algebra
$\Ad$. Let $\Adf$ be the algebra generated over $\C$ by the
symbols $\{ s(v) \r | \r v \in \Lmmf \}$ and quotient out by the
relations:
\begin{equation} \label{relations}
 s(u + v) = \psi(\half \omega(u,v)) s(u)s(v).
\end{equation}
The algebra $\Adf$ is non-trivial and the vector space $\Lmmf$ gives on it a standard basis.
These facts will be proven in the sequel. We have the following map:
\begin{equation*} 
 s : \Lmmf \longrightarrow \Adf, \r v \mapsto s(v).
\end{equation*}
The group $\Gf$ acts on $\Adf$ by automorphisms through its action
on $\Lmmf$. We have a homomorphism of algebras:
\begin{equation} \label{AdtoAdf}
q : \Ad \longrightarrow \Adf.
\end{equation}
The homomorphism (\ref{AdtoAdf}) respects the actions of the group of symmetries $\G$ and $\Gf$ respectively. This is summarized in the following commutative diagram:
\begin{equation}\label{respect}
\begin{CD}
   \G \times \Ad   @>>>   \Ad \\
        @V(p,q)VV       @VqVV \\
   \Gf \times \Adf @>>>   \Adf,
\end{CD}
\end{equation}
where $p : \G \longrightarrow \Gf$ is the canonical quotient map.\\\\
\textbf{Step 3.} Next, we construct an explicit representation of
$\Adf$:
$$\Pif : \Adf \longrightarrow \End(\H).$$
Let $\Lmmf = \VI \bigoplus \VII$ be a Lagrangian decomposition of
$\Lmmf$. In our case $\Lmmf$ is two-dimensional, therefore $\VI$
and $\VII$ are linearly independent lines in $\Lmmf$. Take $\H =
\FSVI$ to be the vector space of complex valued functions on
$\VI$. For an element $ v \in \Lmmf$ define:
\begin{equation} \label{frep}
\Pif(v) = \psi(\half \omega(v_{1},v_{2}))  \bL _{v_{1}} \bM_{v_2},
\end{equation}
where $ v = v_{1} + v_{2}$ is a direct decomposition $v_{1} \in \VI,
\; v_{2} \in \VII $,  $\bL_{v_{1}}$ is the translation operator
defined by $v_{1}$:
$$ \bL_{v_{1}}(f)(x) = f(x+v_{1}), \rev  f \in \FSVI $$
and $\bM_{v_2}$ 
is a notation for the operator of multiplication by the function $
\bM_{v_2}(x) = \psi(\omega(v_{2}, x))$. Next, we show that the
formulas given in (\ref{frep}) satisfy the relations
(\ref{relations}) and thus constitute a representation of the
algebra $\Adf$. Let $u , v \in \Lmmf$. We have to show:
\begin{equation*}     
\Pif(u+v) = \psi(\half \omega(u,v)) \Pif(u) \Pif(v).
\end{equation*}
Compute:
\begin{equation*} 
  \Pif(u+v) = \Pif( (u_{1} + u_{2}) + (v_{1} + v_{2})),
\end{equation*}
where $u =u_{1} + u_{2}$ and $v = v_{1} + v_{2} $ are
decompositions of $u$ and $v$ correspondingly.
\\
\\
Then:
\begin{equation} \label{eq02_2}
   \Pif( (u_{1} + v_{1}) + (u_{2} + v_{2})) = \psi(\half \omega(u_{1} + v_{1}, u_{2} +v_{2}))
   \bL_{u_{1} + v_{1}} \bM_{u_2+v_2}. 
\end{equation}
This is by definition of $\Pif$ (cf. (\ref{frep})). Now use the
following formulas:
\begin{eqnarray*} 
\bL_{u_{1} + v_{1}} & = & \bL_{u_{1}} \bL_{v_{1}}, \\
\bL_{v_{1}} \bM_{u_2} & = & \bM_{u_2}(v_1) \bL_{v_{1}} 
\end{eqnarray*}
to obtain that the right-hand side of (\ref{eq02_2}) is equal to:
\begin{equation*} 
 \psi(\half \omega(u_{1} + v_{1}, u_{2} + v_{2}) + \omega(u_{2} , v_{1}))
 \bL_{u_{1}} \bM_{u_2} \bL_{v_{1}} \bM_{v_2}. 
\end{equation*}
Now use:
\begin{equation*}  
   \half  \omega(u_{1} + v_{1}, u_{2} + v_{2}) + \omega(u_{2} , v_{1}) = \half \omega(u_{1} + u_{2}, v_{1} + v_{2})
   + \half \omega(u_{1},u_{2}) + \half \omega(v_{1},v_{2}),
\end{equation*}
to obtain the result:
\begin{equation*} 
  \psi(\half\omega(u, v))  \Pif(u) \Pif(v),
\end{equation*}
which completes the argument.\\\\
As a consequence of constructing $\Pif$ we automatically proved
that $\Adf$ is non-trivial. It is well known that all linear
operators on $\FSVI$ are linear combinations of translation
operators and multiplication by characters. Therefore, $\Pif: \Adf
\longrightarrow \End(\H)$ is surjective. But, $\dim(\Adf) \leq
p^2$ therefore $\Pif$ is a bijection. This means that $\Adf$ is
isomorphic to a matrix algebra
$\Adf \simeq M_{p}(\C)$.\\\\
\textbf{Step 4.} Completing the proof of existence. The group
$\Gf$ acts on $\Adf$. Therefore, it acts on the category of its
representations. However, $\Adf$ is isomorphic to a matrix
algebra, therefore it has unique irreducible representation, up to
isomorphism. This is the standard representation being of
dimension $p$. But $\dim(\H) = p$, therefore $\Pif$ is an
irreducible representation and  its isomorphism class is fixed by
$\Gf$ meaning that we have a pair:
\begin{eqnarray*}
\Pif : \Adf & \longrightarrow & \End(\H),
\\
\rhof :\Gf  & \longrightarrow & \PGL(\H)
\end{eqnarray*}
satisfying the Egorov identity:
$$\rhof(B) \Pif(v) \rhof(B^{-1}) = \Pif(Bv),$$
where $B \in \Gf$ and $v \in \Adf$.\\\\
It is a well known general fact (attributed to I. Schur) that the
group $\Gf$, where $p$ is an odd prime, has no non-trivial
projective representations. This means that $\rhof$ can be
linearized\footnote{see Appendix \ref{metaplectique} for an
independent proof based on {\it ``The method of canonical Hilbert
space''}.} to give:
$$ \rhof : \Gf \longrightarrow \GL(\H).$$
Now take:
\begin{eqnarray*}
\Hh & = & \H,
\\
\Pih & = & \Pif \circ q,
\\
\rhoh &  = & \rhof \circ p.
\end{eqnarray*}
Because $q$ intertwines the actions of $\G$ and $\Gf$ (cf. diagram
(\ref{respect})) we see that $\Pih$ and $\rhoh$ are compatible,
namely, the Egorov identity is satisfied:
$$\rhoh(B) \Pih(f) \rhoh(B^{-1}) = \Pih(f^B),$$
where $B \in \Gf$ and $f \in \Ad$. Here the notation $\Pih(f^B)$
means to apply any preimage $\overline{B} \in \G$ of $B \in \Gf$
on $f$. In particular, this implies that the isomorphism class of
$\Pih$ is fixed by $\G$. Knowing that such representation $\Pih$
is unique up to an isomorphism, (Theorem \ref {GH}), our desired
object has been obtained. $\EProof$
%
%
%
%
%
%
%
%
%
\subsection{Proof of Lemma \ref{factorization}}\label{proofFACT}
{\it Basic set-up}:  let $ (\Pih,\Hh)$ be a representation of
$\Ad$, which is a representative of the unique irreducible class
which is fixed by $\G$ (cf. Theorem \ref{GH}). Let $\rhoh : \Gf
\longrightarrow \GL(\Hh)$ be the associated honest representation
of the quotient group $\Gf$ (see Theorem \ref{GH2} and Proof
\ref{proofGH}). Recall the notation $\Y = \Gf \times \Lm$. We
consider the function $F: \Y \longrightarrow \C$ defined by the
following formula:
\begin{equation} \label{func1}
F(B,\xi) = \Tr(\rhoh(B) \Pih(\xi)),
\end{equation}
where $\xi \in \Lm$ and $B \in \Gf$. We want to show that $F$
factors through the quotient set $\YY = \Gf \times \V$:
\[
\qtriangle[\Y`\YY`\C;`F`F]
\]
The proof is immediate, taking into account the construction given
in section \ref{proofGH}. Let $\Pif$ be the unique (up to
isomorphism) representation of the quotient algebra $\Adf$. As was
stated in \ref{proofGH}, $\Pih$ is isomorphic to $\Pif \circ q$,
where $q:\Ad \longrightarrow \Adf$ is the quotient homomorphism
between the algebras. This means that $\Pih(\xi) = \Pif(q(\xi))$
depends only on the image $ q(\xi) \in \Lmmf$, and formula
(\ref{func1}) solves the problem. $\EProof$
%
%
%
%
\subsection {Proof of Theorem \ref{deligne}} \label{proofdeligne}
{\it Basic set-up:} in this section we use the notations of
section \ref{proofGH} and Appendix \ref{metaplectique}. Set $\YY =
{\Sp} \times \V$ and let $\alpha : \Sp \times \YY \lto \YY$ denote
the associated action map. Let $F:\YY \lto \C$ be the function
appearing in the statement of Theorem \ref{deligne}, i.e., $F(B,v)
= \Tr(\rhof(B) \Pif(v) )$, where $B \in \Sp$ and $v \in \V$. We
use the notations $\AV$, $\ASp$ and $\AYY$ to denote the
corresponding algebraic varieties. For the convenience of the
reader we repeat here the formulation of the theorem:\\\\
%
%
\begin{theorem}[Geometrization Theorem]
There exists a Weil object $\SF \in \Db(\AYY) $ satisfying the
following properties:
\end{theorem}
%
%
\begin{enumerate}

\item (Perversity) The object $\SF$ is geometrically
irreducible $[\dim(\AYY)]$-perverse of pure weight $w(\SF) = 0$.

\item (Function) The function $F$ is associated to $\SF$ via \textit{sheaf-to-function correspondence}:
\begin{equation*}
  f^{\SF} = F.
\end{equation*}
\item (Equivariance)  For every element $S \in \ASp$ there exists an
isomorphism:

  $$\alpha_S ^* \SF \simeq \SF.$$

\item (Formula) Restricting the sheaf $\SF$ on the closed subvariety $\AT^{\times} \times
\AV$. Using the identifications (\ref {identification1}),
(\ref{identification2}), we have an explicit formula:
\begin{equation*}
  \SF_{|_{\AT^{\times} \times \AV}} ( \left (\begin{smallmatrix} a & 0
\\ 0 & a^{-1} \end{smallmatrix} \right ) , \lambda, \mu)  \simeq \SL_{\lgn(a)} \otimes
  \SL_{\psi(\half \frac{a+1}{a-1} \lambda \cdot \mu)}.
\end{equation*}
\end{enumerate}
\textbf{Construction of the sheaf $\SF$}. We use the notations of
Appendix \ref{metaplectique}. Let $\gH$ be the Heisenberg group.
As a set we have $\gH= \V \times \Fq$. The group structure is
given by the multiplication rule $(v,\lambda) \cdot (v',\lambda')
= (v+v',\lambda+\lambda' + \half \ome(v,v')) $. We fix a section $
s: \V \dashrightarrow \gH $, $s(v) =(v,0)$. The group $\Sp$ acts
by automorphisms on the group $\gH$, through its tautological
action on the vector space $\V$, i.e.,  $ g \cdot (v,\lambda) = (g
v , \lambda) $. We define the semi-direct product $\semi = \Sp
\ltimes \gH$. We consider the map $(\mathrm{Id},s) : \YY \lto
\semi$. We use the notations $\Aheiz$ and $\Asemi$ to denote the
corresponding algebraic varieties.
\\
\\
Let $\SK$ be the Weil representation sheaf (see Theorem
\ref{main_thm}). \textit{Define}:
%
%
\begin{equation*}
\SF = \Tr(\SK_{|_{\AYY}}),
\end{equation*}

where $\Tr$ is defined in complete analogue to the operation of
taking trace in the set-theoretic framework, that is, we take:
\begin{equation*}
\SF(g,v) = \int_{x \in \bA^1} \SK(g,v,x,x),
\end{equation*}

where we use the notation $\int_{x\in \bA^1}$ to denote
integration with compact support along the $x$-variable. We prove
that the sheaf $\SF$ satisfies Properties \ref{prop_del1} -
\ref{prop_del4}.
\\
\\
\textbf{Proof of Property \ref{prop_del2}.}  Property
\ref{prop_del2} follows easily. One should observe that the
collection of operators $\{\Pif(v)\}_{v \in \V}$ extends to a
representation of the group $\gH$, which we will also denote by
$\Pif$ . Both the representations $\rhof$ and $\Pif$ glue to a
single representation $\rhof \ltimes \Pif$ of the semi-direct
product $\semi$. It is a direct verification, that the
representation $\rhof \ltimes \Pif$ is isomorphic to the
representation $\rho \ltimes \pi$ constructed in Appendix
\ref{metaplectique}. Hence we can write:

\begin{equation*}
f^{\SF} = f^{\Tr(\SK_{|_\AYY})} = \Tr(f^{\SK_{|_\AYY}}) =
\Tr(K_{|_\AYY}) = F.
\end{equation*}

In the above equation we use the fact that the operation of taking
geometric trace commutes with sheaf-to-function correspondence.
This proves Property \ref{prop_del2}.
\\
\\
%
%
\textbf{Proof of Property \ref{prop_del3}.} Principally follows
from the multiplication property of the sheaf $\SK$ (Theorem
\ref{main_thm}, Property \ref{prop2}). More precisely, using the
multiplication property we obtain the following isomorphism:
\begin{equation*}
\SK_{|_S} * \SK * \SK_{|_{S^{-1}}} \simeq L_S^* R_{S^{-1}}^* \SK,
\end{equation*}

where $L_S$, $R_{S^{-1}}$ denotes left multiplication by $S$ and
right multiplication by $S^{-1}$ on the group $\Asemi$
respectively. Next, we have:
\begin{equation*}
\alpha_S^* \SF \simeq \Tr ( \alpha_S ^* \SK_{|_\AYY} ) \simeq
\Tr(L_S ^* R_{S^{-1}}^* \SK_{|_\AYY}) \simeq \Tr ( \SK_{|_S} *
\SK_{|_\AYY} * \SK_{|_{S^{-1}}}).
\end{equation*}
Finally, we have the following isomorphisms:
\begin{equation*}
\Tr( \SK_{|_S} * \SK_{|_\AYY} * \SK_{|_{S^{-1}}} ) \simeq  \Tr (
\SK_{|_{S^{-1}}} * \SK_{|_S} * \SK_{|_\AYY}),
\end{equation*}

and:

\begin{equation*}
\SK_{|_{S^{-1}}} * \SK_{|_S} \simeq  \SI,
\end{equation*}
where the first isomorphism is the basic property of the trace.
Its proof in the geometric setting, is a result of a direct
diagram chasing. Structurally, it follows the same lines as in the
usual set-theoretic setting. The second isomorphism is a
consequence of the multiplication property of $\SK$. This
completes the proof of Property \ref{prop_del3}.
\\
\\
\textbf{Proof of Property \ref{prop_del4}.} This property is
directly verified using the explicit formulas appearing in
\ref{realization}.
\\
\\
\textbf{Proof of Property \ref{prop_del1}.} We will use the
notations $\SK_{\rho}$, and $\SK_{\pi}$ to denote the restriction
of the sheaf $\SK$ to the subgroups $\ASp$, and $\Aheiz$
respectively. We recall that we have the formula $\SK_{\pi}
(h,x,y) = \SL_{\psi(\frac{1}{2}\cp \cq + \cp x + \lambda)} \otimes
\delta_{y=x+\cq}$, where $h=(\cq,\cp,\lambda)$. Moreover, it is
easy to verify that the sheaf $\SK_{\rho}$ is irreducible
$[\dim(\ASp)+ 1 ]$-perverse Weil sheaf of pure weight zero.
Finally, we have that:

\begin{equation} \label{mult}
 \SK \simeq \SK_{\rho} * \SK_{\pi}.
\end{equation}

Let $h = (\cq,\cp,0)$. Now write:

\begin{eqnarray*} \label{compute}
\SF(g,h) & \simeq & \Tr(\SK(g,h))
\\
         & \simeq & \int_{x \in \bA^1} \SK_{\rho}(g,x,x-\cq) \otimes
         \SL_{\psi(-\frac{\cp\cq}{2} + \cp x)}
\\
         & \simeq & \int_{x \in \bA^1} \SK_{\rho}(g,x,x-\cq) \otimes
         \SL_{\psi(\frac{1}{2}\cp(\frac{x}{2}-\cq))}
\\
         & \simeq &  \int_{x \in \bA^1} \SK_{\rho}(g,\frac{x+\cq}{2},\frac{x-\cq}{2}) \otimes
         \SL_{\psi(\frac{1}{2}\cp x)}.
\end{eqnarray*}

Where the first isomorphism is a direct consequence of the
formulas of the sheaves involved and isomorphism (\ref{mult}). The
third isomorphism is a change of variables $ (\cq,\frac{x}{2} -
\cq) \rightarrow (\cq,x) $. Consider the isomorphism $\beta : \ASp
\times \bA^1 \times \bA^1 \rightarrow \ASp \times \bA^1 \times
\bA^1$, defined by $\beta(g,\cq,x) =
(g,\frac{x+\cq}{2},\frac{x-\cq}{2})$. The last term in
(\ref{compute}) is equivalent to taking (non-normalized)
fiber-wise Fourier transform of $\beta^* \SK_{\rho}$ considered as
a sheaf on the line bundle $ (\ASp \times \bA^1) \times \bA^1
\rightarrow (\ASp \times \bA^1)$. This implies using the theory of
the $\ell$-adic Fourier transform (cf. \cite{KL}) that $\SF$ is
geometrically irreducible $[\dim(\AYY)]$-perverse of pure weight
zero. This concludes the proof of Property \ref{prop_del1}, and of
Theorem \ref{deligne}. $\r\EProof$
\\
\\
\subsection {Computations for the Vanishing Lemma (Lemma \ref{vanishing})} \label{compvanishing}
In the computations we use some finer technical tools from the
theory of $\ell$-adic cohomology. The interested reader can find a
systematic study of this material in \cite{K, KW, L, BBD, BL}.
\\
\\
We identify the standard torus $\AT \subset \AGG$ with the group
$\Gm$. Fix a non-trivial character sheaf\footnote{That is, a
1-dimensional local system on $\Gm$ that satisfies the property
$m^*\Skummer \iso \Skummer\boxtimes \Skummer$, where $m:\Gm\times
\Gm \to \Gm$ is the multiplication morphism.} $\Skummer$ on $\Gm$.
Denote by $\Sartin$ a non-trivial character sheaf on $\Ga$. Fix
$\lambda, \, \mu \in \bA^1$ with $\lambda\cdot\mu\neq 0$. Consider
the variety $\AX = \Gm - \{1\}$, the sheaf:
\begin{eqnarray}\label{SE}
\SE = \SL_{\psi(\half\frac{a+1}{a-1} \lambda \cdot \mu)} \otimes
\Skummer,
\end{eqnarray}
on $\AX$ and the canonical projection $pr: \AX \lto pt$. Note that
$\SE$ is a non-trivial 1-dimensional local system on $\AX$. The
proof of the Lemma will be given in several steps:
\\
\\
\textbf{Step 1.} \textbf{Vanishing}. We want to show that
$\coH^i(pr_!\SE) = 0$ for $i=0, \, 2$.
\\
\\
By definition:
$$
\coH^0(pr_!\SE) = \Gamma(\AYY,j_!\SE),
$$
where $j:\AX \hookrightarrow \AYY$ is the imbedding of $\AX$ into
a compact curve $\AYY$. The statement follows since:
$$
\Gamma(\AYY,j_!\SE) = \Hom(\Qlb,j_!\SE)
$$
and it is easy to see that any non-trivial morphism $\Qlb \to
j_!\SE$ should be an isomorphism, hence $\Hom(\Qlb,j_!\SE) = 0$.
\\
\\
For the second cohomology we have:
$$
\coH^2(pr_!\SE) = \coH^{-2}(D pr_!\SE)^* = \coH^{-2}(pr_\ast
D\SE)^* = \Gamma(\AX,D\SE[-2])^*,
$$
where $D$ denotes the Verdier duality functor\footnote{In the
vector bundle interpretation, one might think on $D$ as the
operation of taking the dual vector bundle with the dual
connection.} and $[-2]$ means translation functor. The first
equality follows from the definition of $D$, the second equality
is the Poincar\'{e} duality and the third equality easily follows
from the definitions. Again, since the sheaf $D\SE[-2]$ is a
non-trivial 1-dimensional local system on $\AX$ then:
$$
\Gamma(\AX,D\SE[-2]) = \Hom(\Qlb,D\SE[-2]) = 0.
$$
\textbf{Step 2.} \textbf{Dimension}. We claim that dim
$\mathrm{H}^1(pr_!\SE) = 2$.
\\
\\
The (topological) \textit{Euler characteristic} $\chi(pr_!\SE)$ of
the sheaf $pr_!\SE$ is the integer defined by the formula:
$$
\chi(pr_!\SE) = \sum_i  (-1)^i \dim \; \coH^i(pr_!\SE).
$$
Hence from the vanishing of cohomologies (Step 1) we deduce:
\\
\\
\textbf{Substep 2.1.} It is enough to show that $\chi(pr_!\SE) =
-2$.
\\
\\
The actual computation of the Euler characteristic $\chi(pr_!\SE)$
is done using the \textit{Ogg-Shafarevich-Grothendieck formula}
\cite{D3}:
\begin{eqnarray}\label{OSG}
\mathrm{rk}(\SE)\cdot\chi(pr_!\Qlb) - \chi(pr_!\SE) = \sum_{y \in
\AYY\smallsetminus\AX} \Swan_y(\SE).
\end{eqnarray}
Here $\mathrm{rk}(\SE) = 1$ is the rank of the sheaf $\SE$, $\Qlb$
denotes the constant sheaf on $\AX$, and $\AYY$ is some compact
curve containing $\AX$. In other words, this formula expresses the
difference of $\mathrm{rk}(\SE)\cdot\chi(pr_!\Qlb)$ from
$\chi(pr_!\SE)$ as a sum of local contributions, called
\textit{Swan conductors}. We will not to give here the formal
definition of the Swan conductor (see \cite{K, KL, L}), but
instead we will formulate some of the properties and known results
that are needed for our calculations.
\\
\\
Next, we take $\AYY = \P1$. Having that, $\chi(pr_!\Qlb) = -1$,
and using formula (\ref{OSG}) we get:
\\
\\
\textbf{Substep 2.2.} It is enough to show that: $\Swan_0(\SE) +
\Swan_1(\SE) + \Swan_\infty(\SE) = 1.$
\\
\\
We would like now to understand the affect of tensor product on
the Swan conductor. Choose a point $y\in \AYY \smallsetminus \AX$,
and let ${\cal L}_1 ,\, {\cal L}_2 \in \Db(\AX)$ be two sheaves
with ${\cal L}_1$ being \textit{lisse} (smooth) in a neighborhood
of $y$. We have:

\begin{equation}\label{Swan-tensor}
\Swan_y({\cal L}_1 \otimes {\cal L}_2) = \mathrm{rk}({\cal
L}_1)\cdot\Swan_y({\cal L}_2).
\end{equation}

In particular, using property (\ref{Swan-tensor}), and the
explicit formula (\ref{SE}) of the sheaf $\SE$, we deduce that:
\begin{eqnarray*}
\Swan_1(\SE) & = & \Swan_\infty (\Sartin),\\
\Swan_\infty (\SE) & = & \Swan_\infty (\Skummer), \\
\Swan_0(\SE) & = & \Swan_0(\Skummer).
\end{eqnarray*}

\textbf{Substep 2.3.} We have $\Swan_\infty (\Sartin) = 1$, and
$\Swan_\infty (\Skummer) + \Swan_0(\Skummer) = 0$.

Applying the Ogg-Shafarevich-Grothendieck formula to the
Artin-Schreier sheaf $\Sartin$ on $\bA^1$ and the projection
$pr:\bA^1 \lto pt$ we find that:
\begin{eqnarray}\label{SwanArtin}
\Swan_\infty (\Sartin) = \chi(pr_!\Qlb)  - \chi(pr_!\Sartin)  = 1
- 0 = 1.
\end{eqnarray}
Finally, we apply the formula (\ref{OSG}) to the sheaf $\Skummer$
on $\Gm$ and the projection $pr:\Gm \lto pt$ and conclude:
\begin{eqnarray}\label{SwanKummer}
\Swan_\infty(\Skummer) + \Swan_0(\Skummer) = \chi(pr_!\Qlb) -
\chi(pr_!\Skummer)
 = 0 - 0 = 0.
\end{eqnarray}
Note that, in (\ref{SwanArtin}) and (\ref{SwanKummer}) we use the
fact that $pr_!\Sartin$ and $pr_!\Skummer$ are the $0-$objects in
$\Db(pt)$.
\\
\\
This completes the computations of the Vanishing Lemma. $\EProof$
\end{appendix}
%
%

\bigskip
\noindent Department of Mathematics, Technion - Israel Institute
of Technology, Haifa 32000, Israel.\\
{\it E-mail address}: shamgar@math.tau.ac.il
\\
\\
\noindent Department of Mathematics, Tel Aviv University, Tel Aviv
69978, Israel.\\
{\it E-mail address}: hadani@post.tau.ac.il

\bigskip
\center \textsf{Courant Institute, September 15, 2004.}
\end{document}